\documentclass[a4paper,11pt]{article}
\usepackage{jheppub}

\usepackage{amsmath,amssymb}
\usepackage{physics}
\usepackage{mathrsfs}
\usepackage{tikz}
\usetikzlibrary{cd}

\usepackage{slashed}

\usepackage{array}
\usepackage{tabularx}
\usepackage{booktabs}
\usepackage{subcaption}

\title{The Helical SYK Model and Emergent Infrared Integrability}

\author[a,b]{Gustavo Valdivia-Mera,}
\author[a]{Bhavay Tyagi,}
\author[a]{Eric R. Bittner}
\author[a,b]{and Pavan Hosur}

\affiliation[a]{Department of Physics, University of Houston,
Houston, TX 77204, USA}
\affiliation[b]{Texas Center for Superconductivity, University of Houston,
Houston, TX 77204, USA}

\emailAdd{gvaldiviamera@uh.edu}
\emailAdd{btyagi@uh.edu}
\emailAdd{bittner@uh.edu}
\emailAdd{phosur@uh.edu}

\abstract{We construct a helical generalization of the Sachdev--Ye--Kitaev (SYK) model in \(1+1\) dimensions, built from left- and right-moving Majorana fermions with local quartic interactions and random couplings in flavor-chirality space. These interactions organize into a symmetry-controlled hierarchy of quartic chirality sectors. At the most restrictive end of this hierarchy, symmetry forces the quartic structure into a density--density form, which admits an exact solution using bosonization, rendering the theory integrable. Once the full quartic helical interaction space is allowed, including purely chiral, chirality-balanced, and chirality-imbalanced sectors, this symmetry-protected integrable structure is lost. Nevertheless, the large-\(N\) infrared limit remains analytically tractable through short-distance selection rules and disorder averaging. Using conformal perturbation theory about the free fixed point, we show that the entire interaction space is marginally irrelevant, and the theory thus becomes free and integrable in the IR.
}

\begin{document}
    \maketitle
\flushbottom

\newpage
\section{Introduction}
Minimal models of quantum chaos have become central theoretical laboratories for probing many-body dynamics in strongly interacting quantum systems. A widely acknowledged characterization of quantum chaos is provided by the \textit{Bohigas--Giannoni--Schmit} (BGS) conjecture \cite{bohigas1984characterization}, which relates chaotic quantum spectra to the universal spectral statistics of random matrix theory \cite{haake1991quantum,mehta2004random}. Furthermore, by isolating universal features of scrambling \cite{xu2024scrambling}, entanglement \cite{wang2004entanglement}, thermalization \cite{srednicki1994chaos,deutsch2018eigenstate}, complexity \cite{parker2019universal}, free probability \cite{fava2025designs,jindal2024generalized}, and emergent low-energy structure, such models have shaped developments across quantum field theory, quantum gravity, and condensed matter physics. One of the most extensively studied examples is the Sachdev--Ye--Kitaev (SYK) model \cite{sachdev1992gapless,kitaev2015simple1,kitaev2015simple2,maldacena2016remarks}, a \(0+1\)-dimensional quantum mechanical system of interacting Majorana fermions with Gaussian random all-to-all interactions. For even \(q\), its action may be written as
\begin{align}
    S_{\text{SYK}}
    =
    \int dt
    \left[
    \frac{i}{2}\sum_{j=1}^{N}\psi_j \partial_t \psi_j
    -
    i^{q/2}
    \sum_{1\leq i_1<\cdots<i_q\leq N}
    J_{i_1\cdots i_q}
    \psi_{i_1}\cdots \psi_{i_q}
    \right].
\end{align}
A distinctive feature of the SYK model is its analytic tractability in the large-\(N\) limit, where \(N\) denotes the number of Majorana fermions. This unique feature, together with chaotic dynamics and its emergent infrared (IR) conformal symmetry, has made SYK a useful toy model for many-body quantum chaos and holography \cite{shenker2014black,shenker2014multiple,maldacena2016bound,shenker2015stringy,reynolds2016butterflies,almheiri2015models,sachdev2015bekenstein,maldacena2016conformal,jensen2016chaos,sekino2008fast,engelsoy2016investigation,cvetivc2016ads2}. Comprehensive reviews of the model and its applications can be found in Refs.~\cite{sarosi2017ads,rosenhaus2019introduction,chowdhury2022sachdev,jha2025introduction}. 

In this paper, we consider a \(1+1\)-dimensional generalization of the SYK model built from counter-propagating left- and right-moving Majorana fermions. This left--right structure is what we refer to as helical. Higher-dimensional generalizations of SYK have been explored from several perspectives \cite{berkooz2017higher,berkooz2017comments,lian2019chiral,turiaci2017towards,murugan2017more}, and provide part of the motivation for our analysis. In particular, two important \(1+1\)-dimensional models have been studied separately: the purely chiral SYK (CSYK) model \cite{lian2019chiral}, and the chirality-balanced Random Thirring (RT) model \cite{berkooz2017comments}. Our aim is to place these previously isolated studies inside a single helical framework, where the most general local quartic interaction can involve purely chiral, chirality-balanced, and chirality-imbalanced structures. Such counter-propagating Majorana modes also arise naturally in condensed-matter settings, for example as helical Majorana modes in class-DIII topological superconductors; see the subsection ``Class DIII: Helical Majorana fermions'' in Ref.~\cite{teo2010topological}. Superconducting vortices in Weyl semimetals provide another closely related setting where Majorana modes can arise \cite{lu2011supersymmetric,giwa2021fermi}. Related connections between \(1+1\)-dimensional theories and black-hole physics have also been studied in Refs.~\cite{guica2009kerr,berkooz2007fermi,berkooz2015chiral}.

For our model we allow the most general local quartic interaction compatible with this helical field content, with Gaussian random couplings in flavor-chirality space. These couplings have zero mean, and their sector-dependent disorder strengths are denoted by $
J_{4L},\; J_{4R},\; J_{2L2R},\; J_{3L1R},\; J_{1L3R}$,
corresponding respectively to the purely chiral, chirality-balanced, and chirality-imbalanced quartic sectors. Unlike spatially disordered generalizations \cite{gu2017local,gross2017generalization,berkooz2017higher}, note that randomness here is only in the flavor-chirality coupling tensor and the theory remains translationally invariant in spacetime. Since a Majorana fermion in \(1+1\) dimensions has scaling dimension \(\Delta_{\psi}=1/2\), every local quartic interaction has classical scaling dimension $\Delta_{\psi^4}=2$, equal to the spacetime dimension. The interactions in the helical SYK model therefore perturb the free chiral fixed point by a large set of classically marginal operators, and the fate of the theory in the IR is determined by quantum corrections. This is qualitatively different from the original $0+1$-dimensional SYK model, where the random interaction is relevant and drives the theory to a strongly interacting and maximally chaotic IR regime.

The helical model highlights two distinct ways in which integrability can appear, and separates them within a unified framework. The first comes from the purely CSYK model \cite{lian2019chiral}. There, the \(q=4\) chiral interaction is exactly marginal, but integrability via bosonization appears only in small finite-\(N\) cases, such as \(N=4,5,6\), where \(N\) denotes the number of Majorana flavors. At large \(N\), the theory instead exhibits signatures of chaotic behavior. In the helical model, the organizing principle is not simply the number of flavors, but the symmetry structure of the interaction space. For even \(N\), the Majorana flavors can be paired into \(SO(2)\)-invariant blocks, and in the maximally constrained symmetry sector the allowed quartic interactions are forced into density--density form. Bosonization then maps the theory to a quadratic chiral-boson theory that can be canonically diagonalized. Thus, unlike purely chiral SYK, the finite-coupling integrable structure in the helical model persists for arbitrary even \(N\), but only inside the symmetry-protected density--density subspace. Once these symmetry constraints are relaxed and generic flavor--chirality-random quartic tensors are allowed, the density--density structure is lost, and the finite-coupling diagonalization mechanism no longer applies.

The second form of integrability appears in the disorder-averaged large-\(N\) RG flow. The RT model \cite{berkooz2017comments} shows that the chirality-balanced random interaction is marginally irrelevant after disorder averaging, with the first nonvanishing averaged flow appearing beyond leading order. For the full helical theory, we ask what happens to this RG structure when purely chiral, chirality-balanced, and chirality-imbalanced random quartic tensors are all present. Remarkably, the coupled flow is dominated by the chirality-balanced \(\{2L2R\}\) sector and our final result is elegantly simple. In the conventions of this paper, the leading nonvanishing beta functions for all quartic chirality sectors are
\begin{equation}
    \mu\frac{dJ_{\mathcal S}}{d\mu}
    =
    \frac{n_{\mathcal S}}{16\pi^2}\,
    J_{\mathcal S}J_{2L2R}^{\,2}
    +O(J^5),
    \qquad
    \mathcal S\in\{4L,3L1R,2L2R,1L3R,4R\}.
\end{equation}
Here \(\mu\) is the RG scale and \(J_{\mathcal S}\) is the disorder strength in sector \(\mathcal S\). The integer \(n_{\mathcal S}\) equals one for the purely chiral sectors \(\{4L\}\) and \(\{4R\}\), and equals two for the chirality-balanced sector \(\{2L2R\}\) and the chirality-imbalanced sectors \(\{3L1R\}\) and \(\{1L3R\}\). It simply counts the leading topologies that contribute to the flow. Despite the many possible quartic tensors and contraction channels, the disorder-averaged RG flow is governed by a single mechanism. The chirality-balanced \(\{2L2R\}\) sector drives every quartic disorder strength through \(J_{2L2R}^{\,2}\). Thus, for non-zero values of \(J_{2L2R}\), all quartic disorder strengths flow logarithmically toward zero in the IR. The generic finite-coupling helical theory is not integrable. Yet in the disorder-averaged large-\(N\) IR limit, the theory flows back to the free helical fixed point, so integrability re-emerges.

The remainder of the paper is organised as follows. In Sec.~\ref{sec:model}, we define the \(1+1\)-dimensional helical SYK model, introduce the five quartic chirality sectors, and specify the corresponding Gaussian disorder ensemble. In Sec.~\ref{sec:symmetry-hierarchy}, we classify the quartic interaction space using internal flavor-chirality symmetries, obtaining a hierarchy in which relaxing the symmetry enlarges the allowed interaction space from an integrable density--density subspace to the full helical quartic theory. In Sec.~\ref{sec:int-mso2lr}, we analyze the maximally constrained density--density theory using bosonization and canonical diagonalization. This establishes an exactly integrable sector of the helical theory and shows that the integrability mechanism is symmetry-protected, while also explaining why this mechanism fails once generic quartic interactions are allowed. In Sec.~\ref{sec:RG-flow}, we study the random quartic helical theory using conformal perturbation theory (CPT) around the free chiral fixed point. We derive the short-distance operator product expansion (OPE) selection rules, show that the disorder-averaged beta functions vanish at second order, compute the leading nonvanishing cubic flow, and analyze the resulting large-\(N\) IR behavior. Finally, in Sec.~\ref{sec:conclusions}, we make some concluding remarks. Relevant technical details are collected in the appendices. A slightly less technical summary of the main results is given in Table~\ref{tab:main-results}.
\begin{table}[h]
\centering
\renewcommand{\arraystretch}{1.25}
\begin{tabular}{p{0.30\textwidth}|p{0.62\textwidth}}
\hline
\textbf{Section} & \textbf{Main result} \\
\hline
Secs.~\ref{sec:model} and \ref{sec:symmetry-hierarchy}: 
Helical SYK model and symmetry hierarchy
&
We define the helical SYK model where the most general local quartic interaction decomposes into five sectors,
\(\{4L,3L1R,2L2R,1L3R,4R\}\). Internal symmetry constraints organize these sectors into a hierarchy, where relaxing the symmetry enlarges the allowed interaction space from a density--density subspace to the full helical quartic theory. \\
\hline
Sec.~\ref{sec:int-mso2lr}: 
Symmetry protected exact integrability and its loss
&
In the maximally constrained theory, the allowed interactions are forced into density--density form. Bosonization maps this sector to a quadratic chiral-boson theory, which can be canonically diagonalized into decoupled chiral normal modes. This gives an exactly integrable structure. Once generic quartic interactions are allowed, the density--density structure is lost and the diagonalization mechanism, ensuring exact solvability, no longer applies. \\
\hline
Sec.~\ref{sec:RG-flow}: 
Disorder-averaged RG flow and the emergent IR integrability
&
Using CPT around the free helical fixed point, we show that chirality selection rules and Gaussian disorder averaging cause the first nonvanishing averaged flow to appear at cubic order. The chirality-balanced \(2L2R\) sector runs autonomously and drives the remaining sector flows. For \(J_{2L2R}\neq0\), the theory flows logarithmically back toward the free helical conformal fixed point in the deep IR and hence, integrability re-emerges. \\
\hline
\end{tabular}
\caption{Summary of the main results of the helical SYK model studied in this work.}
\label{tab:main-results}
\end{table}

\section{The \texorpdfstring{\(1+1\)}{1+1}-dimensional helical SYK model}
\label{sec:model}

We consider a helical extension of the SYK model in \(1+1\) dimensions. The model is formulated in terms of left- and
right-moving Majorana fermions with the most general local quartic
helical interactions and random couplings in
flavor-chirality space. While the purely chiral and chirality-balanced
quartic interactions have been studied separately
\cite{lian2019chiral,berkooz2017comments}, we
incorporate all quartic chirality sectors, including the purely chiral,
chirality-balanced, and chirality-imbalanced sectors.

The fundamental degrees of freedom are Majorana fermions
\(\psi_{i,\nu}(t,x)\), labeled by a flavor index \(i=1,\dots,N\) and a
chirality label \(\nu\in\{L,R\}\). In Lorentzian signature, their free
chiral action is given by
\begin{equation}
S_0
=
\int dt\,dx\,
\frac{i}{2}\left(\sum_{i=1}^N
\psi_{i,R}(\partial_t+\partial_x)\psi_{i,R}
+
\sum_{i=1}^N
\psi_{i,L}(\partial_t-\partial_x)\psi_{i,L}
\right),
\end{equation}
where the fermions satisfy the canonical equal-time anticommutation relations
\begin{equation}
\{\psi_{i,\nu}(t,x),\psi_{j,\nu'}(t,y)\}
=
\delta_{ij}\delta_{\nu\nu'}\delta(x-y),
\end{equation}
and have scaling dimension \(\Delta_\psi=1/2\) at the free chiral fixed
point. A local quartic interaction, schematically of the form
\(\mathcal{O}\sim\psi\psi\psi\psi\), therefore has scaling dimension $\Delta_\mathcal{O}=2$, equal to the
spacetime dimension of the theory, making such interactions classically
marginal deformations of the free fixed point.

To treat all quartic chirality sectors uniformly, we combine flavor and
chirality into a composite, multi-index label given by 
\begin{equation}
A
\equiv
(i_1,\nu_1;\,i_2,\nu_2;\,i_3,\nu_3;\,i_4,\nu_4).
\end{equation}
The quartic interaction can then be written in the compact form
\begin{equation}
S_{\mathrm{int}}
=
\int dt\,dx\,
\sum_A
J_A\,
\mathcal O_A(t,x).
\label{eq:model-Sint}
\end{equation}
For each multi-index \(A\), the corresponding coupling component and
local quartic Majorana monomial are defined by
\begin{align}
J_A
&\equiv
J_{(i_1,\nu_1)(i_2,\nu_2)(i_3,\nu_3)(i_4,\nu_4)},\\
\mathcal O_A(t,x)
&\equiv
\psi_{i_1,\nu_1}(t,x)
\psi_{i_2,\nu_2}(t,x)
\psi_{i_3,\nu_3}(t,x)
\psi_{i_4,\nu_4}(t,x).
\end{align}
Since the coupling components are real and totally antisymmetric under
exchange of the composite labels \((i,\nu)\), note that only local quartic
terms involving four distinct composite labels contribute.

Although the compact notation in Eq.~\eqref{eq:model-Sint} is useful, each local quartic monomial has a definite
chirality content. The interaction therefore admits a decomposition into quartic
chirality sectors \(\mathcal S\), classified by their left- and
right-moving field content:
\begin{equation}
\mathcal S\in\{4L,3L1R,2L2R,1L3R,4R\},
\end{equation}
and so the interaction can be compactly written as
\begin{equation}
S_{\mathrm{int}}
=
\sum_{\mathcal S}S_{\mathcal S}.
\label{eq:sector-decomposition}
\end{equation}

To avoid overcounting, each sector contribution is written in terms of
canonical representatives of the quartic monomials. In the purely chiral
sectors, this amounts to ordering the four flavor indices. In the presence of both
chiralities, we first fix a canonical chirality order and
then order the flavor indices within each chiral subset. The signs generated by permuting fermions are compensated by the total antisymmetry of the coupling tensor, so each unordered set of composite labels corresponds to a single canonical monomial.   The numerical permutation factors associated with passing from the 
unrestricted tensor notation to this canonical basis are absorbed into the definition of 
the canonical coupling components.

With this convention, the contribution from the purely left-moving
sector \(\{4L\}\) is
\begin{equation}
S_{4L}
=
\int dt\,dx\;
\sum_{1\le i_1<i_2<i_3<i_4\le N}
J_{(i_1,L)(i_2,L)(i_3,L)(i_4,L)}\,
\psi_{i_1,L}\psi_{i_2,L}\psi_{i_3,L}\psi_{i_4,L},
\end{equation}
and the corresponding
contribution for the purely right-moving sector \(\{4R\}\) is obtained by the exchange \(L\leftrightarrow R\). For the
chirality-imbalanced \(\{3L1R\}\) sector, one has
\begin{equation}
S_{3L1R}
=
\int dt\,dx\;
\sum_{\substack{1\le i_1<i_2<i_3\le N\\ i_4=1,\dots,N}}
J_{(i_1,L)(i_2,L)(i_3,L)(i_4,R)}\,
\psi_{i_1,L}\psi_{i_2,L}\psi_{i_3,L}\psi_{i_4,R},
\end{equation}
and the \(\{1L3R\}\) sector is treated analogously. Finally, for the
chirality-balanced \(\{2L2R\}\) sector, one obtains
\begin{equation}
S_{2L2R}
=
\int dt\,dx\;
\sum_{\substack{1\le i_1<i_2\le N\\ 1\le i_3<i_4\le N}}
J_{(i_1,L)(i_2,L)(i_3,R)(i_4,R)}\,
\psi_{i_1,L}\psi_{i_2,L}\psi_{i_3,R}\psi_{i_4,R}.
\end{equation}

This canonical organization also fixes the number of independent
coupling components in each quartic chirality sector. Counting these canonical representatives gives
\begin{equation}
d_{4L}=d_{4R}=\binom{N}{4},
\qquad
d_{3L1R}=d_{1L3R}=\binom{N}{3}N,
\qquad
d_{2L2R}=\binom{N}{2}^{2}.
\end{equation}

Once the canonical representatives have been fixed, the disorder
ensemble is specified directly on the corresponding coupling components.
We take them to be centered Gaussian random variables, statistically independent between distinct chirality sectors, with covariance
\begin{equation}
\overline{J_A}=0,
\qquad
\overline{J_AJ_B}
=
C_{\mathcal S}\,
J_{\mathcal S}^{2}\,
\delta_{AB},
\qquad
A,B\in\mathcal S.
\end{equation}
Here \(J_{\mathcal S}\) is the positive disorder strength associated with the sector 
\(\mathcal S\). The constants \(C_{\mathcal S}\) fix the sector-dependent normalization of the disorder ensemble. 
They are chosen so that all microscopic covariances scale uniformly as \(N^{-3}\), while the 
remaining numerical factors define the convention used to compare disorder strengths across 
different quartic chirality sectors. For the purely chiral sectors, we take
\begin{equation}
C_{4L}=C_{4R}=\frac{3!}{N^3},
\end{equation}
whereas for the chirality-balanced and imbalanced sectors we take
\begin{equation}
C_{2L2R}=C_{3L1R}=C_{1L3R}=\frac{2}{N^3}.
\end{equation}

The canonical sector decomposition and the sector-wise Gaussian ensemble therefore define the full local quartic helical interaction space considered in this work. With this space fixed, in the next section, we look at how successive symmetry constraints restrict the allowed quartic interactions.
\section{Symmetry hierarchy of quartic helical interactions}
\label{sec:symmetry-hierarchy}

Having defined the full local quartic helical interaction space, we now ask how much of this space remains once internal symmetries are imposed. Stronger symmetries restrict the model to smaller invariant subspaces and as those symmetries are relaxed, the allowed interaction space expands. We begin at the most constrained end of this hierarchy and progressively loosen the symmetry requirements until the allowed interaction space reaches the full quartic helical interaction space introduced above. Throughout this section, by a symmetry of an interaction subspace we mean a symmetry of each fixed-coupling interaction in that subspace. This should not be confused with statistical symmetries that may emerge only after disorder averaging in later sections.

The relevant transformations are spacetime independent global flavor rotations acting independently on the two chiralities and are given by
\begin{equation}
\psi'_{i,\nu}
=
\sum_j O^{(\nu)}_{ij}\,\psi_{j,\nu},
\qquad
O^{(\nu)}\in SO(N),
\qquad
\nu\in\{L,R\}.
\end{equation}
where \(\nu\) indicates 
independently chosen flavour rotation for each chirality. Since the free action is invariant under these transformations, all nontrivial symmetry constraints arise from the quartic interaction, and so a generic quartic monomial transforms as
\begin{align}
\psi'_{i_1,\nu_1}
\psi'_{i_2,\nu_2}
\psi'_{i_3,\nu_3}
\psi'_{i_4,\nu_4}
&=
\sum_{j_1,j_2,j_3,j_4}
O^{(\nu_1)}_{i_1j_1}
O^{(\nu_2)}_{i_2j_2}
O^{(\nu_3)}_{i_3j_3}
O^{(\nu_4)}_{i_4j_4}
\psi_{j_1,\nu_1}
\psi_{j_2,\nu_2}
\psi_{j_3,\nu_3}
\psi_{j_4,\nu_4}.
\label{eq:quartic-transform}
\end{align}
The invariant quartic structures are obtained by applying this transformation law to each symmetry choice in the hierarchy. We begin with the maximally constraining case, where the flavor rotations act independently on two-dimensional Majorana pairs of each chirality.

\subsection{The maximally constraining symmetry group:
\([SO(2)]_L^{N/2}\times [SO(2)]_R^{N/2}\)}
 This is defined for even \(N\), so that the Majorana
flavors of each chirality can be grouped into \(N/2\) pairs. Introducing
a pair index \(a=1,\dots,N/2\), we can write
\begin{equation}
\Psi_{a,\nu}^{T}
=
(\psi_{2a-1,\nu},\psi_{2a,\nu}).
\end{equation}
Each doublet transforms under an independent \(SO(2)\) matrix
\(O_{a,\nu}\) acting separately on each pair-chirality
block \((a,\nu)\).

Within a given doublet, the natural invariant building block is the
bilinear
\begin{equation}
n_{a,\nu}
=
i\,\psi_{2a-1,\nu}\psi_{2a,\nu},
\end{equation}
which we call the fermionic density associated with the Majorana pair, whose
invariance follows directly from Eq.~\eqref{eq:quartic-transform}
specialized to the two-dimensional flavor subspace associated with index
\(a\). Indeed,
\begin{align}
\label{MajoranaSO2transform}
\psi'_{2a-1,\nu}\psi'_{2a,\nu}
&=
\sum_{c,d\in\{2a-1,2a\}}
O^{(\nu)}_{2a-1,c}
O^{(\nu)}_{2a,d}\,
\psi_{c,\nu}\psi_{d,\nu}.
\end{align}
Since the fermionic fields anticommute, the product
\(\psi_{c,\nu}\psi_{d,\nu}\) is antisymmetric in \(c,d\). Within a
two-dimensional pair subspace, the antisymmetric product of two Majoranas is unique. It is therefore proportional to the
Levi--Civita tensor. Note that with \(\epsilon_{2a-1,2a}=1\) we have
\begin{equation}
\psi_{c,\nu}\psi_{d,\nu}
=
\epsilon_{cd}\,
\psi_{2a-1,\nu}\psi_{2a,\nu},
\qquad
c,d\in\{2a-1,2a\}.
\end{equation}
Substituting this into Eq. \eqref{MajoranaSO2transform}, we have
\begin{align}
\psi'_{2a-1,\nu}\psi'_{2a,\nu}
&=
\left[
\sum_{c,d\in\{2a-1,2a\}}
O^{(\nu)}_{2a-1,c}
O^{(\nu)}_{2a,d}\,
\epsilon_{cd}
\right]
\psi_{2a-1,\nu}\psi_{2a,\nu},
\end{align}
where the expression in brackets is the determinant of the corresponding
\(2\times2\) matrix \(O_{a,\nu}\). This
determinant is one, since \(O_{a,\nu}\in SO(2)\), giving us
\begin{equation}
\psi'_{2a-1,\nu}\psi'_{2a,\nu}
=
\psi_{2a-1,\nu}\psi_{2a,\nu}.
\end{equation}
Thus the bilinears \(n_{a,\nu}\) are invariant, under the independent \(SO(2)\) rotation
acting on the pair-chirality block \((a,\nu)\), and generate a distinguished class of
density--density quartic structures. Given two distinct pair-chirality
labels \((a,\nu_a)\) and \((b,\nu_b)\), one has
\begin{equation}
(\psi_{2a-1,\nu_a}\psi_{2a,\nu_a})
(\psi_{2b-1,\nu_b}\psi_{2b,\nu_b})
=
-\,n_{a,\nu_a}n_{b,\nu_b},
\end{equation}
Thus, any density--density interaction built from these products is manifestly invariant under the independent \(SO(2)\) rotations acting on the corresponding pair-chirality blocks.

The full symmetry group in this maximally restrictive case is
\begin{equation}
[SO(2)]_L^{N/2}\times [SO(2)]_R^{N/2}
=
\left(\prod_{a=1}^{N/2}[SO(2)]_{(a,L)}\right)
\times
\left(\prod_{b=1}^{N/2}[SO(2)]_{(b,R)}\right),
\end{equation}
which acts independently on every pair-chirality block.

The resulting constraints are stronger than a simple chirality-parity condition. Since each pair-chirality block transforms independently, an invariant quartic monomial must contain either zero fermions from a given block or the full antisymmetric pair \(\psi_{2a-1,\nu}\psi_{2a,\nu}\). Thus the chirality-imbalanced sectors \(\{3L1R\}\) and \(\{1L3R\}\) are excluded, and even within the \(\{4L\}\), \(\{4R\}\), and \(\{2L2R\}\) sectors only density--density structures built from the invariant bilinears \(n_{a,\nu}\) survive.

\subsection{Symmetry reduction and enlargement of the quartic interaction space}

We now relax the symmetry constraints. Starting from the pairwise
\(SO(2)\) symmetry, we first restrict each continuous rotation to the
discrete angles \(\theta_{a,\nu}\in\{0,\pi\}\). At the level of each
two-dimensional flavor block, this gives
\begin{equation}
O_{a,\nu}
=
\pm I_2,
\end{equation}
so the two Majorana fermions in the same pair acquire the same sign.

This intermediate step still acts independently on each pair-chirality
block. The resulting group is
\begin{equation}
\mathbb Z_{2,L}^{N/2}\times \mathbb Z_{2,R}^{N/2}
=
\left(\prod_{a=1}^{N/2}\mathbb Z_2^{(a,L)}\right)
\times
\left(\prod_{b=1}^{N/2}\mathbb Z_2^{(b,R)}\right).
\end{equation}
Passing to the corresponding global chirality parities amounts to
removing the dependence on the pair index. Thus all fermions of a given
chirality transform with the same sign. Introducing two independent
parameters \(\eta_L,\eta_R\in\{\pm1\}\), we take
\begin{equation}
\psi_{i,\nu}
\longrightarrow
\eta_\nu\,\psi_{i,\nu},
\qquad
\nu\in\{L,R\}.
\end{equation}
This defines a global \(\mathbb Z_{2,L}\times\mathbb Z_{2,R}\)
symmetry. If \(\mathcal O_{p_L,p_R}\) denotes a quartic monomial
containing \(p_L\) left-moving and \(p_R\) right-moving fields, then
\begin{equation}
\mathcal O_{p_L,p_R}
\longrightarrow
\eta_L^{\,p_L}\eta_R^{\,p_R}\,
\mathcal O_{p_L,p_R}.
\end{equation}
Invariance for arbitrary \(\eta_L\) and \(\eta_R\) requires
\begin{equation}
p_L\in 2\mathbb Z,
\qquad
p_R\in 2\mathbb Z.
\end{equation}
Thus the purely chiral \(\{4L\}\), \(\{4R\}\), and chirality-balanced \(\{2L2R\}\) sectors are allowed,
whereas the chirality-imbalanced sectors \(\{3L1R\}\) and
\(\{1L3R\}\) are forbidden. In contrast with the maximally restrictive
case, the restriction to density--density interactions is no longer
enforced: general quartic interactions within the allowed sectors are
compatible with the global \(\mathbb Z_{2,L}\times\mathbb Z_{2,R}\)
symmetry.

The final reduction identifies the two signs,
\(\eta_L=\eta_R\equiv\eta\), so that all fermions transform uniformly,
\begin{equation}
\psi_{i,\nu}
\longrightarrow
\eta\,\psi_{i,\nu},
\qquad
\eta\in\{\pm1\}.
\end{equation}
This leaves a single global \(\mathbb Z_2\) symmetry. A quartic monomial
then transforms as
\begin{equation}
\mathcal O_{p_L,p_R}
\longrightarrow
\eta^{\,p_L+p_R}\,
\mathcal O_{p_L,p_R}.
\end{equation}
Since every quartic monomial contains four fermion fields, it is automatically even under this diagonal \(Z_2\). Therefore the diagonal \(Z_2\) imposes no further restriction on the quartic interaction space.

The symmetry hierarchy relevant for classifying quartic chirality
sectors is therefore
\begin{equation}
[SO(2)]_L^{N/2}\times [SO(2)]_R^{N/2}
\;\longrightarrow\;
\mathbb Z_{2,L}\times \mathbb Z_{2,R}
\;\longrightarrow\;
\mathbb Z_2.
\end{equation}
Along this hierarchy, the invariant interaction space expands from the
density--density subspace selected by the maximally restrictive symmetry
group to the full \(\{4L\}\), \(\{4R\}\), and \(\{2L2R\}\) quartic
chirality sectors under \(\mathbb Z_{2,L}\times\mathbb Z_{2,R}\).
Under the diagonal \(\mathbb Z_2\), it becomes the complete quartic
interaction space, including the chirality-imbalanced sectors
\(\{3L1R\}\) and \(\{1L3R\}\). This structure is summarized in
Table~\ref{tab:symmetry-hierarchy} and provides the symmetry
classification used in the dynamical analysis below.

\begin{table}[t]
\centering
\small
\renewcommand{\arraystretch}{1.35}

\begin{tabularx}{\textwidth}{
>{\raggedright\arraybackslash}p{4.3cm}
>{\centering\arraybackslash}p{2cm}
>{\raggedright\arraybackslash}X}
\toprule
\textbf{Symmetry group} &
\textbf{Defined for} &
\textbf{Invariant quartic interactions} \\
\midrule

$[SO(2)]^{N/2}_L\times [SO(2)]^{N/2}_R$
&
even $N$
&
density--density subspace within \(\{4L\}\), \(\{4R\}\), and
\(\{2L2R\}\):
\(n_{a,\nu_a}n_{b,\nu_b}\)
\\[6pt]

$\mathbb Z_{2,L}\times \mathbb Z_{2,R}$
&
any $N$
&
\(\{4L\},\,\{4R\},\,\{2L2R\}\)
\\[6pt]

$\mathbb Z_2$
&
any $N$
&
\(\{4L\},\,\{4R\},\,\{2L2R\},\,\{3L1R\},\,\{1L3R\}\)
\\

\bottomrule
\end{tabularx}

\caption{
Hierarchy of symmetry constraints on the quartic interaction. As the
symmetry is reduced, the invariant interaction space expands from
density--density interaction terms to the full quartic interaction
space.
}
\label{tab:symmetry-hierarchy}
\end{table}
\section{Symmetry-protected integrability and its loss in the full helical theory}
\label{sec:int-mso2lr}

We now turn from the symmetry classification of the previous section to its dynamical consequences. The maximally constrained subspace is not merely smaller, it is also algebraically simpler because its quartic
interactions are forced into density--density form. This structure is special because density--density interactions become quadratic after bosonization, opening the possibility of an exact canonical diagonalization. The mechanism is already manifest in the minimal realization \(N=2\).
For two flavors, the three symmetry choices discussed above,
\begin{equation*}
[SO(2)]_L^{N/2}\times [SO(2)]_R^{N/2},
\qquad
\mathbb Z_{2,L}\times\mathbb Z_{2,R},
\qquad
\mathbb Z_2,
\end{equation*}
all lead to the same admissible quartic interaction space. The
antisymmetry of the coupling tensor forbids any quartic term with
three or four fermions of the same chirality, since there are not
enough distinct flavor labels. The quartic interaction space therefore
collapses to a single nonvanishing structure belonging to the chirality-balanced
sector \(\{2L2R\}\). With the density variables
\(n_{1,L}=i\psi_{1,L}\psi_{2,L}\) and
\(n_{1,R}=i\psi_{1,R}\psi_{2,R}\), this interaction takes the form
\begin{equation}
S_{\mathrm{int}}
=
-\int dt\,dx\,
J_{(1,L)(2,L)(1,R)(2,R)}\,n_{1,L}\,n_{1,R}.
\label{eq:intro-N2-hint}
\end{equation}
This minimal case exposes the relevant structure without any matrix algebra. The only admissible quartic interaction is the product of a left-moving and a right-moving density. Under \(1+1\)-dimensional
bosonization, these densities become gradients of chiral bosonic fields, so the quartic fermionic interaction becomes quadratic in the bosonic variables, and so the resulting theory can then be diagonalized exactly.

The minimal realization shows that the integrable structure is tied to the density--density form of the interaction. We now show that this mechanism persists for arbitrary even \(N=2m\) in the  \([SO(2)]_L^{N/2}\times [SO(2)]_R^{N/2}\)-invariant theory. In the maximally constrained theory, the same symmetry requirement forces all allowed quartic interactions to be built from invariant density bilinears. This reduces the interacting fermionic problem to the canonical diagonalization of a quadratic chiral-boson theory.

\subsection{Exact solution in the maximally constrained theory}
\label{subsec:general-proof-integrability}
Consider the general even-\(N\) theory invariant under
\([SO(2)]_L^{N/2}\times [SO(2)]_R^{N/2}\). Writing \(N=2m\), the symmetry
group factorizes as
\begin{equation}
[SO(2)]_L^{N/2}\times [SO(2)]_R^{N/2}
=
\left(\prod_{a=1}^{m}[SO(2)]_{(a,L)}\right)
\times
\left(\prod_{b=1}^{m}[SO(2)]_{(b,R)}\right).
\end{equation}
Thus the Majorana fermions are organized into independent
pair-chirality blocks \((a,\nu)\), each transforming under its own
\(SO(2)\). This general pairing structure was already
visible at \(N=2\), and it provides the natural variables in which the density--density mechanism becomes explicit.

For each pair-chirality block, we introduce a complex fermion and the
corresponding invariant density bilinear,
\begin{align}
\chi_{a,L}
&=
\frac{1}{\sqrt2}(\psi_{2a-1,L}+i\psi_{2a,L}),
&
n_{a,L}
&=
i\psi_{2a-1,L}\psi_{2a,L},
\nonumber\\[4pt]
\chi_{a,R}
&=
\frac{1}{\sqrt2}(\psi_{2a-1,R}+i\psi_{2a,R}),
&
n_{a,R}
&=
i\psi_{2a-1,R}\psi_{2a,R}.
\label{eq:chi-density-general}
\end{align}
Equivalently, \(n_{a,\nu}\) is the normal-ordered density of the
complex fermion \(\chi_{a,\nu}\), up to the conventional additive
constant. The bilinears \(n_{a,\nu}\) are singlets under the
corresponding \(SO(2)_{(a,\nu)}\) rotations. They are therefore the
elementary invariant building blocks from which the symmetry-allowed
quartic interactions are built.

With the canonical free helical kinetic term fixed, the general Lorentzian action in this maximally constrained subspace can therefore be written as
\begin{align}
S
&=
\int dt\,dx\,
\Bigg[
i\sum_{a=1}^{m}\chi_{a,R}^{\dagger}(\partial_t+\partial_x)\chi_{a,R}
+
i\sum_{a=1}^{m}\chi_{a,L}^{\dagger}(\partial_t-\partial_x)\chi_{a,L}
\Bigg]
\nonumber\\
&\quad
-\int dt\,dx\,
\Bigg[
\sum_{1\le a<b\le m}
J_{(a,L)(b,L)}\,n_{a,L}n_{b,L}
\nonumber\\
&\hspace{4cm}
+
\sum_{1\le a<b\le m}
J_{(a,R)(b,R)}\,n_{a,R}n_{b,R}
\nonumber\\
&\hspace{5.5cm}
+
\sum_{a,b=1}^{m}
J_{(a,L)(b,R)}\,n_{a,L}n_{b,R}
\Bigg].
\label{eq:action-general-complex}
\end{align}
The couplings in Eq.~\eqref{eq:action-general-complex} are simply the
original quartic couplings rewritten in the pair-chirality basis. We
define
\begin{equation*}
J_{(a,\nu_1)(b,\nu_2)}
\equiv
J_{(2a-1,\nu_1)(2a,\nu_1)(2b-1,\nu_2)(2b,\nu_2)},
\end{equation*}
which is the coupling associated with the density product
\(n_{a,\nu_1}n_{b,\nu_2}\). In the purely chiral terms one has \(a<b\),
whereas in the mixed chirality term the pair labels run independently
because \((a,L)\) and \((b,R)\) are distinct pair-chirality blocks.

Since bosonization is most naturally implemented at the operator level
\cite{gogolin1999bosonization,fradkin2013field,vonDelft:1998}, we pass
to the Hamiltonian formulation before applying the bosonization rules.
Performing the standard Legendre transform for the first-order fermionic
action, one obtains
\begin{align}
\mathcal H
&=
i\sum_{a=1}^{m}\chi_{a,L}^{\dagger}\partial_x\chi_{a,L}
-
i\sum_{a=1}^{m}\chi_{a,R}^{\dagger}\partial_x\chi_{a,R}
\nonumber\\
&\quad
+
\sum_{1\le a<b\le m}
J_{(a,L)(b,L)}\,n_{a,L}n_{b,L}
\nonumber\\
&\hspace{2.5cm}
+
\sum_{1\le a<b\le m}
J_{(a,R)(b,R)}\,n_{a,R}n_{b,R}
\nonumber\\
&\hspace{4.5cm}
+
\sum_{a,b=1}^{m}
J_{(a,L)(b,R)}\,n_{a,L}n_{b,R},
\label{eq:ham-general-complex}
\end{align}
where the density--density form now makes the bosonization procedure particularly transparent.

We represent each complex fermion as a Hermitian Klein factor multiplying a chiral
vertex operator,
\begin{equation}
\chi_{a,L}(t,x)
=
F_{a,L}\,:\!e^{-i\phi_{a,L}(t,x)}\!:,
\qquad
\chi_{a,R}(t,x)
=
F_{a,R}\,:\!e^{i\phi_{a,R}(t,x)}\!:,
\label{eq:chiLR-bos-general}
\end{equation}
with Klein factors satisfying
\[
\{F_{a,\nu},F_{b,\nu'}\}
=
2\,\delta_{ab}\delta_{\nu\nu'},
\qquad
F_{a,\nu}^{2}=1.
\]
With the chiral conventions used throughout this work, the bosonic
fields obey
\begin{align}
[\phi_{a,L}(t,x),\partial_{x'}\phi_{b,L}(t,x')]
&=
2\pi i\,\delta_{ab}\delta(x-x'),
\nonumber\\
[\phi_{a,R}(t,x),\partial_{x'}\phi_{b,R}(t,x')]
&=
-2\pi i\,\delta_{ab}\delta(x-x').
\label{eq:comm-bos-general}
\end{align}
Equivalently,
\begin{align}
[\phi_{a,L}(t,x),\phi_{b,L}(t,x')]
&=
-\delta_{ab}\,i\pi\,\operatorname{sgn}(x-x'),
\nonumber\\
[\phi_{a,R}(t,x),\phi_{b,R}(t,x')]
&=
\delta_{ab}\,i\pi\,\operatorname{sgn}(x-x').
\label{eq:comm-bos-integrated-general}
\end{align}
The relative signs encode the opposite chiral propagation of the two
chiralities. Using these conventions, the normal-ordered bosonization identities are
\begin{align}
i:\chi_{a,L}^{\dagger}\partial_x\chi_{a,L}:
&=
\frac{1}{4\pi}:(\partial_x\phi_{a,L})^{2}:,
&
:n_{a,L}:
&=
\frac{\partial_x\phi_{a,L}}{2\pi},
\nonumber\\[4pt]
-i:\chi_{a,R}^{\dagger}\partial_x\chi_{a,R}:
&=
\frac{1}{4\pi}:(\partial_x\phi_{a,R})^{2}:,
&
:n_{a,R}:
&=
-\frac{\partial_x\phi_{a,R}}{2\pi}.
\label{eq:bos-identities-general}
\end{align}
Normal-ordering constants only shift the vacuum energy and will be
omitted. The Hamiltonian density then becomes
\begin{align}
\mathcal H
&=
\frac{1}{4\pi}
\sum_{a=1}^{m}
\Big[
(\partial_x\phi_{a,L})^{2}
+
(\partial_x\phi_{a,R})^{2}
\Big]
\nonumber\\
&\quad
+
\frac{1}{4\pi^{2}}
\Bigg[
\sum_{1\le a<b\le m}
J_{(a,L)(b,L)}\,
\partial_x\phi_{a,L}\partial_x\phi_{b,L}
\nonumber\\
&\hspace{4cm}
+
\sum_{1\le a<b\le m}
J_{(a,R)(b,R)}\,
\partial_x\phi_{a,R}\partial_x\phi_{b,R}
\nonumber\\
&\hspace{5.5cm}
-
\sum_{a,b=1}^{m}
J_{(a,L)(b,R)}\,
\partial_x\phi_{a,L}\partial_x\phi_{b,R}
\Bigg].
\label{eq:Hdens-general-m}
\end{align}
The central simplification is now manifest. For arbitrary even \(N\), the interacting quartic fermionic theory has been mapped to a quadratic system of coupled chiral bosonic fields. The remaining task is to diagonalize this quadratic form while preserving the chiral canonical algebra. This requires keeping track of two separate structures, one
encoding the interaction data and one encoding the canonical commutation relations. To make this explicit, we collect the bosonic fields into the vector
\begin{equation}
\Phi^{T}
=
(\phi_{1,L},\dots,\phi_{m,L},\phi_{1,R},\dots,\phi_{m,R}).
\label{eq:Phi-general}
\end{equation}
The Hamiltonian density can then be written as
\begin{equation}
\mathcal H
=
\frac{1}{4\pi}
(\partial_x\Phi)^{T}\mathcal V(\partial_x\Phi),
\label{eq:ham-general-matrix}
\end{equation}
where the real symmetric matrix \(\mathcal V\) contains all dependence on
the density--density couplings,
\begin{equation}
\mathcal V=
\begin{pmatrix}
\mathcal V_{LL} & \mathcal V_{LR}\\
\mathcal V_{LR}^{T} & \mathcal V_{RR}
\end{pmatrix},
\label{eq:V-block-general}
\end{equation}
with
\begin{align}
(\mathcal V_{LL})_{ab}
&=
\delta_{ab}
+
(1-\delta_{ab})\frac{J_{(a,L)(b,L)}}{2\pi},
\nonumber\\[4pt]
(\mathcal V_{RR})_{ab}
&=
\delta_{ab}
+
(1-\delta_{ab})\frac{J_{(a,R)(b,R)}}{2\pi},
\nonumber\\[4pt]
(\mathcal V_{LR})_{ab}
&=
-\frac{J_{(a,L)(b,R)}}{2\pi}.
\label{eq:V-entries-general}
\end{align}

The canonical structure to be preserved is fixed independently by the
equal-time commutation relations. In vector notation,
\begin{equation}
[\Phi(t,x),\partial_{x'}\Phi^{T}(t,x')]
=
2\pi i\,\mathcal K\,\delta(x-x'),
\qquad
\mathcal K=\operatorname{diag}(I_m,-I_m).
\label{eq:comm-general-K}
\end{equation}
The matrix \(\mathcal K\) is not an additional dynamical input. It is the
compact representation of the opposite symplectic signs carried by the
left- and right-moving fields. The same structure appears in the
first-order kinetic term of the bosonic action,
\begin{equation}
S
=
\frac{1}{4\pi}\int dt\,dx\,
\Big[
(\partial_t\Phi)^{T}\mathcal K(\partial_x\Phi)
-
(\partial_x\Phi)^{T}\mathcal V(\partial_x\Phi)
\Big].
\label{eq:action-general-m}
\end{equation}
The diagonalization must therefore find a canonical basis in which the
quadratic form \(\mathcal V\) is diagonal, while the equal-time algebra
encoded by \(\mathcal K\) remains unchanged. Accordingly, the relevant
linear transformations are not ordinary orthogonal rotations, but
canonical transformations of the form
\begin{equation}
\Phi=\mathcal M\widetilde{\Phi},
\qquad
\mathcal M^{T}\mathcal K \mathcal M=\mathcal K,
\label{eq:canonical-transform-general}
\end{equation}
such that \(\mathcal M\) belongs to \(O(m,m)\), the group of linear
transformations preserving the bilinear form defined by \(\mathcal K\).

The normal-mode basis and the corresponding velocities are obtained
from the equations of motion derived from Eq.~\eqref{eq:action-general-m},
\begin{equation}
\mathcal K\,\partial_t\partial_x\Phi
=
\mathcal V\,\partial_x^{2}\Phi.
\label{eq:EOM-general}
\end{equation}
For plane-wave configurations
\(\Phi(x,t)\sim \mathbf v\,e^{i(kx-\omega t)}\), this equation becomes
\[
\mathcal V\mathbf v=-(\omega/k)\mathcal K\mathbf v .
\]
Defining \(u\equiv-\omega/k\), one obtains the generalized eigenvalue
problem
\begin{equation}
\mathcal V\mathbf v=u\,\mathcal K\mathbf v.
\label{eq:gen-eig-general-m}
\end{equation}
The eigenvectors determine the canonical normal-mode basis, while the
eigenvalues \(u\) distinguish the two propagation directions. Positive
eigenvalues correspond to left-moving normal modes in the conventions
used here, whereas negative eigenvalues correspond to right-moving
normal modes. The physical speeds are the corresponding positive
magnitudes.

The Hamiltonian is bounded from below precisely when the quadratic form
is positive definite, \(\mathcal V>0\). Equivalently, using the block
decomposition \eqref{eq:V-block-general}, the stable region may be
characterized by
\begin{equation}
\mathcal V_{LL}>0,
\qquad
\mathcal V_{RR}
-
\mathcal V_{LR}^{T}\mathcal V_{LL}^{-1}\mathcal V_{LR}
>0.
\label{eq:schur-general}
\end{equation}
Within this stable region, the generalized eigenvalue problem has the
spectral structure required for a canonical diagonalization. Since
\(\mathcal V>0\), the matrix
\(\mathcal V^{1/2}\mathcal K\mathcal V^{1/2}\) is real symmetric and is
similar to \(\mathcal K\mathcal V\). It therefore has real eigenvalues.
By Sylvester's law of inertia \cite{sylvester1852xix}, it has the same number of positive and
negative eigenvalues as \(\mathcal K\), namely \(m\) of each sign.

Let \(u_{a,L}>0\) denote the positive eigenvalues and \(u_{a,R}<0\) the
negative eigenvalues. The corresponding eigenvectors can be chosen to be
\(\mathcal K\)-orthonormal and assembled into a matrix
\(\mathcal M\in O(m,m)\). In this basis,
\begin{equation}
\mathcal M^{T}\mathcal V \mathcal M
=
\operatorname{diag}(c_{1,L},\dots,c_{m,L},c_{1,R},\dots,c_{m,R}),
\label{eq:diag-V-general}
\end{equation}
where
\begin{equation}
c_{a,L}=u_{a,L}>0,
\qquad
c_{a,R}=-u_{a,R}>0
\label{eq:normal-speeds-general}
\end{equation}
are the positive propagation speeds of the diagonal normal modes.

In terms of the transformed fields
\[
\widetilde{\Phi}^{T}
=
(\widetilde{\phi}_{1,L},\dots,\widetilde{\phi}_{m,L},
\widetilde{\phi}_{1,R},\dots,\widetilde{\phi}_{m,R}),
\]
the Hamiltonian becomes
\begin{equation}
\mathcal H
=
\frac{1}{4\pi}
\sum_{a=1}^{m}
\left[
c_{a,L}(\partial_x\widetilde{\phi}_{a,L})^{2}
+
c_{a,R}(\partial_x\widetilde{\phi}_{a,R})^{2}
\right],
\label{eq:Hdiag-general}
\end{equation}
and the action factorizes into independent chiral normal modes,
\begin{equation}
S
=
\frac{1}{4\pi}
\sum_{a=1}^{m}\int dt\,dx\,
\left[
\partial_x\widetilde{\phi}_{a,L}
(\partial_t-c_{a,L}\partial_x)\widetilde{\phi}_{a,L}
-
\partial_x\widetilde{\phi}_{a,R}
(\partial_t+c_{a,R}\partial_x)\widetilde{\phi}_{a,R}
\right].
\label{eq:action-factorized-general}
\end{equation}

The \([SO(2)]^{N/2}_L\times [SO(2)]^{N/2}_R\)-invariant theory has therefore been reduced to \(m\) independent left-moving and \(m\) independent right-moving chiral normal modes. Equivalently, the interacting fermionic model is mapped to a set of decoupled chiral Luttinger modes. For every even \(N=2m\), this canonical
diagonalization establishes exact solvability throughout the stable
region.

This conclusion is also supported by a simple counting check. The number
of independent couplings compatible with the symmetry is
\begin{equation}
\binom{m}{2}+\binom{m}{2}+m^{2}=m(2m-1),
\end{equation}
which matches the dimension of the canonical group,
\begin{equation}
\dim O(m,m)=m(2m-1).
\end{equation}
This counting agreement does not by itself prove diagonalizability,
which follows from the positivity of \(\mathcal V\) and the spectral
argument above. Rather, it provides a useful consistency check that the
number of symmetry-allowed density couplings matches the dimension of
the canonical transformations preserving \(\mathcal K\).

\subsubsection{\(N=2\): minimal realization}
\label{subsubsec:N2-example}

Having established the general construction, we can now revisit the
minimal realization in the canonical language. For \(N=2\), one has
\(m=1\), so the bosonized theory contains one left-moving and one
right-moving chiral field. The canonical diagonalization therefore
reduces to a two-dimensional quadratic problem.

In the ordered basis
\begin{equation}
\Phi^{T}=(\phi_{L},\phi_{R}),
\qquad
\mathcal K=\operatorname{diag}(1,-1),
\end{equation}
the interaction matrix is
\begin{equation}
\mathcal V=
\begin{pmatrix}
1 & -\dfrac{J_{(1,L)(1,R)}}{2\pi}\\[6pt]
-\dfrac{J_{(1,L)(1,R)}}{2\pi} & 1
\end{pmatrix}.
\label{eq:N2-V}
\end{equation}
The stability condition is simply the positivity of \(\mathcal V\).
Since
\begin{equation}
\det\mathcal V
=
1-\left(\frac{J_{(1,L)(1,R)}}{2\pi}\right)^2,
\end{equation}
the stable region is therefore
\begin{equation}
\left|\frac{J_{(1,L)(1,R)}}{2\pi}\right|<1.
\label{eq:N2-stability}
\end{equation} 
Within this region, the generalized eigenvalue problem
\(\mathcal V\mathbf v=u\,\mathcal K\mathbf v\) is completely explicit.
Equivalently, the eigenvalues of \(\mathcal K\mathcal V\) obey
\begin{equation}
u^{2}
=
1-\left(\frac{J_{(1,L)(1,R)}}{2\pi}\right)^2.
\end{equation}
Thus there is one positive and one negative eigenvalue,
\begin{equation}
u_{L}
=
+\sqrt{1-\left(\frac{J_{(1,L)(1,R)}}{2\pi}\right)^2},
\qquad
u_{R}
=
-\sqrt{1-\left(\frac{J_{(1,L)(1,R)}}{2\pi}\right)^2}.
\end{equation}
Following the general prescription, the physical propagation speeds are
positive quantities,
\begin{equation}
c_{L}=u_{L},
\qquad
c_{R}=-u_{R}.
\end{equation}
Therefore
\begin{equation}
c_{L}=c_{R}\equiv c_{\circ}
=
\sqrt{1-\left(\frac{J_{(1,L)(1,R)}}{2\pi}\right)^2},
\label{eq:N2-speed}
\end{equation}
where the stability condition ensures that the normal-mode speed $c_\circ$ is real and positive. 
The equality of magnitudes is a special feature of the minimal
realization. The two normal modes nevertheless remain physically
distinct, because they carry opposite chirality.

The canonical transformation \(\mathcal M\in O(1,1)\), constructed from
the eigenvectors of \(\mathcal K\mathcal V\), brings the Hamiltonian to
\begin{equation}
\mathcal H
=
\frac{1}{4\pi}
\left[
c_{\circ}(\partial_x\widetilde{\phi}_{L})^{2}
+
c_{\circ}(\partial_x\widetilde{\phi}_{R})^{2}
\right],
\end{equation}
and the corresponding action factorizes as
\begin{equation}
S
=
\frac{1}{4\pi}\int dt\,dx\,
\left[
\partial_x\widetilde{\phi}_{L}
(\partial_t-c_{\circ}\partial_x)\widetilde{\phi}_{L}
-
\partial_x\widetilde{\phi}_{R}
(\partial_t+c_{\circ}\partial_x)\widetilde{\phi}_{R}
\right],
\end{equation}
realizing the general mechanism in its simplest form. The
density--density interaction bosonizes to a quadratic theory, and its
canonical diagonalization gives two decoupled chiral normal modes.

\subsubsection{\(N=4\): first nontrivial multicomponent realization}
\label{subsubsec:N4-example}

The first qualitatively richer realization appears at \(N=4\), or
\(m=2\). In contrast with the minimal case, each chirality now contains
two \(SO(2)\)-invariant densities. After bosonization, the interaction
is encoded in a nontrivial \(4\times4\) quadratic form coupling the
corresponding chiral bosonic fields. This is the first case in which
the mode-mixing structure is genuinely multicomponent, while still
remaining controlled by the canonical diagonalization developed above.

In the ordered basis
\begin{equation}
\Phi^{T}
=
(\phi_{1,L},\phi_{2,L},\phi_{1,R},\phi_{2,R}),
\qquad
\mathcal K=\operatorname{diag}(1,1,-1,-1),
\end{equation}
the quadratic Hamiltonian is determined by
\begin{equation}
\mathcal V=
\begin{pmatrix}
1
&
\dfrac{J_{(1,L)(2,L)}}{2\pi}
&
-\dfrac{J_{(1,L)(1,R)}}{2\pi}
&
-\dfrac{J_{(1,L)(2,R)}}{2\pi}
\\[8pt]
\dfrac{J_{(1,L)(2,L)}}{2\pi}
&
1
&
-\dfrac{J_{(2,L)(1,R)}}{2\pi}
&
-\dfrac{J_{(2,L)(2,R)}}{2\pi}
\\[8pt]
-\dfrac{J_{(1,L)(1,R)}}{2\pi}
&
-\dfrac{J_{(2,L)(1,R)}}{2\pi}
&
1
&
\dfrac{J_{(1,R)(2,R)}}{2\pi}
\\[8pt]
-\dfrac{J_{(1,L)(2,R)}}{2\pi}
&
-\dfrac{J_{(2,L)(2,R)}}{2\pi}
&
\dfrac{J_{(1,R)(2,R)}}{2\pi}
&
1
\end{pmatrix}.
\label{eq:N4-V}
\end{equation}
The quadratic form contains left-left, right-right, and left-right
density--density coupling blocks. Their simultaneous presence produces
a genuinely multicomponent bosonic problem.

The stability condition remains \(\mathcal V>0\). Positivity of the
purely chiral \(LL\) and \(RR\) principal blocks gives the necessary
bounds
\begin{equation}
\left|\frac{J_{(1,L)(2,L)}}{2\pi}\right|<1,
\qquad
\left|\frac{J_{(1,R)(2,R)}}{2\pi}\right|<1.
\label{eq:N4-chiral-bounds}
\end{equation}
The mixed couplings \(J_{(a,L)(b,R)}\) further restrict the stable region
through the full positivity condition \(\mathcal V>0\), equivalently
through the Schur complement condition.

The normal modes are determined, as in the general analysis, by
\begin{equation}
\mathcal V\,\mathbf v=u\,\mathcal K\,\mathbf v.
\label{eq:N4-eig}
\end{equation}
Equivalently, one may solve the ordinary eigenvalue problem for
\(\mathcal K\mathcal V\). In the
present case the characteristic polynomial is quartic,
\begin{equation}
\det(\mathcal K\mathcal V-u\,I_4)
=
u^{4}
+
\mathfrak a_{2}\,u^{2}
-
\mathfrak a_{3}\,u
+
\mathfrak a_{4}
=
0 .
\label{eq:N4-quartic}
\end{equation}
For the matrix in Eq.~\eqref{eq:N4-V}, the \(u^3\) coefficient vanishes
because \(\operatorname{tr}(\mathcal K\mathcal V)=0\). The remaining
coefficients \(\mathfrak a_{2},\mathfrak a_{3},\mathfrak a_{4}\) are
fixed by the spectral invariants of \(\mathcal K\mathcal V\), or
equivalently by traces of its higher powers. Their explicit evaluation,
together with the canonical eigenvector prescription that constructs
the \(O(2,2)\) normal-mode basis, is given in
Appendix~\ref{app:N4-canonical-diagonalization}.

Although the algebra is richer than in the minimal case, the qualitative
structure is fixed by the general argument. Whenever \(\mathcal V>0\),
the spectrum contains two positive and two negative eigenvalues,
\begin{equation}
u_{1,L},u_{2,L}>0,
\qquad
u_{1,R},u_{2,R}<0.
\end{equation}
The physical propagation speeds are defined by
\begin{equation}
c_{a,L}=u_{a,L},
\qquad
c_{a,R}=-u_{a,R},
\qquad
a=1,2.
\end{equation}
For generic values of the six independent couplings, these speeds are
all distinct. Thus, unlike the minimal realization, the \(N=4\) theory
already displays the splitting of normal-mode velocities produced by the
full multicomponent quadratic form.

Once the canonical transformation \(\mathcal M\in O(2,2)\) is
constructed, the Hamiltonian takes the diagonal form
\begin{equation}
\mathcal H
=
\frac{1}{4\pi}
\sum_{a=1}^{2}
\left[
c_{a,L}(\partial_x\widetilde{\phi}_{a,L})^{2}
+
c_{a,R}(\partial_x\widetilde{\phi}_{a,R})^{2}
\right],
\label{eq:N4-Hdiag}
\end{equation}
and the action factorizes into four independent chiral normal modes,
\begin{equation}
S
=
\frac{1}{4\pi}
\sum_{a=1}^{2}\int dt\,dx\,
\left[
\partial_x\widetilde{\phi}_{a,L}
(\partial_t-c_{a,L}\partial_x)\widetilde{\phi}_{a,L}
-
\partial_x\widetilde{\phi}_{a,R}
(\partial_t+c_{a,R}\partial_x)\widetilde{\phi}_{a,R}
\right].
\label{eq:N4-Sdiag}
\end{equation}
Therefore the exact solvability does not rely on the
special features of the minimal realization. In the \(N=4\) theory,
multiple invariant densities per chirality produce nontrivial mixing
among chiral bosonic fields, but the problem remains a quadratic
canonical diagonalization problem. The diagonal normal-mode basis now
provides the starting point for computing local observables.

\subsubsection{Correlation functions and thermodynamics in the diagonal basis}
\label{subsec:physical-observables}

Once the canonical diagonalization has been performed, observables in
the maximally constrained theory can be computed directly in the
normal-mode basis. We present the formulas explicitly for \(N=4\),
where the normal-mode sums are finite. The extension to arbitrary
even \(N\) follows by replacing the sum over two modes of each chirality
by a sum over \(m=N/2\) modes.

The diagonal action gives the equations of motion
\begin{equation}
(\partial_t-c_{a,L}\partial_x)\widetilde{\phi}_{a,L}=0,
\qquad
(\partial_t+c_{a,R}\partial_x)\widetilde{\phi}_{a,R}=0,
\qquad
a=1,2.
\label{eq:obs-eom}
\end{equation}
Thus the left-moving normal modes depend on \(x+c_{a,L}t\), while the
right-moving normal modes depend on \(x-c_{a,R}t\). The spatial zero
mode does not affect the local observables considered here.

\paragraph{Bosonic two-point functions.}

We place the system on a circle of length \(L_{\mathrm{sys}}\), with
momenta
\begin{equation}
q=\frac{2\pi}{L_{\mathrm{sys}}}m_q,
\qquad
m_q\in\mathbb Z_{>0}.
\end{equation}
A mode expansion consistent with Eq.~\eqref{eq:obs-eom} is
\begin{align}
\widetilde{\phi}_{a,L}(t,x)
&=
-\sum_{q>0}\frac{1}{\sqrt{m_q}}
\left[
e^{-iq(x+c_{a,L}t)}b_{q,a,L}
+
e^{iq(x+c_{a,L}t)}b^\dagger_{q,a,L}
\right]e^{-\epsilon q/2},
\nonumber\\[4pt]
\widetilde{\phi}_{a,R}(t,x)
&=
-\sum_{q>0}\frac{1}{\sqrt{m_q}}
\left[
e^{iq(x-c_{a,R}t)}b_{q,a,R}
+
e^{-iq(x-c_{a,R}t)}b^\dagger_{q,a,R}
\right]e^{-\epsilon q/2}.
\label{eq:obs-mode}
\end{align}
In the local thermodynamic limit, the oscillator sums give
\begin{align}
\big\langle
\widetilde{\phi}_{a,L}(t,x)\widetilde{\phi}_{b,L}(0,0)
\big\rangle
&=
-\delta_{ab}
\log\!\left[
\frac{2\pi i}{L_{\mathrm{sys}}}
\left(c_{a,L}t+x-i0^+\right)
\right],
\nonumber\\[4pt]
\big\langle
\widetilde{\phi}_{a,R}(t,x)\widetilde{\phi}_{b,R}(0,0)
\big\rangle
&=
-\delta_{ab}
\log\!\left[
\frac{2\pi i}{L_{\mathrm{sys}}}
\left(c_{a,R}t-x-i0^+\right)
\right],
\label{eq:zeroT-boson-prop}
\end{align}
while mixed correlators vanish. The factor of \(L_{\mathrm{sys}}\)
makes the logarithm dimensionless and contributes only an additive
normalization to local observables.

At finite temperature \(T=\beta^{-1}\), the standard map to the thermal
cylinder gives
\begin{align}
\big\langle
\widetilde{\phi}_{a,L}(t,x)\widetilde{\phi}_{b,L}(0,0)
\big\rangle_\beta
&=
-\delta_{ab}
\log\!\left[
\frac{2i\beta c_{a,L}}{L_{\mathrm{sys}}}
\sinh\!\left(
\frac{\pi}{\beta}
\left(
t+\frac{x}{c_{a,L}}-i0^+
\right)
\right)
\right],
\nonumber\\[6pt]
\big\langle
\widetilde{\phi}_{a,R}(t,x)\widetilde{\phi}_{b,R}(0,0)
\big\rangle_\beta
&=
-\delta_{ab}
\log\!\left[
\frac{2i\beta c_{a,R}}{L_{\mathrm{sys}}}
\sinh\!\left(
\frac{\pi}{\beta}
\left(
t-\frac{x}{c_{a,R}}-i0^+
\right)
\right)
\right].
\label{eq:finiteT-boson-prop}
\end{align}
These propagators encode the local fluctuations of the diagonal normal
modes and provide the basic input for both thermodynamic quantities and
fermionic correlation functions.

\paragraph{Thermodynamic observables.}

The energy density is the diagonal Hamiltonian density,
\begin{equation}
T^{0}{}_{0}
=
\frac{1}{4\pi}
\sum_{a=1}^{2}
\left[
c_{a,L}(\partial_x\widetilde{\phi}_{a,L})^2
+
c_{a,R}(\partial_x\widetilde{\phi}_{a,R})^2
\right].
\label{eq:energy-density-diag}
\end{equation}
The associated energy current follows from local conservation,
\(\partial_t T^{0}{}_{0}+\partial_xT^{x}{}_{0}=0\), together with
Eq.~\eqref{eq:obs-eom}. This gives
\begin{equation}
T^{x}{}_{0}
=
\frac{1}{4\pi}
\sum_{a=1}^{2}
\left[
-c_{a,L}^{2}(\partial_x\widetilde{\phi}_{a,L})^2
+
c_{a,R}^{2}(\partial_x\widetilde{\phi}_{a,R})^2
\right],
\label{eq:energy-current-diag}
\end{equation}
where the relative sign reflects the opposite propagation of the two
chiralities.

The thermal expectation value of the energy density is evaluated by
point-splitting the composite operator in the spatial direction at equal
time. For the relative separation \(x=\epsilon\),
\begin{equation}
\big\langle
(\partial_x\widetilde{\phi}_{a,\nu})^2
\big\rangle_\beta
=
-\partial_x^2
\big\langle
\widetilde{\phi}_{a,\nu}(t,x)
\widetilde{\phi}_{a,\nu}(0,0)
\big\rangle_\beta
\Big|_{t\to0,\;x=\epsilon}.
\label{eq:point-splitting}
\end{equation}
Using Eq.~\eqref{eq:finiteT-boson-prop}, the short-distance expansion is
\begin{equation}
\big\langle
(\partial_x\widetilde{\phi}_{a,\nu})^2
\big\rangle_\beta
=
-\frac{1}{\epsilon^2}
+
\frac{\pi^2}{3\beta^2 c_{a,\nu}^{2}}
+ O(\epsilon^2).
\label{eq:derivative-prop-expansion}
\end{equation}
The first term is the temperature-independent vacuum divergence, while
the second term is the finite thermal contribution.

Substituting Eq.~\eqref{eq:derivative-prop-expansion} into
Eq.~\eqref{eq:energy-density-diag} gives
\begin{align}
\varepsilon
&=
-\frac{1}{4\pi\epsilon^2}
\sum_{a=1}^{2}(c_{a,L}+c_{a,R})
+
\frac{\pi}{12\beta^2}
\sum_{a=1}^{2}
\left(
\frac{1}{c_{a,L}}+\frac{1}{c_{a,R}}
\right)
+ O(\epsilon^2).
\end{align}
After subtracting the vacuum contribution, the physical thermal energy
density is therefore
\begin{equation}
\varepsilon_{\mathrm{phys}}
=
\frac{\pi}{12\beta^2}
\sum_{a=1}^{2}
\left(
\frac{1}{c_{a,L}}+\frac{1}{c_{a,R}}
\right).
\label{eq:energy-density-phys}
\end{equation}

The same subtraction applied to Eq.~\eqref{eq:energy-current-diag} gives
a vanishing energy current in thermal equilibrium,
\begin{equation}
j_{\varepsilon,\mathrm{phys}}=0.
\label{eq:energy-current-zero}
\end{equation}
Indeed, the finite left-moving contribution is
\(-\pi/(12\beta^2)\) for each normal mode, while the corresponding
right-moving contribution is \(+\pi/(12\beta^2)\). The cancellation
reflects the equilibrium balance between the two chiralities.

Writing \(T=\beta^{-1}\), the entropy density follows from
\begin{equation}
T\,\frac{\partial s}{\partial T}
=
\frac{\partial\varepsilon_{\mathrm{phys}}}{\partial T}.
\end{equation}
With the integration constant fixed by requiring the entropy to vanish
at zero temperature, one obtains
\begin{equation}
s
=
\frac{\pi}{6\beta}
\sum_{a=1}^{2}
\left(
\frac{1}{c_{a,L}}+\frac{1}{c_{a,R}}
\right).
\label{eq:entropy-density}
\end{equation}
Thus, each chiral normal mode contributes additively to the thermal
response. The interaction enters only through the propagation speeds
\(c_{a,\nu}\); once the theory is written in the diagonal basis, the
thermodynamic structure is that of free chiral modes with
interaction-renormalized velocities.

\paragraph{Fermionic two-point functions.}

The bosonic formulation also reconstructs the correlation functions of
the original fermionic degrees of freedom. For \(N=4\), the Majoranas of
each chirality are grouped into two complex fermions
\(\chi_{a,\nu}\), with \(a=1,2\). We consider the flavor-averaged
thermal Majorana correlator
\begin{equation}
G_{\nu}(t,x)
=
\frac{1}{4}
\sum_{i=1}^{4}
\big\langle
\psi_{i,\nu}(t,x)\psi_{i,\nu}(0,0)
\big\rangle_\beta .
\label{eq:Majorana-average-def}
\end{equation}
Using
\[
\psi_{2a-1,\nu}
=
\frac{1}{\sqrt2}
(\chi_{a,\nu}+\chi_{a,\nu}^{\dagger}),
\qquad
\psi_{2a,\nu}
=
\frac{1}{i\sqrt2}
(\chi_{a,\nu}-\chi_{a,\nu}^{\dagger}),
\]
together with the fact that anomalous correlators such as
\(\langle\chi_{a,\nu}\chi_{a,\nu}\rangle_\beta\) vanish by the
\(SO(2)^{(a,\nu)}\) symmetry, one obtains
\begin{equation}
G_{\nu}(t,x)
=
\frac{1}{4}
\sum_{a=1}^{2}
\left[
G^{(+)}_{a,\nu}(t,x)
+
G^{(-)}_{a,\nu}(t,x)
\right],
\label{eq:Majorana-average-Gpm}
\end{equation}
where
\begin{align}
G^{(+)}_{a,\nu}(t,x)
&\equiv
\big\langle
\chi_{a,\nu}(t,x)\chi_{a,\nu}^{\dagger}(0,0)
\big\rangle_\beta,
&
G^{(-)}_{a,\nu}(t,x)
&\equiv
\big\langle
\chi_{a,\nu}^{\dagger}(t,x)\chi_{a,\nu}(0,0)
\big\rangle_\beta .
\label{eq:Gpm-def}
\end{align}

For the two-point functions considered here, the Klein factors do not
affect same-\((a,\nu)\) correlators, and the relevant dependence is
carried by the vertex-operator part of the bosonized fermions. The
original bosonic fields entering these vertex operators are related to
the diagonal normal modes by the canonical transformation
\begin{equation}
\Phi=\mathcal M\,\widetilde{\Phi},
\qquad
\mathcal M\in O(2,2).
\end{equation}
Writing
\begin{equation}
\mathcal M=
\begin{pmatrix}
\mathcal M^{(LL)} & \mathcal M^{(LR)}\\
\mathcal M^{(RL)} & \mathcal M^{(RR)}
\end{pmatrix},
\end{equation}
one has
\begin{align}
\phi_{a,L}
&=
\sum_{b=1}^{2}
(\mathcal M^{(LL)})_{ab}\widetilde{\phi}_{b,L}
+
\sum_{b=1}^{2}
(\mathcal M^{(LR)})_{ab}\widetilde{\phi}_{b,R},
\nonumber\\
\phi_{a,R}
&=
\sum_{b=1}^{2}
(\mathcal M^{(RL)})_{ab}\widetilde{\phi}_{b,L}
+
\sum_{b=1}^{2}
(\mathcal M^{(RR)})_{ab}\widetilde{\phi}_{b,R}.
\label{eq:phi-original-normal}
\end{align}
Although the diagonal normal modes are free chiral bosons, the original
fermions are vertex operators built from these linear combinations. Since
the diagonal theory is Gaussian, their two-point functions factorize into
products of normal-mode vertex correlators, with exponents determined by
the canonical transformation \(\mathcal M\). Up to the standard
nonuniversal normalization of vertex operators, one finds
\begin{align}
G^{(+)}_{a,L}(t,x)
&\sim
\prod_{b=1}^{2}
\left[
\frac{2i\beta c_{b,L}}{L_{\mathrm{sys}}}
\sinh\!\left(
\frac{\pi}{\beta}
\left(
t+\frac{x}{c_{b,L}}-i0^+
\right)
\right)
\right]^{-\bigl[(\mathcal M^{(LL)})_{ab}\bigr]^2}
\nonumber\\
&\quad\times
\prod_{b=1}^{2}
\left[
\frac{2i\beta c_{b,R}}{L_{\mathrm{sys}}}
\sinh\!\left(
\frac{\pi}{\beta}
\left(
t-\frac{x}{c_{b,R}}-i0^+
\right)
\right)
\right]^{-\bigl[(\mathcal M^{(LR)})_{ab}\bigr]^2}.
\label{eq:Gplus-left-scaling}
\end{align}
The correlator \(G^{(-)}_{a,L}\) has the same universal power-law
envelope, with the conjugate real-time prescription associated with the
conjugate vertex operators. After suppressing these prescription
differences and the nonuniversal vertex-operator normalization, one may
write \(G^{(+)}_{a,\nu}\sim G^{(-)}_{a,\nu}\) at the level of universal
scaling forms.

Combining Eq.~\eqref{eq:Majorana-average-Gpm} with this universal
equivalence gives the left-moving Majorana scaling form
\begin{align}
G_{L}(t,x)
&\sim
\frac{1}{2}
\sum_{a=1}^{2}
\prod_{b=1}^{2}
\left[
\frac{2i\beta c_{b,L}}{L_{\mathrm{sys}}}
\sinh\!\left(
\frac{\pi}{\beta}
\left(
t+\frac{x}{c_{b,L}}-i0^+
\right)
\right)
\right]^{-\bigl[(\mathcal M^{(LL)})_{ab}\bigr]^2}
\nonumber\\
&\quad\times
\prod_{b=1}^{2}
\left[
\frac{2i\beta c_{b,R}}{L_{\mathrm{sys}}}
\sinh\!\left(
\frac{\pi}{\beta}
\left(
t-\frac{x}{c_{b,R}}-i0^+
\right)
\right)
\right]^{-\bigl[(\mathcal M^{(LR)})_{ab}\bigr]^2}.
\label{eq:GL-scaling}
\end{align}
Similarly, the right-moving Majorana scaling form is
\begin{align}
G_{R}(t,x)
&\sim
\frac{1}{2}
\sum_{a=1}^{2}
\prod_{b=1}^{2}
\left[
\frac{2i\beta c_{b,L}}{L_{\mathrm{sys}}}
\sinh\!\left(
\frac{\pi}{\beta}
\left(
t+\frac{x}{c_{b,L}}-i0^+
\right)
\right)
\right]^{-\bigl[(\mathcal M^{(RL)})_{ab}\bigr]^2}
\nonumber\\
&\quad\times
\prod_{b=1}^{2}
\left[
\frac{2i\beta c_{b,R}}{L_{\mathrm{sys}}}
\sinh\!\left(
\frac{\pi}{\beta}
\left(
t-\frac{x}{c_{b,R}}-i0^+
\right)
\right)
\right]^{-\bigl[(\mathcal M^{(RR)})_{ab}\bigr]^2}.
\label{eq:GR-scaling}
\end{align}

These expressions make explicit how the original Majorana fields inherit
the normal-mode structure of the diagonal theory. Each original chiral
boson is a linear combination of all diagonal normal modes, so the
fermionic two-point functions factorize into products of chiral
normal-mode correlators. The interaction dependence enters through two
pieces of diagonal data, the normal-mode speeds \(c_{a,\nu}\) and the
canonical transformation matrix \(\mathcal M\). This completes the
description of local observables in the symmetry-protected
density--density sector.

\subsection{Breakdown of the symmetry-protected diagonalization mechanism}
\label{subsec:breakdown-integrability}

The preceding analysis identifies the precise mechanism behind the exact
solution of the maximally constrained theory. It requires three
ingredients. The Majorana flavors must be organized into disjoint
pair-chirality blocks, the allowed quartic interactions must be products
of the corresponding invariant densities, and these densities must
bosonize to independent chiral gradients. Under these conditions the
fermionic interaction becomes a quadratic bosonic form, which can be
diagonalized by a canonical transformation. Once this density structure
is lost, the finite-coupling diagonalization mechanism no longer applies.

The minimal case \(N=2\) is useful as a benchmark. There is only one
nonvanishing quartic structure, and it necessarily lies in the
\(\{2L2R\}\) sector. The interaction is therefore automatically of
density--density type. This shows that exact solvability is not tied to
small \(N\) by itself, but to the presence of a structure that forces the
interaction into independent density products.

The first case in which this structure ceases to be automatic is
\(N=3\). For three flavors, the purely chiral sectors \(\{4L\}\) and
\(\{4R\}\) vanish identically, but the sectors \(\{2L2R\}\),
\(\{3L1R\}\), and \(\{1L3R\}\) can be formed. The role of the symmetry is
therefore to determine which of these sectors are present in the
interaction.

Consider first the \(\mathbb Z_{2,L}\times\mathbb Z_{2,R}\)-invariant
theory, where only the chirality-balanced sector survives. The corresponding
\(\{2L2R\}\) contribution is
\begin{equation}
S_{\mathrm{int}}^{(2L2R)}
=
\int dt\,dx
\sum_{i_1<i_2,\;i_3<i_4}
J_{(i_1,L)(i_2,L)(i_3,R)(i_4,R)}\,
\psi_{i_1,L}\psi_{i_2,L}\psi_{i_3,R}\psi_{i_4,R}.
\end{equation}
For \(N=3\), the antisymmetric flavor pairs are \((12)\), \((13)\), and
\((23)\). We use these pairs to define
\begin{equation}
\widetilde n_{1,\nu}=\psi_{1,\nu}\psi_{2,\nu},
\qquad
\widetilde n_{2,\nu}=\psi_{1,\nu}\psi_{3,\nu},
\qquad
\widetilde n_{3,\nu}=\psi_{2,\nu}\psi_{3,\nu}.
\label{eq:N3-tilde-bilinears}
\end{equation}
Note that this tilde is important. These bilinears are not densities of disjoint
Majorana pairs. They are overlapping antisymmetric flavor bilinears,
since different \(\tilde n_{a,\nu}\) share Majorana fields. In this antisymmetric-pair basis, the
chirality-balanced contribution can be written as
\begin{equation}
S_{\mathrm{int}}^{(2L2R)}
=
\int dt\,dx
\sum_{a,b=1}^{3}
\widetilde J_{(a,L)(b,R)}\,
\tilde n_{a,L}\,
\tilde n_{b,R}.
\label{eq:N3-balanced-tilde}
\end{equation}
The matrix \(\widetilde J_{(a,L)(b,R)}\) is written in a form parallel to
the pair-chirality couplings of the even-\(N\) theory, but its labels now
refer to overlapping antisymmetric flavor pairs.

A chirality-preserving change of flavor basis acts as
\begin{equation}
\psi_{i,\nu}
\longrightarrow
\sum_j O^{(\nu)}_{ij}\psi_{j,\nu},
\qquad
O^{(\nu)}\in SO(3),
\end{equation}
and induces a linear transformation on the antisymmetric bilinears,
\begin{equation}
\widetilde n_{a,\nu}
\longrightarrow
\sum_{a'}
\widetilde O^{(\nu)}_{aa'}\,
\widetilde n_{a',\nu}.
\end{equation}
For \(SO(3)\), the antisymmetric two-index representation is equivalent
to the vector representation, \(\wedge^2\mathbf 3\simeq\mathbf 3\).
Thus \(\widetilde O^{(\nu)}\) is the induced three-dimensional rotation
on the space of antisymmetric bilinears. This is a basis change in the
space of overlapping bilinears, not a construction of independent
density variables.

Under these transformations, Eq.~\eqref{eq:N3-balanced-tilde} becomes
\begin{align}
S_{\mathrm{int}}^{(2L2R)}
&\longrightarrow
\int dt\,dx
\sum_{a,b}
\widetilde J_{(a,L)(b,R)}
\left(
\sum_{a'}
\widetilde O^{(L)}_{aa'}\,\widetilde n_{a',L}
\right)
\left(
\sum_{b'}
\widetilde O^{(R)}_{bb'}\,\widetilde n_{b',R}
\right)
\nonumber\\
&=
\int dt\,dx
\sum_{a',b'}
\widetilde J'_{(a',L)(b',R)}\,
\widetilde n_{a',L}\widetilde n_{b',R},
\end{align}
with
\begin{equation}
\widetilde J'_{(a',L)(b',R)}
=
\sum_{a,b}
\widetilde O^{(L)}_{aa'}\,
\widetilde J_{(a,L)(b,R)}\,
\widetilde O^{(R)}_{bb'}.
\end{equation}
Using independent rotations in the left and right antisymmetric-pair
spaces, the real matrix \(\widetilde J_{(a,L)(b,R)}\) can be brought to
diagonal singular-value form,
\begin{equation}
\widetilde J'_{(a',L)(b',R)}
=
\widetilde J_{a'}\,\delta_{a'b'}.
\end{equation}
With \(SO(3)\) rather than \(O(3)\) rotations, one may choose a convention
in which a possible orientation sign is absorbed into one of the
diagonal entries. The entries \(\widetilde J_{a'}\) are couplings in the
rotated antisymmetric-pair basis. They should not be confused with the
independent pair-chirality couplings multiplying invariant densities in
the even-\(N\) maximally constrained theory.

The crucial distinction is that diagonalizing the matrix
\(\widetilde J_{(a,L)(b,R)}\) simplifies the coupling data, but does not
produce the independent \(SO(2)\)-invariant density variables
\(n_{a,\nu}\) that underlie the canonical diagonalization in the
maximally constrained theory. As is clear from
Eq.~\eqref{eq:N3-tilde-bilinears}, different \(\tilde n_{a,\nu}\) share
Majorana fields and therefore cannot define independent density
variables associated with disjoint \(SO(2)\) Majorana doublets.
Consequently, even after the diagonal form is reached, the chirality-balanced
\(N=3\) theory does not acquire the density--density organization that
made the even-\(N\) maximally constrained theory quadratic after
bosonization.

The obstruction can equivalently be expressed in terms of representation theory language. The
bilinears \(\widetilde n_{a,\nu}\) transform as the vector
representation of the corresponding flavor rotation group. Hence the
chirality-balanced sector transforms as
\[
(\mathbf 3,\mathbf 3)
\quad
\text{under}
\quad
SO(3)_L\times SO(3)_R.
\]
If the symmetry is reduced further to the diagonal \(\mathbb Z_2\), the
chirality-imbalanced sectors are also allowed. The sector \(\{3L1R\}\)
contains the \(SO(3)_L\) singlet
\(\psi_{1,L}\psi_{2,L}\psi_{3,L}\) multiplied by a right-moving
Majorana, and therefore transforms as \((\mathbf 1,\mathbf 3)\), and similarly the
sector \(\{1L3R\}\) transforms as \((\mathbf 3,\mathbf 1)\). We know that the chirality-preserving flavor rotations cannot mix these sectors.

Relaxing the symmetry constraints therefore enlarges the interaction
space in a way that disrupts the density--density organization of the
maximally constrained even-\(N\) theory. Even within the
chirality-balanced sector \(\{2L2R\}\), diagonalizing
\(\widetilde J_{(a,L)(b,R)}\) only simplifies the coupling matrix. It
does not reproduce the independent density variables \(n_{a,\nu}\)
required by the canonical diagonalization mechanism. Once the
chirality-imbalanced sectors \(\{3L1R\}\) and \(\{1L3R\}\) are included,
the obstruction becomes unavoidable because the new operators belong to
inequivalent chirality sectors. Consequently, the symmetry-protected exact solvability of the maximally
constrained theory does not survive once these constraints are removed.
The canonical density--density diagonalization mechanism is lost in the
full quartic interaction space. The key question is then whether this
loss is the end of the story, or whether a different organizing principle
appears at large-\(N\). This is the question we turn to next.

\section{Disorder-averaged RG flow and emergent infrared integrability}
\label{sec:RG-flow}

Once the symmetry constraints responsible for the exact solvability of
the maximally constrained theory are relaxed, the enlarged helical
quartic interaction space is naturally probed through its
renormalization-group flow. This viewpoint is especially natural in an
SYK-like large-\(N\) setting, where the IR behavior is expected to
be governed by collective disorder-averaged couplings rather than by
individual microscopic realizations
\cite{sachdev1992gapless,kitaev2015simple1,kitaev2015simple2,maldacena2016remarks,berkooz2017comments,lian2019chiral}.
For the helical theory constructed in Sec.~\ref{sec:model}, the
ultraviolet fixed point is the free helical Majorana conformal theory.
The local quartic Majorana monomials have classical scaling dimension equal to
two at this fixed point, and the corresponding terms
\(J_A\mathcal O_A\) are therefore marginal deformations. We work in the
perturbative regime \(|J_A|\ll1\), where conformal perturbation theory
gives a controlled expansion of the scale dependence of the couplings
around a two-dimensional conformal fixed point
\cite{zamolodchikov1987renormalization,francesco2012conformal,amoretti2017conformal,gaberdiel2009conformal}.
The resulting flow determines whether these local quartic deformations
drive the theory toward an interacting IR regime, or whether the
system is instead driven back toward the free helical conformal fixed
point.

We implement this perturbative renormalization-group analysis in
Euclidean signature, obtained by \(t\to -i\tau\). The perturbative
expansion is then naturally formulated in terms of the Euclidean path
integral,
\begin{equation}
Z
=
\int \mathcal D\psi\,e^{-S_E}
=
\int \mathcal D\psi\,e^{-S_0}e^{-S_{\mathrm{int}}},
\label{eq:general-partition}
\end{equation}
where \(S_0\) and \(S_{\mathrm{int}}\) denote the Euclidean
continuations of the free and interaction contributions to the action.
In this representation, the interaction is written as
\begin{equation}
S_{\mathrm{int}}
=
-\int d\tau\,dx\,
\sum_A
J_A^{\mathrm{bare}}\,
\mathcal O_A(\tau,x),
\label{eq:general-Sint}
\end{equation}
with the same microscopic labeling conventions introduced in
Sec.~\ref{sec:model}. The superscript ``bare'' emphasizes that these
microscopic couplings are defined before ultraviolet logarithmic
contributions have been isolated and absorbed into renormalized
scale-dependent couplings.

Expanding \(e^{-S_{\mathrm{int}}}\) turns the perturbative problem into
a short-distance expansion of products of quartic insertions in the free
helical theory. Wick contractions separate these products into singular
coordinate kernels multiplying residual local monomials, accompanied by
algebraic contraction tensors that encode the flavor, chirality,
fermionic sign, and combinatorial structure of each topology. In this path-integral implementation, what we call OPE selection rules
are equivalently selection rules for these short-distance kernels. The
OPE language refers to the local expansion of coincident operator
insertions. In the Euclidean path-integral formulation used here, the
same expansion is realized by Wick contracting the insertions and
integrating the resulting singular kernels over their relative
separations. Near coincident insertion points, only the component that
projects back onto the quartic basis can renormalize the marginal
couplings. Higher-dimensional local monomials and nonlocal terms do not
contribute to their beta functions. The running is controlled by the
logarithmic scale dependence of the coefficients generated by these
short-distance contractions \cite{gaberdiel2009conformal}. We denote the extraction of local logarithmic terms by
\((\cdots)_{\mathrm{local,log}}\) and encode their effect in an
effective interaction,
\begin{equation}
S_{\mathrm{int}}^{\mathrm{eff}}
=
S_{\mathrm{int}}
+
\sum_{n\geq2}\Delta S_{\mathrm{local}}^{(n)} .
\label{eq:general-Seff-definition}
\end{equation}
The order-\(n\) correction to the local effective interaction is
obtained by extracting the local logarithmic part of the corresponding
term in the expansion of \(e^{-S_{\mathrm{int}}}\),
\begin{equation}
\Delta S_{\mathrm{local}}^{(n)}
=
-
\left[
\frac{(-1)^n}{n!}\,
S_{\mathrm{int}}^n
\right]_{\mathrm{local,log}} .
\label{eq:general-deltaSn}
\end{equation}
In the quartic basis, this local contribution has the form
\begin{equation}
\Delta S_{\mathrm{local}}^{(n)}
=
-\int d\tau\,dx\,
\sum_A
\delta J_A^{(n)}(\mu)\,
\mathcal O_A(\tau,x),
\label{eq:general-deltaS-local}
\end{equation}
where \(\delta J_A^{(n)}(\mu)\sim O(J^n)\). The renormalization scale
\(\mu\) is introduced when the logarithmic short-distance contribution
is separated from the long-distance matching scale.

The effective interaction can therefore be written as
\begin{equation}
S_{\mathrm{int}}^{\mathrm{eff}}
=
-\int d\tau\,dx\,
\sum_A
\left[
J_A^{\mathrm{bare}}
+
\sum_{n\geq2}\delta J_A^{(n)}(\mu)
\right]
\mathcal O_A(\tau,x).
\label{eq:general-Seff-couplings}
\end{equation}
Equivalently, the renormalized microscopic coupling \(J_A(\mu)\) is
defined by
\begin{equation}
J_A(\mu)
=
J_A^{\mathrm{bare}}
+
\sum_{n\geq2}\delta J_A^{(n)}(\mu),
\end{equation}
or
\begin{equation}
J_A^{\mathrm{bare}}
=
J_A(\mu)
-
\sum_{n\geq2}\delta J_A^{(n)}(\mu).
\label{eq:general-bare-coupling}
\end{equation}
Since the bare coupling is independent of \(\mu\), differentiating
Eq.~\eqref{eq:general-bare-coupling} gives the microscopic beta
function
\begin{equation}
\hat\beta_A
\equiv
\mu\frac{dJ_A(\mu)}{d\mu}
=
\mu\frac{d}{d\mu}
\sum_{n\geq2}\delta J_A^{(n)}(\mu).
\label{eq:general-micro-beta}
\end{equation}
This quantity describes the running of an individual microscopic
coupling for a fixed disorder realization.

The physically relevant large-\(N\) variables are the
disorder-averaged sector strengths introduced in
Sec.~\ref{sec:model}. The positive disorder strength
\(J_{\mathcal S}\) associated with a quartic chirality sector
\(\mathcal S\) is therefore promoted to a scale-dependent quantity
\(J_{\mathcal S}(\mu)\), defined by
\begin{equation}
\sum_{A\in\mathcal S}
\overline{
J_A(\mu)^2
}
=
d_{\mathcal S}C_{\mathcal S}
J_{\mathcal S}^2(\mu).
\label{eq:general-sector-second-moment}
\end{equation}
Differentiating Eq.~\eqref{eq:general-sector-second-moment} with
respect to \(\mu\) gives the sector beta function
\begin{equation}
\beta_{\mathcal S}
\equiv
\mu\frac{dJ_{\mathcal S}(\mu)}{d\mu}
=
\frac{1}{d_{\mathcal S}C_{\mathcal S}J_{\mathcal S}(\mu)}
\sum_{A\in\mathcal S}
\overline{
J_A(\mu)\hat\beta_A
}.
\label{eq:general-physical-from-micro}
\end{equation}
Thus the collective renormalization-group flow is obtained by
projecting the microscopic running onto the disorder-averaged sector
variables.

Equation~\eqref{eq:general-physical-from-micro} has an immediate
consequence for the perturbative expansion. A microscopic contribution
\(\hat\beta_A^{(n)}\sim O(J^n)\) enters the sector beta function through
an ensemble average of order \(n+1\). Since the disorder ensemble is
centered and Gaussian, all odd moments of the microscopic couplings
vanish. Therefore every even microscopic order gives no contribution
to the disorder-averaged sector flow,
\begin{equation}
\beta_{\mathcal S}^{(n)}=0,
\qquad
\forall\, n \ \mathrm{even}.
\label{eq:even-orders-vanish}
\end{equation}
In particular, the second-order contribution vanishes for all quartic
chirality sectors,
\begin{equation}
\beta_{4L}^{(2)}
=
\beta_{4R}^{(2)}
=
\beta_{2L2R}^{(2)}
=
\beta_{3L1R}^{(2)}
=
\beta_{1L3R}^{(2)}
=
0.
\label{eq:second-physical-vanish}
\end{equation}
The first non-trivial order contributing to the collective flow is therefore
third order. The remaining task is to determine which third-order
short-distance structures produce local logarithms and survive the
large-\(N\) sector projection.

\subsection{First nonvanishing contribution: third-order sector flow}
\label{subsec:CPT-third-order}

We now compute the cubic contribution, which is the first order that
can contribute to the disorder-averaged sector flow. The calculation
proceeds through three successive filters. First, contraction counting
and rotational neutrality determine which coordinate kernels can produce
logarithmic scale dependence. Second, the shape-space analysis separates
primitive logarithms from endpoint and overlapping short-distance
singularities. Finally, the large-\(N\) sector projection selects which
primitive tensor structures survive in the collective disorder-averaged
beta functions.

Using Eq.~\eqref{eq:general-deltaSn}, the \(O(J^3)\) local contribution
is
\begin{equation}
\Delta S_{\mathrm{local}}^{(3)}
=
\left[
\frac{1}{3!}S_{\mathrm{int}}^3
\right]_{\mathrm{local,log}} .
\label{eq:third-deltaS-def}
\end{equation}
Substituting Eq.~\eqref{eq:general-Sint}, one obtains
\begin{equation}
\frac{1}{3!}S_{\mathrm{int}}^3
=
-\frac{1}{3!}
\sum_{B,C,D}
J_B^{\mathrm{bare}}
J_C^{\mathrm{bare}}
J_D^{\mathrm{bare}}
\int d^2X\,d^2Y\,d^2Z\;
\mathcal O_B(X)\,
\mathcal O_C(Y)\,
\mathcal O_D(Z),
\label{eq:third-S3-start}
\end{equation}
where \(X=(\tau_X,x_X)\), \(Y=(\tau_Y,x_Y)\), and
\(Z=(\tau_Z,x_Z)\) denote Euclidean spacetime points.

\paragraph{Cubic local expansion.}

We choose \(Y\) as the local reference point and introduce the relative
coordinates
\begin{equation}
X=Y+r,
\qquad
Z=Y-s,
\label{eq:third-relative-coordinates}
\end{equation}
with \(r=(\tau_r,x_r)\) and \(s=(\tau_s,x_s)\). The three pairwise
separations are then \(r\), \(s\), and \(r+s\), corresponding
respectively to \(X-Y\), \(Y-Z\), and \(X-Z\). We regulate the
short-distance region by imposing the pairwise cutoff
\begin{equation}
|r|>\epsilon,
\qquad
|s|>\epsilon,
\qquad
|r+s|>\epsilon,
\label{eq:third-pairwise-cutoff}
\end{equation}
and bound the overall short-distance neighborhood by a macroscopic
matching scale \(L\). This cutoff excludes the singular coincidence
loci while retaining the near-coincident region responsible for
logarithmic short-distance dependence.

In this regulated short-distance region, the product of three quartic
insertions admits a local expansion organized by contraction topologies,
\begin{equation}
\mathcal O_B(Y+r)\,
\mathcal O_C(Y)\,
\mathcal O_D(Y-s)
=
\sum_A
\sum_\alpha
\mathcal T_{BCD}^{(\alpha)\,A}\,
K^{(\alpha)}(r,s)\,
\mathcal O_A(Y)
+\cdots .
\label{eq:third-short-distance}
\end{equation}
Here \(\mathcal T_{BCD}^{(\alpha)\,A}\) denotes the algebraic tensor
obtained from the Wick contractions prescribed by the topology
\(\alpha\) and from the projection of the remaining fields onto the
local quartic monomial \(\mathcal O_A(Y)\). It contains the flavor
structure, chirality data, fermionic signs, and combinatorial factors
of the contraction. The factor \(K^{(\alpha)}(r,s)\) contains the
coordinate dependence generated by the free propagators. The ellipsis
denotes terms outside the marginal quartic basis, including
higher-dimensional local monomials and nonlocal contributions, which do
not renormalize the marginal couplings.

\paragraph{Kernel homogeneity and rotational neutrality.}

The coordinate dependence is fixed by the free holomorphic and
antiholomorphic propagators. Since the three insertion points are
separated by \(r\), \(s\), and \(r+s\), each contraction topology is
naturally characterized by the distribution of its propagators among
these three pairwise separations. A generic coordinate kernel can
therefore be written as
\begin{align}
K^{(\alpha)}(r,s)
&=
\frac{1}{
(2\pi)^{a_\alpha+b_\alpha+c_\alpha+a'_\alpha+b'_\alpha+c'_\alpha}
}
(\tau_r+i x_r)^{-a_\alpha}
(\tau_s+i x_s)^{-b_\alpha}
\bigl[(\tau_r+\tau_s)+i(x_r+x_s)\bigr]^{-c_\alpha}
\nonumber\\
&\quad\times
(\tau_r-i x_r)^{-a'_\alpha}
(\tau_s-i x_s)^{-b'_\alpha}
\bigl[(\tau_r+\tau_s)-i(x_r+x_s)\bigr]^{-c'_\alpha}.
\label{eq:third-coordinate-kernel}
\end{align}
The unprimed exponents count left-moving contractions, while the primed
exponents count right-moving contractions. More precisely, the pairs
\((a_\alpha,a'_\alpha)\), \((b_\alpha,b'_\alpha)\), and
\((c_\alpha,c'_\alpha)\) measure the propagator content carried by
\(X-Y\), \(Y-Z\), and \(X-Z\), respectively.

Define
\begin{equation}
m_\alpha
=
a_\alpha+b_\alpha+c_\alpha,
\qquad
n_\alpha
=
a'_\alpha+b'_\alpha+c'_\alpha .
\label{eq:third-mn-def}
\end{equation}
The sum \(m_\alpha+n_\alpha\) counts the total number of propagators in
the contraction topology \(\alpha\). Since three quartic monomials
contain twelve fermion fields, reducing them to a single local quartic
monomial requires four pairwise contractions. Hence
\begin{equation}
m_\alpha+n_\alpha=4.
\label{eq:third-counting}
\end{equation}

After the product of insertions has been projected onto a fixed local
quartic monomial \(\mathcal O_A(Y)\), the remaining coordinate kernel is
integrated over the relative configuration. The integral over rigid
rotations of this configuration kills any kernel with nonzero angular
charge. For a given topology, this angular charge is measured by
\(m_\alpha-n_\alpha\), so rotational neutrality requires
\(m_\alpha=n_\alpha\). Together with
Eq.~\eqref{eq:third-counting}, this gives
\begin{equation}
m_\alpha=n_\alpha=2.
\label{eq:third-selection}
\end{equation}
This is the first selection rule of the third-order analysis. Each
chiral block carries two propagators. The unprimed triple
\((a_\alpha,b_\alpha,c_\alpha)\) can therefore be chosen in six ways,
and the primed triple \((a'_\alpha,b'_\alpha,c'_\alpha)\) in six ways.
Thus the neutral coordinate problem contains \(36\) kernel classes. This
count is still kinematical since it identifies the scalar logarithmic
candidates, but does not yet decide which of them define primitive local
logarithms.

\paragraph{Shape-space classification of primitive logarithms.}

At third order, the local logarithmic contribution must be further
decomposed into primitive terms and contributions tied to nested
short-distance regions. To perform this separation, we isolate the
common scale and overall orientation of the three-point configuration
from its relative shape. Introduce polar variables for the two relative
vectors,
\begin{equation}
\tau_r+i x_r=\rho_r e^{i\theta_r},
\qquad
\tau_s+i x_s=\rho_s e^{i\theta_s},
\label{eq:third-polar-rs}
\end{equation}
and define
\begin{equation}
\rho\equiv\rho_r,
\qquad
\lambda\equiv\frac{\rho_s}{\rho_r},
\qquad
\varphi\equiv\theta_r,
\qquad
\phi\equiv\theta_s-\theta_r .
\label{eq:third-scale-shape}
\end{equation}
Here \(\rho\) and \(\varphi\) describe the common scale and overall
orientation, while the dimensionless ratio \(\lambda\) and the relative
angle \(\phi\) encode the shape of the configuration. In these variables,
\begin{align}
\tau_r+i x_r
&=
\rho e^{i\varphi},
\nonumber\\
\tau_s+i x_s
&=
\rho \lambda\,e^{i(\varphi+\phi)},
\nonumber\\
(\tau_r+\tau_s)+i(x_r+x_s)
&=
\rho e^{i\varphi}
\left(1+\lambda e^{i\phi}\right).
\label{eq:third-shape-identities}
\end{align}
The regulated region in Eq.~\eqref{eq:third-pairwise-cutoff} becomes
\begin{equation}
\epsilon\le \rho\le L,
\qquad
\frac{\epsilon}{\rho}\le \lambda\le \frac{L}{\rho},
\qquad
|1+\lambda e^{i\phi}|>\frac{\epsilon}{\rho},
\qquad
0\le\varphi<2\pi,
\qquad
0\le\phi<2\pi .
\label{eq:third-domain}
\end{equation}
The first two inequalities regulate the pairwise collapses \(X\to Y\)
and \(Y\to Z\), while the third regulates the approach to the
coincidence locus \(X\to Z\).

Substituting Eq.~\eqref{eq:third-shape-identities} into
Eq.~\eqref{eq:third-coordinate-kernel}, the kernel factorizes as
\begin{equation}
K^{(\alpha)}(\rho,\lambda,\varphi,\phi)
=
\frac{
e^{-i(m_\alpha-n_\alpha)\varphi}
}{
(2\pi)^{m_\alpha+n_\alpha}\rho^{m_\alpha+n_\alpha}
}
F^{(\alpha)}(\lambda,\phi),
\label{eq:third-factorized-kernel}
\end{equation}
where
\begin{equation}
F^{(\alpha)}(\lambda,\phi)
=
\lambda^{-(b_\alpha+b'_\alpha)}
e^{-i(b_\alpha-b'_\alpha)\phi}
\left(1+\lambda e^{i\phi}\right)^{-c_\alpha}
\left(1+\lambda e^{-i\phi}\right)^{-c'_\alpha}.
\label{eq:third-shape-function}
\end{equation}
The \(X-Y\) exponents \(a_\alpha\) and \(a'_\alpha\) have been absorbed
into the common radial and orientation factors, whereas
\(b_\alpha,b'_\alpha,c_\alpha,c'_\alpha\) determine the dimensionless
shape integral. The pair \(b_\alpha,b'_\alpha\) fixes the power of
\(\lambda\) and the angular Fourier mode, while \(c_\alpha,c'_\alpha\)
controls the behavior near the \(X-Z\) coincidence locus.

Combining this factorization with
\(d^2r\,d^2s=\rho^3\lambda\,d\rho\,d\lambda\,d\varphi\,d\phi\), and using
\(m_\alpha+n_\alpha=4\), the full coordinate integral becomes
\begin{equation}
\int d^2r\,d^2s\,
K^{(\alpha)}(r,s)
=
\frac{1}{16\pi^4}
\int_{\mathcal D_\epsilon}
\frac{d\rho}{\rho}\,
d\lambda\,d\varphi\,d\phi\;
\lambda\,
e^{-i(m_\alpha-n_\alpha)\varphi}
F^{(\alpha)}(\lambda,\phi),
\label{eq:third-kernel-integral}
\end{equation}
where \(\mathcal D_\epsilon\) denotes the regulated domain specified in
Eq.~\eqref{eq:third-domain}. The integral over the overall orientation
gives
\begin{equation}
\int_0^{2\pi}d\varphi\,
e^{-i(m_\alpha-n_\alpha)\varphi}
=
2\pi\,
\delta_{m_\alpha,n_\alpha},
\label{eq:third-angular}
\end{equation}
thereby enforcing Eq.~\eqref{eq:third-selection} directly at the level
of the regulated coordinate integral.

After the orientation integral imposes \(m_\alpha=n_\alpha=2\), the
radial integral has logarithmic form. The remaining task is to decide
whether the shape integral supplies a finite coefficient,
\begin{equation}
\int d\lambda\,d\phi\,
\lambda F^{(\alpha)}(\lambda,\phi),
\end{equation}
or instead retains dependence on a regulated endpoint. The possible
nontrivial singularity in shape space is controlled by the
\(X-Z\) factors. The combination
\((1+\lambda e^{\pm i\phi})\) becomes small precisely at
\((\lambda,\phi)=(1,\pi)\), corresponding to the coincidence limit
\(X\to Z\).

This structure is exposed most directly by rewriting the \(\phi\)
integral as a contour integral. Setting \(w=e^{i\phi}\), one finds
\begin{equation}
\int_0^{2\pi}d\phi\;\lambda F^{(\alpha)}(\lambda,\phi)
=
\frac{\lambda^{1-(b_\alpha+b'_\alpha)}}{i}
\oint_{|w|=1}dw\;
w^{-1-(b_\alpha-b'_\alpha)+c'_\alpha}
(1+\lambda w)^{-c_\alpha}
(w+\lambda)^{-c'_\alpha}.
\label{eq:shape-contour}
\end{equation}
The moving poles associated with the \(X-Z\) factors are located at
\(w=-1/\lambda\) and \(w=-\lambda\), whenever the corresponding
exponents are nonzero. As \(\lambda\) crosses \(1\), these poles
exchange their position relative to the unit circle. The point
\(\lambda=1\) is therefore the analytic remnant of the \(X-Z\)
coincidence channel.

The classification is controlled by the pair
\((c_\alpha,c'_\alpha)\). If \(c_\alpha=c'_\alpha=0\), the topology does
not probe the \(X-Z\) coincidence channel and is not part of the
primitive one-sided class analyzed here. If both exponents are nonzero,
the regulated shape region becomes singular at
\((\lambda,\phi)=(1,\pi)\). Writing locally
\(\lambda=1+\delta\) and \(\phi=\pi+\theta\), one has
\begin{equation}
|1+\lambda e^{i\phi}|^2
\sim
\delta^2+\theta^2 .
\label{eq:shape-local-singularity}
\end{equation}
Since the pairwise regulator imposes
\(|r+s|=\rho |1+\lambda e^{i\phi}|>\epsilon\), the shape cutoff is
explicitly \(\rho\)-dependent. In the logarithmic case this produces an
additional contribution
\begin{equation}
\int_\epsilon^L\frac{d\rho}{\rho}\log\frac{\rho}{\epsilon}
=
\frac12\log^2\frac{L}{\epsilon},
\end{equation}
which signals an overlapping short-distance singularity rather than a
primitive cubic counterterm. Such terms are excluded from the primitive
coefficient \(\zeta_\alpha\). The detailed classification is given in
Appendix~\ref{appendix:shape-classification}.

The only remaining possibility is that exactly one of \(c_\alpha\) or
\(c'_\alpha\) is nonzero. For \(c_\alpha>0\) and \(c'_\alpha=0\), the
kinematic conditions \(m_\alpha=n_\alpha=2\) give nine candidates,
\begin{equation}
\begin{aligned}
(a_\alpha,b_\alpha,c_\alpha)
&\in
\{(0,0,2),(1,0,1),(0,1,1)\},
\\[3pt]
(a'_\alpha,b'_\alpha,c'_\alpha)
&\in
\{(2,0,0),(1,1,0),(0,2,0)\}.
\end{aligned}
\label{eq:primitive-candidate-left}
\end{equation}
The contour analysis in
Appendix~\ref{appendix:shape-classification} shows that six of these
nine candidates generate primitive logarithmic coefficients,
\begin{equation}
\begin{aligned}
&(0,0,2;\,2,0,0),
\qquad
(0,0,2;\,0,2,0),
\qquad
(1,0,1;\,2,0,0),
\\[3pt]
&(1,0,1;\,0,2,0),
\qquad
(0,1,1;\,2,0,0),
\qquad
(0,1,1;\,0,2,0).
\end{aligned}
\label{eq:primitive-six-one-sided}
\end{equation}
The remaining three one-sided candidates either vanish after contour
integration or produce endpoint logarithms in shape space, which are
subdivergent rather than primitive. Exchanging primed and unprimed
contraction data gives the chirality-conjugate one-sided class. Thus the
full primitive set contains twelve logarithmic kernels, arranged into
six conjugate pairs. Equivalently, the \(36\) neutral kernel classes
reduce to six independent coordinate structures up to chirality
conjugation.

\paragraph{Projection to disorder-averaged sector flow.}

For each surviving primitive kernel, the local logarithmic part of the
coordinate integral takes the form
\begin{equation}
\left[
\int d^2r\,d^2s\;
K^{(\alpha)}(r,s)
\right]_{\mathrm{local,log}}
=
\zeta_\alpha
\log\frac{L}{\epsilon}.
\label{eq:third-zeta-alpha}
\end{equation}
The coefficient \(\zeta_\alpha\) contains the full numerical weight of
the primitive coordinate integral, including the propagator
normalization carried by \(K^{(\alpha)}\). The macroscopic matching
scale may equivalently be identified as \(L\sim\mu^{-1}\), up to
scheme-dependent constants, so that the logarithm is written as
\(\log(1/\mu\epsilon)\).

Substituting the projected local expansion into
Eq.~\eqref{eq:third-S3-start}, the primitive logarithmic contribution
takes the form
\begin{equation}
\Delta S_{\mathrm{local}}^{(3)}
=
-\int d^2Y\,
\sum_A
\delta J_A^{(3)}(\mu)\,
\mathcal O_A(Y),
\label{eq:third-deltaS-local}
\end{equation}
with
\begin{equation}
\delta J_A^{(3)}(\mu)
=
\frac{1}{3!}
\log\frac{1}{\mu\epsilon}
\sum_{B,C,D}
J_B^{\mathrm{bare}}
J_C^{\mathrm{bare}}
J_D^{\mathrm{bare}}
\sum_{\alpha\in\mathcal P_{BCD}^{A}}
\zeta_\alpha\,
\mathcal T_{BCD}^{(\alpha)\,A}.
\label{eq:third-deltaJ-general}
\end{equation}
Here \(\mathcal P_{BCD}^{A}\) denotes the set of primitive contraction
topologies in the local channel that takes the ordered triple
\((B,C,D)\) into the quartic monomial \(A\). The unrestricted sum over
\(B,C,D\) keeps track of the microscopic placements of the three
interaction insertions.

Since
\(J_A^{\mathrm{bare}}=J_A(\mu)-\delta J_A^{(3)}(\mu)+\cdots\) is
independent of the renormalization scale, the bare couplings inside
Eq.~\eqref{eq:third-deltaJ-general} may be replaced by the renormalized
couplings at this order. Using
\(\mu\,d\log(1/\mu\epsilon)/d\mu=-1\), one obtains
\begin{equation}
\hat\beta_A^{(3)}
\equiv
\left.
\mu\frac{dJ_A(\mu)}{d\mu}
\right|_{O(J^3)}
=
-
\frac{1}{3!}
\sum_{B,C,D}
J_B(\mu)\,
J_C(\mu)\,
J_D(\mu)
\sum_{\alpha\in\mathcal P_{BCD}^{A}}
\zeta_\alpha\,
\mathcal T_{BCD}^{(\alpha)\,A}.
\label{eq:third-micro-beta}
\end{equation}
Equation~\eqref{eq:third-micro-beta} gives the running of individual
microscopic couplings. The physical large-\(N\) flow is obtained by
projecting this microscopic running onto the disorder-averaged sector
variables using Eq.~\eqref{eq:general-physical-from-micro}. For a final
quartic chirality sector \(\mathcal S\), this gives
\begin{equation}
\beta_{\mathcal S}^{(3)}
=
-
\frac{1}{3!\,d_{\mathcal S}C_{\mathcal S}J_{\mathcal S}(\mu)}
\sum_{A\in\mathcal S}
\sum_{B,C,D}
\sum_{\alpha\in\mathcal P_{BCD}^{A}}
\zeta_\alpha\,
\mathcal T_{BCD}^{(\alpha)\,A}
\,
\overline{
J_A(\mu)
J_B(\mu)
J_C(\mu)
J_D(\mu)
}.
\label{eq:third-sector-beta-fourpoint}
\end{equation}
For the centered Gaussian ensemble, the fourth moment factorizes as
\begin{align}
\overline{
J_AJ_BJ_CJ_D
}
&=
\overline{J_AJ_B}\;
\overline{J_CJ_D}
+
\overline{J_AJ_C}\;
\overline{J_BJ_D}
+
\overline{J_AJ_D}\;
\overline{J_BJ_C}.
\label{eq:third-wick-pairing}
\end{align}
Since the covariance is diagonal in the canonical coupling basis and
sector-wise, the external coupling \(J_A\), with \(A\in\mathcal S\),
can pair only with a microscopic coupling in the same sector
\(\mathcal S\). The remaining two microscopic couplings must then
belong to a common quartic chirality sector \(\mathcal S'\). The
disorder average therefore organizes the surviving contributions
according to sector pairings of the form
\begin{equation}
\mathcal S\times\mathcal S'\times\mathcal S'
\longrightarrow
\mathcal S .
\label{eq:third-sector-selection}
\end{equation}

After the chirality content of the primitive kernels is imposed, the
only sector pairings that remain in the complete quartic helical
interaction space are those with \(\mathcal S'=\{2L2R\}\). Ordered by
the chirality content of the final sector, the surviving structures are
\begin{align}
\{4L\}\times \{2L2R\}\times \{2L2R\}
&\longrightarrow
\{4L\},
\label{eq:third-pure-sector-4L}
\\[3pt]
\{4R\}\times \{2L2R\}\times \{2L2R\}
&\longrightarrow
\{4R\},
\label{eq:third-pure-sector-4R}
\\[3pt]
\{2L2R\}\times \{2L2R\}\times \{2L2R\}
&\longrightarrow
\{2L2R\},
\label{eq:third-balanced-sector-structure}
\\[3pt]
\{3L1R\}\times \{2L2R\}\times \{2L2R\}
&\longrightarrow
\{3L1R\},
\label{eq:third-imbalanced-sector-3L1R}
\\[3pt]
\{1L3R\}\times \{2L2R\}\times \{2L2R\}
&\longrightarrow
\{1L3R\}.
\label{eq:third-imbalanced-sector-1L3R}
\end{align}
These structures organize the surviving disorder contractions in terms
of one representative contribution for each sector pairing. The
unrestricted sum over the three interaction insertions and the three
Gaussian Wick pairings determines how many equivalent microscopic
realizations contribute to a given representative tensor structure. For
\(\mathcal S'\neq\mathcal S\), the external-sector coupling can occupy
three insertion positions, after which the remaining Wick pairing is
fixed. For \(\mathcal S'=\mathcal S\), the insertion ordering is
indistinguishable at the sector level, while the fourth Gaussian moment
supplies three Wick pairings. These factors, together with the
antisymmetric sector multiplicities already encoded in the sector
normalization, give the net coefficient \(\kappa_{\mathcal S}\) used
below.

The structural third-order beta function can then be written uniformly
as
\begin{equation}
\beta_{\mathcal S}^{(3)}
\simeq
-\frac{\kappa_{\mathcal S}}{N^7}\,
J_{\mathcal S}J_{2L2R}^{2}
\sum_{A\in\mathcal S}
\sum_{B\in\{2L2R\}}
\sum_{\alpha\in\mathcal P_{\mathcal S}^{\mathrm{prim}}}
\zeta_\alpha\,
\mathcal T_{ABB}^{(\alpha)\,A}.
\label{eq:third-beta-structure-general}
\end{equation}
Here \(A\in\mathcal S\) and \(B\in\{2L2R\}\) run over the canonical
independent components of the corresponding antisymmetric sectors. The
set \(\mathcal P_{\mathcal S}^{\mathrm{prim}}\) contains the primitive
kernels compatible with the corresponding sector pairing. The
combinatorial factor is \(\kappa_{\mathcal S}=4!\) for
\(\mathcal S=\{4L\},\{4R\}\), \(\kappa_{\mathcal S}=3!\) for
\(\mathcal S=\{3L1R\},\{1L3R\}\), and \(\kappa_{\mathcal S}=4\) for
\(\mathcal S=\{2L2R\}\). The symbol \(\simeq\) denotes equality at the
level of the structural large-\(N\) form, before the sector-summed
tensor coefficients and their leading large-\(N\) scaling are evaluated
explicitly.

\paragraph{Large-\(N\) graphical selection and final beta functions.}

Although Eq.~\eqref{eq:third-beta-structure-general} gives a uniform
sector-level expression, it still contains sector-summed tensor
structures whose large-\(N\) scaling has not yet been evaluated. The
coordinate analysis has reduced the \(36\) neutral kernel classes to
twelve primitive logarithmic kernels, or six independent structures up
to chirality conjugation. The remaining step is the large-\(N\) flavor
count, which determines which primitive structures survive the overall
\(N^{-7}\) normalization. Table~\ref{tab:topologies-by-beta} records the
sector assignment before this final large-\(N\) selection is imposed.

\begin{table}[tbp]
\centering
\small
\renewcommand{\arraystretch}{1.35}
\begin{tabularx}{\textwidth}{
>{\raggedright\arraybackslash}p{3.3cm}
>{\raggedright\arraybackslash}X}
\toprule
\textbf{Beta function} &
\textbf{Sector-compatible primitive kernels before large-\(N\) selection} \\
\midrule
\(\beta_{4L}^{(3)}\)
&
\((0,0,2;\,0,2,0)\),
\((1,0,1;\,0,2,0)\),
\((0,1,1;\,0,2,0)\)
\\[4pt]
\(\beta_{4R}^{(3)}\)
&
\((0,2,0;\,0,0,2)\),
\((0,2,0;\,1,0,1)\),
\((0,2,0;\,0,1,1)\)
\\[4pt]
\(\beta_{3L1R}^{(3)}\)
&
\((0,0,2;\,0,2,0)\),
\((1,0,1;\,0,2,0)\),
\((0,1,1;\,0,2,0)\),
\((2,0,0;\,0,1,1)\),
\((0,2,0;\,0,1,1)\)
\\[4pt]
\(\beta_{1L3R}^{(3)}\)
&
\((0,1,1;\,2,0,0)\),
\((0,1,1;\,0,2,0)\),
\((0,2,0;\,0,0,2)\),
\((0,2,0;\,1,0,1)\),
\((0,2,0;\,0,1,1)\)
\\[4pt]
\(\beta_{2L2R}^{(3)}\)
&
all twelve primitive logarithmic kernels
\\
\bottomrule
\end{tabularx}
\caption{
Sector assignment of primitive logarithmic kernels before the
large-\(N\) flavor-counting selection.
}
\label{tab:topologies-by-beta}
\end{table}

The large-\(N\) selection is a flavor-counting problem. For a final
sector \(\mathcal S\) and a primitive topology \(\alpha\), define
\begin{equation}
\sum_{A\in\mathcal S}
\sum_{B\in\{2L2R\}}
\mathcal T_{ABB}^{(\alpha)\,A}
\sim
N^{p_{\mathcal S,\alpha}} .
\label{eq:largeN-counting-definition}
\end{equation}
Here again the sector sums run over canonical independent components.
Because the structural beta function in
Eq.~\eqref{eq:third-beta-structure-general} carries an overall
\(N^{-7}\) normalization, only sector-topology entries with
\(p_{\mathcal S,\alpha}=7\) survive in the strict large-\(N\) limit.

\begin{figure}[tbp]
\centering
\small
\setlength{\tabcolsep}{8pt}
\renewcommand{\arraystretch}{1.3}
\begin{tabular}{cc}
\begin{tikzpicture}[
    scale=0.85,
    transform shape,
    every node/.style={font=\scriptsize},
    line/.style={thick, black},
    ext/.style={thin, black},
    vtx/.style={circle, fill=yellow!35, draw=black, inner sep=1.6pt}
]
\path[use as bounding box] (-1.5,-2.2) rectangle (7.0,2.2);
\node at (2.6,1.8)
{\((a)\qquad\boldsymbol{(0,0,2;\,2,0,0)\leftrightarrow(2,0,0;\,0,0,2)}\)};
\node[vtx,label=below:\(Z\)] (Z) at (0,0) {};
\node[vtx,label=below:\(X\)] (X) at (2.6,0) {};
\node[vtx,label=below:\(Y\)] (Y) at (5.2,0) {};
\draw[line,bend left=28] (Z) to node[above] {\(\delta_{i_1k_1}\)} (X);
\draw[line,bend right=28] (Z) to node[below] {\(\delta_{i_2k_2}\)} (X);
\draw[line,bend left=28] (X) to node[above] {\(\delta_{i_3j_3}\)} (Y);
\draw[line,bend right=28] (X) to node[below] {\(\delta_{i_4j_4}\)} (Y);
\draw[ext] (Z) -- ++(-0.85,0.35) node[left] {\(k_3\)};
\draw[ext] (Z) -- ++(-0.85,-0.35) node[left] {\(k_4\)};
\draw[ext] (Y) -- ++(0.85,0.35) node[right] {\(j_1\)};
\draw[ext] (Y) -- ++(0.85,-0.35) node[right] {\(j_2\)};
\node at (2.6,-1.6)
{\(\displaystyle \sum_{A,B}\mathcal T_{ABB}^{(\alpha)\,A}\sim N^4\)};
\end{tikzpicture}
&
\begin{tikzpicture}[
    scale=0.85,
    transform shape,
    every node/.style={font=\scriptsize},
    line/.style={thick, black},
    ext/.style={thin, black},
    vtx/.style={circle, fill=yellow!35, draw=black, inner sep=1.6pt}
]
\path[use as bounding box] (-1.5,-2.2) rectangle (7.0,2.2);
\node at (2.6,1.8)
{\((b)\qquad\boldsymbol{(0,0,2;\,0,2,0)\leftrightarrow(0,2,0;\,0,0,2)}\)};
\node[vtx,label=below:\(X\)] (X) at (0,0) {};
\node[vtx,label=below:\(Z\)] (Z) at (2.6,0) {};
\node[vtx,label=below:\(Y\)] (Y) at (5.2,0) {};
\draw[line,bend left=28] (X) to node[above] {\(\delta_{i_1k_1}\)} (Z);
\draw[line,bend right=28] (X) to node[below] {\(\delta_{i_2k_2}\)} (Z);
\draw[line,bend left=28] (Z) to node[above] {\(\delta_{k_3j_3}\)} (Y);
\draw[line,bend right=28] (Z) to node[below] {\(\delta_{k_4j_4}\)} (Y);
\draw[ext] (X) -- ++(-0.85,0.35) node[left] {\(i_3\)};
\draw[ext] (X) -- ++(-0.85,-0.35) node[left] {\(i_4\)};
\draw[ext] (Y) -- ++(0.85,0.35) node[right] {\(j_1\)};
\draw[ext] (Y) -- ++(0.85,-0.35) node[right] {\(j_2\)};
\node at (2.6,-1.6)
{\(\displaystyle \sum_{A,B}\mathcal T_{ABB}^{(\alpha)\,A}\sim N^6\)};
\end{tikzpicture}
\\[1.8em]

\begin{tikzpicture}[
    scale=0.85,
    transform shape,
    every node/.style={font=\scriptsize},
    line/.style={thick, black},
    ext/.style={thin, black},
    vtx/.style={circle, fill=yellow!35, draw=black, inner sep=1.6pt}
]
\path[use as bounding box] (-2.8,-2.2) rectangle (2.8,3.6);
\node at (0,3.2)
{\((c)\qquad\boldsymbol{(0,1,1;\,2,0,0)\leftrightarrow(2,0,0;\,0,1,1)}\)};
\node[vtx,label=above:\(Z\)] (Z) at (0,1.45) {};
\node[vtx,label=below left:\(Y\)] (Y) at (-1.35,0.25) {};
\node[vtx,label=below right:\(X\)] (X) at (1.35,0.25) {};
\draw[line,bend right=18] (Z) to node[left] {\(\delta_{j_1k_1}\,\)} (Y);
\draw[line,bend left=18] (Z) to node[right] {\(\delta_{i_1k_2}\)} (X);
\draw[line,bend left=42] (Y) to node[below] {\(\delta_{i_2j_3}\)} (X);
\draw[line,bend right=42] (Y) to node[below] {\(\delta_{i_3j_4}\)} (X);
\draw[ext] (Z) -- ++(-0.62,0.76) node[above left] {\(k_3\)};
\draw[ext] (Z) -- ++(0.62,0.76) node[above right] {\(k_4\)};
\draw[ext] (Y) -- ++(-0.72,-0.18) node[left] {\(j_2\)};
\draw[ext] (X) -- ++(0.72,-0.18) node[right] {\(i_4\)};
\node at (0,-1.6)
{\(\displaystyle \sum_{A,B}\mathcal T_{ABB}^{(\alpha)\,A}\sim N^5\)};
\end{tikzpicture}
&
\begin{tikzpicture}[
    scale=0.85,
    transform shape,
    every node/.style={font=\scriptsize},
    line/.style={thick, black},
    ext/.style={thin, black},
    vtx/.style={circle, fill=yellow!35, draw=black, inner sep=1.6pt}
]
\path[use as bounding box] (-2.8,-2.2) rectangle (2.8,3.6);
\node at (0,3.2)
{\((d)\qquad\boldsymbol{(1,0,1;\,0,2,0)\leftrightarrow(0,2,0;\,1,0,1)}\)};
\node[vtx,label=above:\(X\)] (X) at (0,1.45) {};
\node[vtx,label=below left:\(Y\)] (Y) at (-1.35,0.25) {};
\node[vtx,label=below right:\(Z\)] (Z) at (1.35,0.25) {};
\draw[line,bend right=18] (X) to node[left] {\(\delta_{i_1j_1}\,\)} (Y);
\draw[line,bend left=18] (X) to node[right] {\(\delta_{i_2k_1}\)} (Z);
\draw[line,bend left=42] (Y) to node[below] {\(\delta_{j_3k_3}\)} (Z);
\draw[line,bend right=42] (Y) to node[below] {\(\delta_{j_4k_4}\)} (Z);
\draw[ext] (X) -- ++(-0.62,0.76) node[above left] {\(i_3\)};
\draw[ext] (X) -- ++(0.62,0.76) node[above right] {\(i_4\)};
\draw[ext] (Y) -- ++(-0.72,-0.18) node[left] {\(j_2\)};
\draw[ext] (Z) -- ++(0.72,-0.18) node[right] {\(k_2\)};
\node at (0,-1.6)
{\(\displaystyle \sum_{A,B}\mathcal T_{ABB}^{(\alpha)\,A}\sim N^6\)};
\end{tikzpicture}
\\[1.8em]

\begin{tikzpicture}[
    scale=0.85,
    transform shape,
    every node/.style={font=\scriptsize},
    line/.style={thick, black},
    ext/.style={thin, black},
    vtx/.style={circle, fill=yellow!35, draw=black, inner sep=1.6pt}
]
\path[use as bounding box] (-1.5,-2.2) rectangle (7.0,2.4);
\node at (2.6,2.0)
{\((e)\qquad\boldsymbol{(1,0,1;\,2,0,0)\leftrightarrow(2,0,0;\,1,0,1)}\)};
\node[vtx,label=below:\(Z\)] (Z) at (0,0) {};
\node[vtx,label=below:\(X\)] (X) at (2.6,0) {};
\node[vtx,label=below:\(Y\)] (Y) at (5.2,0) {};
\draw[line] (Z) to node[above] {\(\delta_{i_1k_1}\)} (X);
\draw[line,bend left=45] (X) to node[above] {\(\delta_{i_2j_3}\)} (Y);
\draw[line,bend right=28] (X) to node[below] {\(\delta_{i_3j_4}\)} (Y);
\draw[line,bend left=25] (X) to node[below] {\(\delta_{i_4j_1}\)} (Y);
\draw[ext] (Z) -- ++(-0.85,0.55) node[left] {\(k_2\)};
\draw[ext] (Z) -- ++(-0.95,0.00) node[left] {\(k_3\)};
\draw[ext] (Z) -- ++(-0.85,-0.55) node[left] {\(k_4\)};
\draw[ext] (Y) -- ++(0.85,0.00) node[right] {\(j_2\)};
\node at (2.6,-1.6)
{\(\displaystyle \sum_{A,B}\mathcal T_{ABB}^{(\alpha)\,A}\sim N^4\)};
\end{tikzpicture}
&
\begin{tikzpicture}[
    scale=0.85,
    transform shape,
    every node/.style={font=\scriptsize},
    line/.style={thick, black},
    ext/.style={thin, black},
    vtx/.style={circle, fill=yellow!35, draw=black, inner sep=1.6pt}
]
\path[use as bounding box] (-1.5,-2.2) rectangle (7.0,2.4);
\node at (2.6,2.0)
{\((f)\qquad\boldsymbol{(0,1,1;\,0,2,0)\leftrightarrow(0,2,0;\,0,1,1)}\)};
\node[vtx,label=below:\(X\)] (X) at (0,0) {};
\node[vtx,label=below:\(Z\)] (Z) at (2.6,0) {};
\node[vtx,label=below:\(Y\)] (Y) at (5.2,0) {};
\draw[line] (X) to node[above] {\(\delta_{i_1k_1}\)} (Z);
\draw[line,bend left=45] (Z) to node[above] {\(\delta_{j_3k_3}\)} (Y);
\draw[line,bend left=28] (Z) to node[below] {\(\delta_{j_1k_2}\)} (Y);
\draw[line,bend right=28] (Z) to node[below] {\(\delta_{j_4k_4}\)} (Y);
\draw[ext] (X) -- ++(-0.85,0.55) node[left] {\(i_2\)};
\draw[ext] (X) -- ++(-0.95,0.00) node[left] {\(i_3\)};
\draw[ext] (X) -- ++(-0.85,-0.55) node[left] {\(i_4\)};
\draw[ext] (Y) -- ++(0.85,0.00) node[right] {\(j_2\)};
\node at (2.6,-1.6)
{\(\displaystyle \sum_{A,B}\mathcal T_{ABB}^{(\alpha)\,A}\sim N^7\)};
\end{tikzpicture}

\end{tabular}

\caption{
(a)-(f) correspond to the six graphical large-\(N\) counting classes for the primitive tensor
structures. Only the conjugate pair shown in panel (f) scales as
\(N^7\). All other representative classes are subleading.
}
\label{fig:largeN-graphical-selection}
\end{figure}
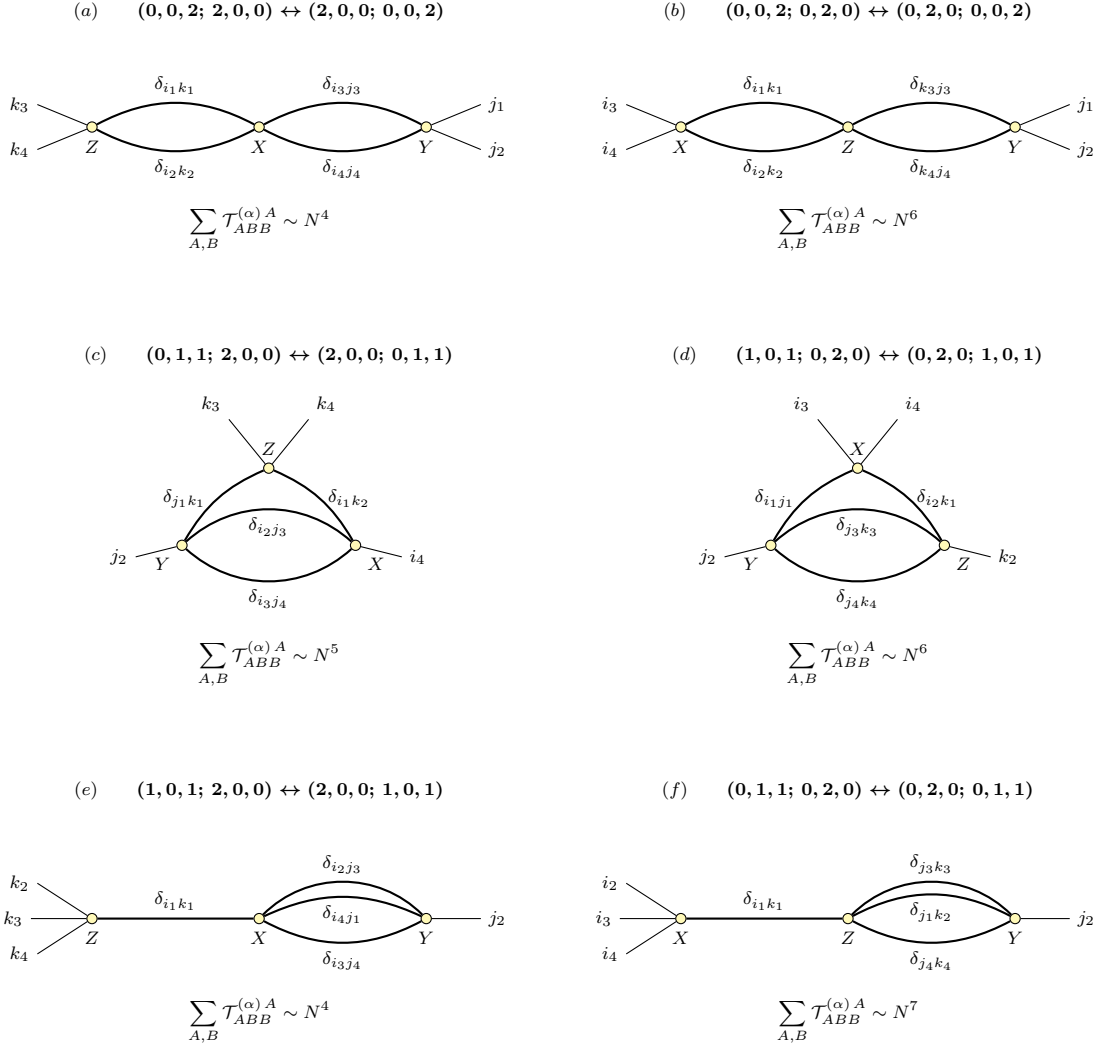

The exponent \(p_{\mathcal S,\alpha}\) is obtained by imposing the
flavor identifications encoded in a primitive topology, together with
the Gaussian covariance identifications between the two \(\{2L2R\}\)
insertions, and then counting the flavor labels that remain independent.
In the graphical representation, the sector labels are suppressed in
order to display the representative covariance-paired tensor structures
that control the flavor count. Figure~\ref{fig:largeN-graphical-selection}
summarizes this counting for the six representative graphical classes.
The panels display only the large-\(N\) scaling information. The
finite-\(N\) antisymmetrization factors and fermionic signs are part of
the tensor-sum calculation in
Appendix~\ref{appendix:tensor-sums-leading-topologies}.

The graphical counting performs the final selection. Among the six
primitive structures independent up to chirality conjugation, only one
retains enough independent flavor sums to contribute at order \(N^7\).
Its two chirality-conjugate representatives are
\begin{equation}
\alpha_1\equiv(0,1,1;\,0,2,0),
\qquad
\alpha_2\equiv(0,2,0;\,0,1,1),
\label{eq:third-leading-alpha-def}
\end{equation}
with \(\alpha_2\) obtained from \(\alpha_1\) by exchanging primed and
unprimed contraction data. All remaining sector-allowed primitive
kernels scale at most as \(N^6\), and are therefore suppressed by the
overall \(N^{-7}\) normalization in
Eq.~\eqref{eq:third-beta-structure-general}.

With the leading topology pair fixed, the remaining coefficients are
purely algebraic flavor-index sums. As throughout the sector-level
analysis, the sums over sector labels are understood to run over the
canonical independent components of the corresponding antisymmetric
coupling tensors. We denote the leading sector sums by
\begin{equation}
\Sigma_{\mathcal S}^{(r)}
\equiv
\sum_{A\in\mathcal S}
\sum_{B\in\{2L2R\}}
\mathcal T_{ABB}^{(\alpha_r)\,A}.
\label{eq:third-leading-Sigma-def}
\end{equation}
The leading large-\(N\) evaluation of these sums, including the relation
between canonical sector sums and redundant unordered sums, is given in
Table~\ref{tab:app-leading-sector-sums} in
Appendix~\ref{appendix:tensor-sums-leading-topologies}. Only
sector-compatible entries are displayed below. The purely chiral sectors
receive a leading contribution from one member of the pair, whereas the
chirality-balanced and chirality-imbalanced sectors receive leading
contributions from both,
\begin{align}
\Sigma_{\{4L\}}^{(1)}
=
\Sigma_{\{4R\}}^{(2)}
&=
-\frac{1}{48}N^7+O(N^6),
\label{eq:third-leading-T-pure}
\\[4pt]
\Sigma_{\{3L1R\}}^{(1)}
=
\Sigma_{\{3L1R\}}^{(2)}
=
\Sigma_{\{1L3R\}}^{(1)}
=
\Sigma_{\{1L3R\}}^{(2)}
&=
-\frac{1}{12}N^7+O(N^6),
\label{eq:third-leading-T-imbalanced}
\\[4pt]
\Sigma_{\{2L2R\}}^{(1)}
=
\Sigma_{\{2L2R\}}^{(2)}
&=
-\frac{1}{8}N^7+O(N^6).
\label{eq:third-leading-T-balanced}
\end{align}
The equalities in
Eqs.~\eqref{eq:third-leading-T-pure}--
\eqref{eq:third-leading-T-balanced} are leading large-\(N\)
statements.

The contour classification in
Appendix~\ref{appendix:shape-classification} gives the geometric weights
of the surviving leading pair,
\begin{equation}
\zeta_{\alpha_1}
=
\zeta_{\alpha_2}
=
\frac{1}{8\pi^2}.
\label{eq:third-leading-zeta-values}
\end{equation}
Substituting Eqs.~\eqref{eq:third-leading-T-pure}--
\eqref{eq:third-leading-T-balanced} and
Eq.~\eqref{eq:third-leading-zeta-values} into
Eq.~\eqref{eq:third-beta-structure-general}, and retaining only the
leading large-\(N\) contribution, gives
\begin{equation}
\beta^{(3)}_{\mathcal S}
=
\frac{n_{\mathcal S}}{16\pi^2}\,
J_{\mathcal S}J_{2L2R}^{2},
\label{eq:universal-beta-function}
\end{equation}
where $\mathcal S\in\{4L,3L1R,2L2R,1L3R,4R\}$, \(n_{\mathcal S}=1\) for the purely chiral sectors and
\(n_{\mathcal S}=2\) for the chirality-balanced and
chirality-imbalanced sectors.

Thus, the chirality-balanced disorder strength \(J_{2L2R}\) controls
the leading third-order sector-level flow. At leading order in large
\(N\), this sector evolves autonomously, while the purely chiral and
chirality-imbalanced disorder strengths are multiplicatively driven by
its running.

\subsection{Infrared structure of the perturbative RG flow}
\label{subsec:IR-structure-renormalized-theory}

With the leading third-order beta functions in
Eq.~\eqref{eq:universal-beta-function} in hand, we can now study the
flow of the sector disorder strengths toward the IR. We parametrize this flow by
\begin{equation}
\ell\equiv\log\frac{\mu_0}{\mu},
\label{eq:IR-ell-def}
\end{equation}
where \(\mu\) is the running renormalization scale and \(\mu_0\) is a
reference scale in the perturbative ultraviolet regime at which the
initial couplings are specified. Increasing \(\ell\) therefore
corresponds to decreasing \(\mu\), and hence to probing lower energies
or longer distances. Since
\(\beta_{\mathcal S}=\mu\,dJ_{\mathcal S}/d\mu\), the flow equations in
terms of \(\ell\) are
\begin{equation}
\frac{dJ_{\mathcal S}}{d\ell}
=
-\beta_{\mathcal S}.
\label{eq:IR-ell-flow}
\end{equation}

It is useful to set
\[
g(\ell)\equiv J_{2L2R}(\ell).
\]
The autonomous equation for \(g\), together with the logarithmic form of
the remaining sector equations for nonzero sector strengths, gives the
triangular system
\begin{align}
\frac{dg}{d\ell}
&=
-\frac{1}{8\pi^2}g^3,
\label{eq:IR-g-flow}
\\[4pt]
\frac{d\log J_{4L}}{d\ell}
=
\frac{d\log J_{4R}}{d\ell}
&=
-\frac{1}{16\pi^2}g^2,
\label{eq:IR-pure-chiral-log-flow}
\\[4pt]
\frac{d\log J_{3L1R}}{d\ell}
=
\frac{d\log J_{1L3R}}{d\ell}
&=
-\frac{1}{8\pi^2}g^2.
\label{eq:IR-imbalanced-log-flow}
\end{align}
Thus \(g\) evolves autonomously, while the remaining sector variables
are multiplicatively renormalized by its running.

The autonomous equation for \(g\) can be integrated directly. With
initial condition \(g(0)=g_0\), one finds
\begin{equation}
g^2(\ell)
=
\frac{g_0^2}
{1+\dfrac{1}{4\pi^2}g_0^2\ell}.
\label{eq:IR-balanced-solution}
\end{equation}
Equivalently,
\begin{equation}
g^2(\ell)
\sim
\frac{4\pi^2}{\ell},
\qquad
\ell\to\infty.
\label{eq:IR-balanced-asymptotic}
\end{equation}
The chirality-balanced disorder strength \(J_{2L2R}\) is therefore
marginally irrelevant. It decreases logarithmically toward the IR.

This autonomous running controls the remaining sectors
multiplicatively. Defining
\begin{equation}
\mathcal R(\ell)
\equiv
1+\frac{1}{4\pi^2}g_0^2\ell,
\label{eq:IR-R-def}
\end{equation}
the purely chiral sector variables obey
\begin{align}
J_{4L}(\ell)
&=
J_{4L}^{(0)}
\mathcal R(\ell)^{-1/4},
\label{eq:IR-J4L-solution}
\\[4pt]
J_{4R}(\ell)
&=
J_{4R}^{(0)}
\mathcal R(\ell)^{-1/4}.
\label{eq:IR-J4R-solution}
\end{align}
Hence
\begin{equation}
J_{4L}(\ell),\;J_{4R}(\ell)
\sim
\ell^{-1/4},
\qquad
\ell\to\infty.
\label{eq:IR-chiral-asymptotic}
\end{equation}
The chirality-imbalanced sectors instead obey
\begin{align}
J_{3L1R}(\ell)
&=
J_{3L1R}^{(0)}
\mathcal R(\ell)^{-1/2},
\label{eq:IR-J3L1R-solution}
\\[4pt]
J_{1L3R}(\ell)
&=
J_{1L3R}^{(0)}
\mathcal R(\ell)^{-1/2}.
\label{eq:IR-J1L3R-solution}
\end{align}
Consequently,
\begin{equation}
J_{3L1R}(\ell),\;J_{1L3R}(\ell)
\sim
\ell^{-1/2},
\qquad
\ell\to\infty.
\label{eq:IR-imbalanced-asymptotic}
\end{equation}

These solutions give a coherent perturbative picture of the IR
regime. When the chirality-balanced disorder strength is present, it is
marginally irrelevant and controls the approach to the IR. Both
\(J_{2L2R}\) and the chirality-imbalanced sector variables decay as
\(\ell^{-1/2}\), while the purely chiral variables decay more slowly, as
\(\ell^{-1/4}\). This difference reflects the triangular structure of
the flow. The chirality-balanced sector evolves autonomously, whereas
the remaining sectors are multiplicatively renormalized by its running.

If \(J_{2L2R}\) is absent from the outset, the leading third-order beta
functions vanish identically, and the remaining sector variables are not
renormalized within the present approximation. Thus, at leading order in
large-\(N\) and to third order in CPT, the
flow does not produce a nontrivial IR attractor. Whenever the
chirality-balanced disorder strength is present, the sector variables
are instead driven logarithmically back toward the free helical
conformal fixed point.

\section{Conclusions, discussions and future work}
\label{sec:conclusions}
We constructed a \(1+1\)-dimensional helical generalization of the SYK model and analyzed its behavior by decomposing the full local quartic interaction space into the five chirality sectors \(\{4L,3L1R,2L2R,1L3R,4R\}\). The first part of our analysis addressed integrability as a symmetry-protected property. In the maximally constrained symmetry sector, the allowed quartic interactions are forced into density--density form. Bosonization then maps this sector to a quadratic chiral-boson theory, which can be canonically diagonalized. This gives an exactly integrable finite-coupling subspace of the helical theory. Once the symmetry constraints are relaxed, the full quartic helical interaction space opens up, the density--density structure is no longer protected, and this finite-coupling integrable structure is lost.

The second part of the analysis showed that integrability reappears through a different mechanism, namely the disorder-averaged large-\(N\) RG flow. The essential physics is governed by short-distance OPE selection rules and Gaussian disorder averaging. The first nonvanishing contribution to the disorder-averaged beta functions arises at third order. At this order, a logarithmic marginal correction requires \(m+n=4\), where \(m\) and \(n\) are the holomorphic and antiholomorphic singularity weights of the OPE kernel. Rotational invariance around the free helical fixed point imposes Lorentz-spin neutrality, \(m=n\), so only spinless short-distance kernels survive the angular average. Thus the surviving logarithms have \(m=n=2\). After this selection rule and the large-\(N\) flavor counting are imposed, only two primitive leading topologies contribute. For each sector-compatible member of this primitive pair, the sector normalization \(\kappa_{\mathcal S}\) and the leading tensor sum \(\Sigma_{\mathcal S}^{(\alpha)}\) combine in the universal way
\[
\frac{\kappa_{\mathcal S}\Sigma_{\mathcal S}^{(\alpha)}}{N^7}
=
-\frac{1}{2}.
\]
The details of this large-\(N\) cancellation are collected in Appendix \ref{appendix:tensor-sums-leading-topologies}. This cancellation is not accidental. The tensor sum \(\Sigma_{\mathcal S}^{(\alpha)}\) counts the leading contractions allowed in a given sector, while the normalization entering the flow of the RMS disorder strength \(J_{\mathcal S}\) removes the corresponding sector multiplicity. What remains is a sector-blind contribution per primitive leading topology. Thus the sector dependence of the full beta-function coefficient is purely combinatorial and enters only through the integer \(n_{\mathcal S}\) counting the surviving leading topologies. The underlying contribution of each primitive leading topology is universal.

The coupled flow therefore has a robust hierarchical structure. The chirality-balanced coupling \(J_{2L2R}\) runs autonomously and is marginally irrelevant. Whenever it is present, it multiplicatively drives the other quartic disorder strengths toward zero. In the conventions used in this work, the beta functions are positive, so the disorder strengths decrease logarithmically toward the IR. The disorder-averaged theory therefore flows back toward the free helical fixed point, causing integrability to re-emerge asymptotically.

This RG behavior differs qualitatively from the original \(0+1\)-dimensional SYK model, where the random interaction is relevant and drives the theory to a strongly interacting, maximally chaotic IR regime. In the Hamiltonian description of the original SYK model, there is no quadratic spatial kinetic term or dispersion competing with the random interaction. By contrast, in the helical theory studied here, the counter-propagating kinetic terms define a free \(1+1\)-dimensional conformal fixed point, with the quartic interactions appearing as marginal perturbations around it.

It is also useful to clarify what kind of RG data are being described, as discussed in Appendix \ref{appendix:Anomalous-dimensions-and-disorder-strength-beta-functions}. The beta functions computed in this work are not beta functions for individual fixed operator couplings. They describe the running of disorder strengths \(J_{\mathcal S}\), which measure the widths of entire Gaussian ensembles of microscopic couplings in each chirality sector. For this reason, their coefficients cannot be read off from the dressing of a single fermion field or a single operator insertion alone. They are obtained by following the RG flow of the full random ensemble and then projecting it back onto the disorder strength \(J_{\mathcal S}\). After this procedure, the result takes a strikingly simple form. Each primitive leading topology contributes universally, and the only remaining sector dependence is the integer number of such topologies.

The broader value of the helical framework is that it places several previously separate \(1+1\)-dimensional SYK-like constructions inside a single interaction space. The purely chiral sectors connect naturally to CSYK-type models, while the chirality-balanced \(2L2R\) sector contains the RT-type dynamics. The full helical theory also includes chirality-imbalanced sectors, making it possible to ask how these different structures coexist and how they are organized by the disorder-averaged RG flow.

A natural next step is to go beyond the perturbative sector-flow analysis and compute the disorder-averaged fermion two-point functions, together with the exact large-\(N\) Schwinger-Dyson structure, for the full helical interaction space. This would also open the door to dynamical questions not addressed by the beta-function analysis, such as whether the full helical theory exhibits thermalization or operator spreading. The connections to discrete systems, briefly hinted at in the introduction, also provide a natural direction for future work, since they could clarify how the helical interaction structure is realized in lattice or condensed-matter systems. Finally, given the AdS/CFT correspondence, it would be interesting to explore whether the helical model can play a role in an effective holographic construction. The random interactions break conformal invariance, preventing an exact correspondence with \(\mathrm{AdS}_3\) spacetime. On the other hand, the RG analysis shows that the disorder-averaged theory returns logarithmically to the free helical conformal fixed point in the IR, implying a possible approximate correspondence with \(\mathrm{AdS}_3\) at non-zero interactions. We hope to return to these questions in the future.

\section*{Acknowledgements}
G.V.-M. was partially supported by the Army Research Office under grant no.~W911NF-23-1-0202 and gratefully acknowledges the Center for Mexican American and Latino/a Studies at the University of Houston for its generous support through the Lydia Mendoza Fellowship. The work of B.T. and E.R.B. is supported by the National Science Foundation grant no.~CHE-2404788 and the Robert A. Welch Foundation (E-1337). P.H. was supported by the National Science Foundation grant no.~DMR-2047193.

\appendix

\section{Canonical diagonalization of the \(N=4\) maximally constrained theory}
\label{app:N4-canonical-diagonalization}

In this appendix we spell out the canonical diagonalization of the
\(N=4\) maximally constrained theory discussed in
Sec.~\ref{subsubsec:N4-example}. The purpose is to make explicit how the
four signed normal-mode velocities are determined from the interaction
matrix, and how the corresponding eigenvectors assemble into the
canonical \(O(2,2)\) transformation used in the main text.

For \(N=4\), the bosonic fields are ordered as
\begin{equation}
\Phi^{T}
=
(\phi_{1,L},\phi_{2,L},\phi_{1,R},\phi_{2,R}),
\qquad
\mathcal K=\operatorname{diag}(1,1,-1,-1).
\label{app-eq:N4-Phi-K}
\end{equation}
The interaction matrix \(\mathcal V\) is given in
Eq.~\eqref{eq:N4-V}. The normal modes are determined by the generalized
eigenvalue problem
\begin{equation}
\mathcal V\,\mathbf v
=
u\,\mathcal K\,\mathbf v .
\label{app-eq:N4-generalized-eig}
\end{equation}
Since \(\mathcal K^{2}=I_4\), this is equivalently the ordinary
eigenvalue problem
\begin{equation}
\mathcal A\,\mathbf v
=
u\,\mathbf v,
\qquad
\mathcal A\equiv\mathcal K\mathcal V .
\label{app-eq:N4-A-eig}
\end{equation}
Explicitly,
\begin{equation}
\mathcal A
=
\begin{pmatrix}
1
&
\dfrac{J_{(1,L)(2,L)}}{2\pi}
&
-\dfrac{J_{(1,L)(1,R)}}{2\pi}
&
-\dfrac{J_{(1,L)(2,R)}}{2\pi}
\\[8pt]
\dfrac{J_{(1,L)(2,L)}}{2\pi}
&
1
&
-\dfrac{J_{(2,L)(1,R)}}{2\pi}
&
-\dfrac{J_{(2,L)(2,R)}}{2\pi}
\\[8pt]
\dfrac{J_{(1,L)(1,R)}}{2\pi}
&
\dfrac{J_{(2,L)(1,R)}}{2\pi}
&
-1
&
-\dfrac{J_{(1,R)(2,R)}}{2\pi}
\\[8pt]
\dfrac{J_{(1,L)(2,R)}}{2\pi}
&
\dfrac{J_{(2,L)(2,R)}}{2\pi}
&
-\dfrac{J_{(1,R)(2,R)}}{2\pi}
&
-1
\end{pmatrix}.
\label{app-eq:N4-A-explicit}
\end{equation}
The lower two rows differ in sign from the corresponding rows of
\(\mathcal V\) because of the multiplication by \(\mathcal K\). The
upper-right signs are inherited directly from the mixed \(LR\)
density--density convention in Eq.~\eqref{eq:N4-V}.

The signed normal-mode velocities \(u\) are the roots of the
characteristic polynomial
\begin{equation}
\det(\mathcal A-uI_4)
=
u^{4}
+
\mathfrak a_{2}\,u^{2}
-
\mathfrak a_{3}\,u
+
\mathfrak a_{4}
=
0 .
\label{app-eq:N4-characteristic}
\end{equation}
The cubic term is absent because
\begin{equation}
\operatorname{tr}\mathcal A=0.
\label{app-eq:A-trace-zero}
\end{equation}
The remaining coefficients are the elementary spectral invariants of
\(\mathcal A\). Using Newton's identities together with
Eq.~\eqref{app-eq:A-trace-zero}, they may be written as
\begin{align}
\mathfrak a_{2}
&=
-\frac{1}{2}\operatorname{tr}\mathcal A^{2},
\label{app-eq:a2-trace}
\\[4pt]
\mathfrak a_{3}
&=
\frac{1}{3}\operatorname{tr}\mathcal A^{3},
\label{app-eq:a3-trace}
\\[4pt]
\mathfrak a_{4}
&=
\det\mathcal A
=
\det\mathcal V .
\label{app-eq:a4-det}
\end{align}
The last equality follows from \(\det\mathcal K=1\).

Evaluating these invariants for Eq.~\eqref{app-eq:N4-A-explicit} gives
\begin{equation}
\begin{aligned}
\mathfrak a_{2}
&=
-2
-
\left(
\frac{J_{(1,L)(2,L)}}{2\pi}
\right)^{2}
-
\left(
\frac{J_{(1,R)(2,R)}}{2\pi}
\right)^{2}
+
\left(
\frac{J_{(1,L)(1,R)}}{2\pi}
\right)^{2}
+
\left(
\frac{J_{(1,L)(2,R)}}{2\pi}
\right)^{2}
\\[4pt]
&\quad
+
\left(
\frac{J_{(2,L)(1,R)}}{2\pi}
\right)^{2}
+
\left(
\frac{J_{(2,L)(2,R)}}{2\pi}
\right)^{2}.
\end{aligned}
\label{app-eq:a2-explicit}
\end{equation}

\begin{equation}
\begin{aligned}
\mathfrak a_{3}
&=
2\Bigg[
\left(
\frac{J_{(1,L)(2,L)}}{2\pi}
\right)^{2}
-
\left(
\frac{J_{(1,R)(2,R)}}{2\pi}
\right)^{2}
\\[4pt]
&\quad
-
\frac{
J_{(1,L)(2,L)}
J_{(1,L)(1,R)}
J_{(2,L)(1,R)}
}{(2\pi)^{3}}
-
\frac{
J_{(1,L)(2,L)}
J_{(1,L)(2,R)}
J_{(2,L)(2,R)}
}{(2\pi)^{3}}
\\[4pt]
&\quad
+
\frac{
J_{(1,R)(2,R)}
J_{(1,L)(1,R)}
J_{(1,L)(2,R)}
}{(2\pi)^{3}}
+
\frac{
J_{(1,R)(2,R)}
J_{(2,L)(1,R)}
J_{(2,L)(2,R)}
}{(2\pi)^{3}}
\Bigg].
\end{aligned}
\label{app-eq:a3-explicit}
\end{equation}

\begin{equation}
\begin{aligned}
\mathfrak a_{4}
&=
1
-
\left(
\frac{J_{(1,L)(2,L)}}{2\pi}
\right)^{2}
-
\left(
\frac{J_{(1,R)(2,R)}}{2\pi}
\right)^{2}
-
\left(
\frac{J_{(1,L)(1,R)}}{2\pi}
\right)^{2}
-
\left(
\frac{J_{(1,L)(2,R)}}{2\pi}
\right)^{2}
\\[4pt]
&\quad
-
\left(
\frac{J_{(2,L)(1,R)}}{2\pi}
\right)^{2}
-
\left(
\frac{J_{(2,L)(2,R)}}{2\pi}
\right)^{2}
+
\left(
\frac{
J_{(1,L)(2,L)}J_{(1,R)(2,R)}
}{(2\pi)^{2}}
\right)^{2}
\\[4pt]
&\quad
+
\left[
\frac{
J_{(2,L)(2,R)}J_{(1,L)(1,R)}
-
J_{(1,L)(2,R)}J_{(2,L)(1,R)}
}{(2\pi)^{2}}
\right]^{2}
\\[4pt]
&\quad
+
\frac{
2J_{(1,L)(2,L)}
\left[
J_{(1,L)(1,R)}J_{(2,L)(1,R)}
+
J_{(1,L)(2,R)}J_{(2,L)(2,R)}
\right]
}{(2\pi)^{3}}
\\[4pt]
&\quad
+
\frac{
2J_{(1,R)(2,R)}
\left[
J_{(1,L)(1,R)}J_{(1,L)(2,R)}
+
J_{(2,L)(1,R)}J_{(2,L)(2,R)}
\right]
}{(2\pi)^{3}}
\\[4pt]
&\quad
-
\frac{
2J_{(1,L)(2,L)}J_{(1,R)(2,R)}
\left[
J_{(1,L)(1,R)}J_{(2,L)(2,R)}
+
J_{(1,L)(2,R)}J_{(2,L)(1,R)}
\right]
}{(2\pi)^{4}} .
\end{aligned}
\label{app-eq:a4-explicit}
\end{equation}
These expressions are the explicit spectral data entering the
\(N=4\) normal-mode velocities.

It remains to solve the quartic equation
Eq.~\eqref{app-eq:N4-characteristic}. To put it in the standard
depressed-quartic form, define
\begin{equation}
p_{4}\equiv\mathfrak a_{2},
\qquad
q_{4}\equiv-\mathfrak a_{3},
\qquad
r_{4}\equiv\mathfrak a_{4}.
\label{app-eq:pqr4-def}
\end{equation}
Then Eq.~\eqref{app-eq:N4-characteristic} becomes
\begin{equation}
u^{4}+p_{4}u^{2}+q_{4}u+r_{4}=0 .
\label{app-eq:depressed-quartic}
\end{equation}
Ferrari's method begins from the identity
\begin{align}
u^{4}+p_{4}u^{2}+q_{4}u+r_{4}
&=
\left(
u^{2}+\frac{p_{4}}{2}+y
\right)^{2}
\nonumber\\
&\quad
-
\left[
2y\,u^{2}
-
q_{4}u
+
\left(
y^{2}+p_{4}y+\frac{p_{4}^{2}}{4}-r_{4}
\right)
\right].
\label{app-eq:ferrari-identity}
\end{align}
The bracket becomes a perfect square in \(u\) provided that its
discriminant vanishes. This gives the resolvent cubic
\begin{equation}
y^{3}
+
p_{4}y^{2}
+
\left(
\frac{p_{4}^{2}}{4}-r_{4}
\right)y
-
\frac{q_{4}^{2}}{8}
=
0 .
\label{app-eq:resolvent-cubic}
\end{equation}

Removing the quadratic term by writing
\begin{equation}
y=z-\frac{p_{4}}{3},
\label{app-eq:y-z-shift}
\end{equation}
one obtains
\begin{equation}
z^{3}+P_{4}z+Q_{4}=0,
\label{app-eq:depressed-cubic}
\end{equation}
where
\begin{equation}
P_{4}
=
-\frac{p_{4}^{2}}{12}-r_{4},
\qquad
Q_{4}
=
-\frac{p_{4}^{3}}{108}
+
\frac{p_{4}r_{4}}{3}
-
\frac{q_{4}^{2}}{8}.
\label{app-eq:P4-Q4-def}
\end{equation}
Cardano's formula gives a real solution in the form
\begin{equation}
z
=
\left(
-\frac{Q_{4}}{2}
+
\sqrt{
\frac{Q_{4}^{2}}{4}
+
\frac{P_{4}^{3}}{27}
}
\right)^{1/3}
+
\left(
-\frac{Q_{4}}{2}
-
\sqrt{
\frac{Q_{4}^{2}}{4}
+
\frac{P_{4}^{3}}{27}
}
\right)^{1/3}.
\label{app-eq:z-cardano}
\end{equation}
When the two cube-root arguments are real, the real cube roots are
chosen. In the casus irreducibilis, the two arguments are complex
conjugates, and the conjugate cube-root branches are chosen so that
\(z\) is real. The corresponding root of the resolvent cubic is
\begin{equation}
y
=
z-\frac{p_{4}}{3}.
\label{app-eq:y-root}
\end{equation}

For a generic nondegenerate quartic in the stable region, one may choose
the real root of the resolvent cubic for which the factorization below
is nonsingular. Degenerate cases, including the case in which this
intermediate parametrization becomes singular, are obtained by
continuity. With such a choice, Eq.~\eqref{app-eq:ferrari-identity}
factorizes the quartic into two quadratic equations,
\begin{align}
u^{2}
-
\sqrt{2y}\,u
+
\left(
\frac{p_{4}}{2}
+
y
+
\frac{q_{4}}{2\sqrt{2y}}
\right)
&=
0,
\label{app-eq:first-quadratic}
\\[4pt]
u^{2}
+
\sqrt{2y}\,u
+
\left(
\frac{p_{4}}{2}
+
y
-
\frac{q_{4}}{2\sqrt{2y}}
\right)
&=
0.
\label{app-eq:second-quadratic}
\end{align}
Thus the four signed eigenvalues can be written compactly as
\begin{equation}
u_{\eta,\rho}
=
\frac{\eta}{2}\sqrt{2y}
+
\frac{\rho}{2}
\left[
-2(y+p_{4})
-
\eta\,\frac{2q_{4}}{\sqrt{2y}}
\right]^{1/2},
\qquad
\eta,\rho=\pm1 .
\label{app-eq:quartic-roots}
\end{equation}
In the stable region \(\mathcal V>0\), these four roots are real and
split into two positive and two negative eigenvalues. We denote them by
\begin{equation}
u_{1,L},u_{2,L}>0,
\qquad
u_{1,R},u_{2,R}<0.
\label{app-eq:signed-roots}
\end{equation}
The physical propagation speeds are positive quantities,
\begin{equation}
c_{a,L}=u_{a,L},
\qquad
c_{a,R}=-u_{a,R},
\qquad
a=1,2 .
\label{app-eq:physical-speeds}
\end{equation}

The canonical normal-mode basis is obtained from the corresponding
eigenvectors. Let \(\mathbf v_{\rho}\) be an eigenvector of
\(\mathcal A\) with eigenvalue \(u_{\rho}\). Equivalently,
\begin{equation}
(\mathcal A-u_{\rho}I_4)\mathbf v_{\rho}=0.
\label{app-eq:eigenvector-equation}
\end{equation}
For nondegenerate eigenvalues, \(\mathbf v_{\rho}\) may be chosen as any
nonzero column of the adjugate matrix
\(\operatorname{adj}(\mathcal A-u_{\rho}I_4)\). Its normalization is then
fixed by the \(\mathcal K\)-inner product. Indeed, from
\(\mathcal V\mathbf v_{\rho}=u_{\rho}\mathcal K\mathbf v_{\rho}\), one
finds
\begin{equation}
\mathbf v_{\rho}^{T}\mathcal V\mathbf v_{\rho}
=
u_{\rho}\,
\mathbf v_{\rho}^{T}\mathcal K\mathbf v_{\rho}.
\label{app-eq:K-norm-sign}
\end{equation}
Since \(\mathcal V>0\), the sign of
\(\mathbf v_{\rho}^{T}\mathcal K\mathbf v_{\rho}\) is the sign of
\(u_{\rho}\). We therefore choose the eigenvectors so that
\begin{equation}
\mathbf v_{a,L}^{T}\mathcal K\mathbf v_{b,L}
=
\delta_{ab},
\qquad
\mathbf v_{a,R}^{T}\mathcal K\mathbf v_{b,R}
=
-\delta_{ab},
\qquad
\mathbf v_{a,L}^{T}\mathcal K\mathbf v_{b,R}
=
0.
\label{app-eq:K-orthonormality}
\end{equation}
Assembling them as
\begin{equation}
\mathcal M
=
\big(
\mathbf v_{1,L},
\mathbf v_{2,L},
\mathbf v_{1,R},
\mathbf v_{2,R}
\big),
\label{app-eq:M-from-eigenvectors}
\end{equation}
one obtains
\begin{equation}
\mathcal M^{T}\mathcal K\mathcal M
=
\mathcal K,
\label{app-eq:M-preserves-K}
\end{equation}
so that \(\mathcal M\in O(2,2)\). Moreover,
\begin{equation}
\mathcal M^{T}\mathcal V\mathcal M
=
\operatorname{diag}
(c_{1,L},c_{2,L},c_{1,R},c_{2,R}).
\label{app-eq:M-diagonalizes-V}
\end{equation}
This is the canonical diagonalization used in the main text.

As a check, consider the limit in which the mixed \(LR\) couplings are
turned off,
\begin{equation}
J_{(1,L)(1,R)},
\quad
J_{(1,L)(2,R)},
\quad
J_{(2,L)(1,R)},
\quad
J_{(2,L)(2,R)}
\longrightarrow 0.
\label{app-eq:mixed-off-limit}
\end{equation}
The matrix \(\mathcal A\) then separates into a left-moving block and a
right-moving block. The signed left-moving eigenvalues reduce to
\begin{equation}
\{u_{1,L},u_{2,L}\}
\longrightarrow
\left\{
1+\frac{J_{(1,L)(2,L)}}{2\pi},
\;
1-\frac{J_{(1,L)(2,L)}}{2\pi}
\right\}.
\label{app-eq:decoupled-left-limit}
\end{equation}
The positive right-moving speeds are obtained from the negative signed
right-moving eigenvalues,
\begin{equation}
\{-u_{1,R},-u_{2,R}\}
\longrightarrow
\left\{
1+\frac{J_{(1,R)(2,R)}}{2\pi},
\;
1-\frac{J_{(1,R)(2,R)}}{2\pi}
\right\}.
\label{app-eq:decoupled-right-limit}
\end{equation}
Thus the exact quartic solution correctly reproduces the decoupled left-
and right-moving density sectors. The mixed couplings deform these
velocities and rotate the fields by an \(O(2,2)\) canonical
transformation, but the theory remains exactly diagonalizable throughout
the stable region \(\mathcal V>0\).

\section{Primitive logarithmic kernels}
\label{appendix:shape-classification}

This appendix classifies the cubic coordinate kernels that generate
primitive local logarithms in the conformal perturbation theory of
Sec.~\ref{subsec:CPT-third-order}. We restrict throughout to neutral
kernels,
\begin{equation}
m_\alpha=n_\alpha=2,
\label{eq:app-shape-neutrality}
\end{equation}
since only these kernels are compatible with logarithmic radial scaling.
A primitive logarithm is defined as a contribution of the form
\begin{equation}
\left[
\int d^2r\,d^2s\,
K^{(\alpha)}(r,s)
\right]_{\rm prim,log}
=
\zeta_\alpha
\log\frac{L}{\epsilon},
\label{eq:app-primitive-log-def}
\end{equation}
where \(\zeta_\alpha\) is a finite scale-independent coefficient.
Endpoint logarithms in shape space and logarithms produced by
overlapping short-distance regions are not included in
\(\zeta_\alpha\).

After separating the common scale and the overall orientation of the
three-point configuration as in Sec.~\ref{subsec:CPT-third-order}, the
neutral coordinate integral can be written as
\begin{equation}
\int d^2r\,d^2s\,
K^{(\alpha)}(r,s)
=
\frac{1}{(2\pi)^4}
\int_{\mathcal D_\epsilon}
\frac{d\rho}{\rho}\,
d\lambda\,d\varphi\,d\phi\;
\lambda\,
F^{(\alpha)}(\lambda,\phi),
\label{eq:app-neutral-kernel-integral}
\end{equation}
with
\begin{equation}
F^{(\alpha)}(\lambda,\phi)
=
\lambda^{-(b_\alpha+b'_\alpha)}
e^{-i(b_\alpha-b'_\alpha)\phi}
\left(1+\lambda e^{i\phi}\right)^{-c_\alpha}
\left(1+\lambda e^{-i\phi}\right)^{-c'_\alpha}.
\label{eq:app-shape-function}
\end{equation}
The exponents obey
\begin{equation}
a_\alpha+b_\alpha+c_\alpha=2,
\qquad
a'_\alpha+b'_\alpha+c'_\alpha=2.
\label{eq:app-neutrality-expanded}
\end{equation}
The factors involving \(c_\alpha\) and \(c'_\alpha\) are distinguished
because they are attached to the \(X-Z\) separation,
\begin{equation}
z_r+z_s
=
\rho e^{i\varphi}
\left(1+\lambda e^{i\phi}\right),
\label{eq:app-XZ-factor}
\end{equation}
and therefore become singular at the shape-space point
\begin{equation}
\lambda=1,
\qquad
\phi=\pi .
\label{eq:app-XZ-shape-locus}
\end{equation}

The rigid-angle integral gives a factor \(2\pi\), and the remaining
angular integral may be written as a contour integral. Defining
\begin{equation}
I_\alpha(\lambda)
\equiv
\int_0^{2\pi}d\phi\;
\lambda F^{(\alpha)}(\lambda,\phi),
\label{eq:app-Ialpha-def}
\end{equation}
and setting \(w=e^{i\phi}\), one obtains
\begin{equation}
I_\alpha(\lambda)
=
\frac{\lambda^{1-(b_\alpha+b'_\alpha)}}{i}
\oint_{|w|=1}dw\;
w^{-1-(b_\alpha-b'_\alpha)+c'_\alpha}
(1+\lambda w)^{-c_\alpha}
(w+\lambda)^{-c'_\alpha}.
\label{eq:app-master-contour}
\end{equation}
The moving poles associated with the \(X-Z\) denominators are located at
\[
w=-\frac{1}{\lambda},
\qquad
w=-\lambda,
\]
whenever the corresponding exponents are nonzero. Their exchange across
the unit circle at \(\lambda=1\) is the contour representation of the
regulated \(X\to Z\) channel.

The primitive coefficient is obtained when the shape integral has a
finite scale-independent limit,
\begin{equation}
\mathcal I_\alpha
=
\int_0^\infty d\lambda\,
I_\alpha(\lambda),
\label{eq:app-shape-coefficient}
\end{equation}
with no residual dependence on the regulated endpoints. In that case,
\begin{equation}
\zeta_\alpha
=
\frac{1}{8\pi^3}\,
\mathcal I_\alpha .
\label{eq:app-zeta-from-shape-coefficient}
\end{equation}

\subsection{Nonprimitive classes}
\label{app:nonprimitive-classes}

The classification is organized by the pair
\((c_\alpha,c'_\alpha)\). If
\[
c_\alpha=c'_\alpha=0,
\]
the kernel contains no \(X-Z\) denominator. The angular integral reduces
to
\begin{equation}
I_\alpha(\lambda)
=
2\pi\,
\delta_{b_\alpha,b'_\alpha}\,
\lambda^{1-2b_\alpha}.
\label{eq:app-no-XZ-Ialpha}
\end{equation}
The nonzero cases \(b_\alpha=b'_\alpha=b\), with \(b=0,1,2\), give
either power-divergent endpoint contributions or an endpoint logarithm
in \(\lambda\). In particular, the case \(b=1\) produces a logarithm of
the regulated shape endpoint; after the radial integration this becomes
a nested logarithm, not a primitive single logarithm. Thus no kernel with
\(c_\alpha=c'_\alpha=0\) contributes to \(\zeta_\alpha\).

If both \(c_\alpha\) and \(c'_\alpha\) are nonzero, the shape integral is
singular at the \(X-Z\) locus. Locally write
\[
\lambda=1+\delta,
\qquad
\phi=\pi+\theta .
\]
Then
\begin{equation}
|1+\lambda e^{i\phi}|^2
\sim
\delta^2+\theta^2,
\label{eq:app-XZ-local-distance}
\end{equation}
while the pairwise regulator imposes
\begin{equation}
\sqrt{\delta^2+\theta^2}
>
\frac{\epsilon}{\rho}.
\label{eq:app-XZ-local-cutoff}
\end{equation}
The local radial behavior of the shape integral is therefore governed by
\begin{equation}
\int_{\epsilon/\rho}^{O(1)}
dr\;
r^{1-(c_\alpha+c'_\alpha)} .
\label{eq:app-overlap-radial-behavior}
\end{equation}
When \(c_\alpha+c'_\alpha=2\), this produces
\[
\log\frac{\rho}{\epsilon},
\]
and the remaining radial integration gives
\begin{equation}
\int_\epsilon^L
\frac{d\rho}{\rho}
\log\frac{\rho}{\epsilon}
=
\frac12\log^2\frac{L}{\epsilon}.
\label{eq:app-overlap-double-log}
\end{equation}
For \(c_\alpha+c'_\alpha>2\), the shape integral is power divergent in
the cutoff \(\epsilon/\rho\). These contributions are tied to the
regulated \(X-Z\) boundary and do not define finite primitive cubic
coefficients.

It follows that primitive logarithms can arise only from one-sided
\(X-Z\) kernels, for which exactly one of \(c_\alpha\) and
\(c'_\alpha\) is nonzero. This condition is necessary but not sufficient:
within the one-sided class, the remaining exponents
\(b_\alpha,b'_\alpha\) determine whether the contour integral is finite,
vanishes, or gives an endpoint logarithm.

\subsection{One-sided kernels}
\label{app:one-sided-kernels}

We first consider the class
\[
c_\alpha>0,
\qquad
c'_\alpha=0.
\]
The contour integral becomes
\begin{equation}
I_\alpha(\lambda)
=
\frac{\lambda^{1-(b+b')}}{i}
\oint_{|w|=1}dw\;
w^{-1-(b-b')}
(1+\lambda w)^{-c},
\label{eq:app-one-sided-contour}
\end{equation}
where we have suppressed the subscript \(\alpha\) on
\(b_\alpha,b'_\alpha,c_\alpha\). The nine candidates are
\begin{equation}
\begin{aligned}
(a,b,c)
&\in
\{(0,0,2),(1,0,1),(0,1,1)\},
\\
(a',b',c')
&\in
\{(2,0,0),(1,1,0),(0,2,0)\}.
\end{aligned}
\label{eq:app-one-sided-candidates}
\end{equation}
Their contour integrals are summarized in
Table~\ref{tab:app-one-sided-kernels}. Here \(\Theta\) denotes the
Heaviside step function, and only finite endpoint-independent shape
integrals contribute to \(\zeta_\alpha\).

\begin{table}[tbp]
\centering
\small
\renewcommand{\arraystretch}{1.35}
\begin{tabularx}{\textwidth}{
>{\raggedright\arraybackslash}p{3.5cm}
>{\raggedright\arraybackslash}p{4.2cm}
>{\centering\arraybackslash}p{2.5cm}
>{\centering\arraybackslash}X}
\toprule
\textbf{Kernel \(\alpha\)} &
\textbf{\(I_\alpha(\lambda)\)} &
\(\boldsymbol{\mathcal I_\alpha}\) &
\(\boldsymbol{\zeta_\alpha}\) \\
\midrule
\((0,0,2;\,2,0,0)\)
&
\(2\pi\lambda\,\Theta(1-\lambda)\)
&
\(\pi\)
&
\(\dfrac{1}{8\pi^2}\)
\\[4pt]

\((0,0,2;\,1,1,0)\)
&
\(0\)
&
\(0\)
&
\(0\)
\\[4pt]

\((0,0,2;\,0,2,0)\)
&
\(2\pi\lambda^{-3}\,\Theta(\lambda-1)\)
&
\(\pi\)
&
\(\dfrac{1}{8\pi^2}\)
\\[4pt]

\((1,0,1;\,2,0,0)\)
&
\(2\pi\lambda\,\Theta(1-\lambda)\)
&
\(\pi\)
&
\(\dfrac{1}{8\pi^2}\)
\\[4pt]

\((1,0,1;\,1,1,0)\)
&
\(2\pi\lambda^{-1}\,\Theta(\lambda-1)\)
&
endpoint log
&
nonprimitive
\\[4pt]

\((1,0,1;\,0,2,0)\)
&
\(-2\pi\lambda^{-3}\,\Theta(\lambda-1)\)
&
\(-\pi\)
&
\(-\dfrac{1}{8\pi^2}\)
\\[4pt]

\((0,1,1;\,2,0,0)\)
&
\(-2\pi\lambda\,\Theta(1-\lambda)\)
&
\(-\pi\)
&
\(-\dfrac{1}{8\pi^2}\)
\\[4pt]

\((0,1,1;\,1,1,0)\)
&
\(2\pi\lambda^{-1}\,\Theta(1-\lambda)\)
&
endpoint log
&
nonprimitive
\\[4pt]

\((0,1,1;\,0,2,0)\)
&
\(2\pi\lambda^{-3}\,\Theta(\lambda-1)\)
&
\(\pi\)
&
\(\dfrac{1}{8\pi^2}\)
\\
\bottomrule
\end{tabularx}
\caption{
One-sided primitive-kernel classification for
\(c_\alpha>0\), \(c'_\alpha=0\). The finite values of
\(\mathcal I_\alpha=\int d\lambda\,I_\alpha(\lambda)\) determine
\(\zeta_\alpha=\mathcal I_\alpha/(8\pi^3)\). Entries labeled
``endpoint log'' retain explicit dependence on the regulated shape
endpoints and are therefore not primitive.
}
\label{tab:app-one-sided-kernels}
\end{table}

Thus six kernels in the class \(c_\alpha>0\), \(c'_\alpha=0\) generate
primitive logarithmic contributions:
\begin{align}
\zeta_{(0,0,2;\,2,0,0)}
&=
\frac{1}{8\pi^2},
&
\zeta_{(0,0,2;\,0,2,0)}
&=
\frac{1}{8\pi^2},
\nonumber\\
\zeta_{(1,0,1;\,2,0,0)}
&=
\frac{1}{8\pi^2},
&
\zeta_{(1,0,1;\,0,2,0)}
&=
-\frac{1}{8\pi^2},
\nonumber\\
\zeta_{(0,1,1;\,2,0,0)}
&=
-\frac{1}{8\pi^2},
&
\zeta_{(0,1,1;\,0,2,0)}
&=
\frac{1}{8\pi^2}.
\label{eq:app-zeta-one-sided-final}
\end{align}
The kernel \((0,0,2;\,1,1,0)\) vanishes after contour integration,
whereas \((1,0,1;\,1,1,0)\) and \((0,1,1;\,1,1,0)\) give endpoint
logarithms in shape space.

The conjugate one-sided class,
\[
c_\alpha=0,
\qquad
c'_\alpha>0,
\]
is obtained by exchanging primed and unprimed exponents. This maps the
contour problem to its complex conjugate and gives the same real
coefficient pattern for the exchanged kernels. The full neutral
coordinate analysis therefore contains twelve primitive logarithmic
topologies: six in the class \(c_\alpha>0\), \(c'_\alpha=0\), and six in
the conjugate class.

The leading pair retained by the large-\(N\) sector projection in the
main text is
\begin{equation}
\alpha_1=(0,1,1;\,0,2,0),
\qquad
\alpha_2=(0,2,0;\,0,1,1),
\end{equation}
and its primitive coefficients are
\begin{equation}
\zeta_{\alpha_1}
=
\zeta_{\alpha_2}
=
\frac{1}{8\pi^2}.
\label{eq:app-leading-zeta-pair}
\end{equation}
\section{Large-\texorpdfstring{\(N\)}{N} tensor sums for the leading topologies}
\label{appendix:tensor-sums-leading-topologies}

This appendix evaluates the flavor tensor sums associated with the two
primitive logarithmic topologies that survive the large-\(N\) selection
in the cubic sector flow. The coordinate dependence is contained in the
kernels \(K^{(\alpha)}(r,s)\), including the propagator normalization.
The tensor \(\mathcal T^{(\alpha)}\) denotes the corresponding algebraic
structure: flavor contractions, chirality assignments, fermionic signs,
and combinatorial factors.

For a fixed topology \(\alpha\),
\(\mathcal T_{BCD}^{(\alpha)\,A}\) is obtained by performing the Wick
contractions prescribed by \(\alpha\) in the full ordered Majorana
product
\[
\mathcal O_B(Y+r)\,
\mathcal O_C(Y)\,
\mathcal O_D(Y-s),
\]
and then projecting the remaining fields onto the reference local
quartic monomial \(\mathcal O_A(Y)\). Thus all signs associated with the
topology are part of \(\mathcal T^{(\alpha)}\) before the subsequent
large-\(N\) flavor counting is performed.

The relevant disorder-averaged contributions have representative form
\(\mathcal T_{ABB}^{(\alpha)\,A}\), where \(A\) belongs to the final
sector \(\mathcal S\) and \(B\) belongs to the chirality-balanced sector
\(\{2L2R\}\). The two primitive topologies that survive the graphical
large-\(N\) selection are
\begin{equation}
\alpha_1=(0,1,1;\,0,2,0),
\qquad
\alpha_2=(0,2,0;\,0,1,1).
\label{eq:app-leading-alpha-def}
\end{equation}
The first triple records the left-moving contraction data, while the
second records the right-moving data.

The flavor contractions factorize into left- and right-moving blocks
after the topology and its fermionic sign have been fixed. This
factorization does not introduce an additional interchiral sign. Indeed,
if the external quartic sector contains \(k\) left-moving fields and
\(4-k\) right-moving fields, the ordered product has the schematic form
\begin{equation}
X_L^kX_R^{4-k}\,
Y_L^2Y_R^2\,
Z_L^2Z_R^2 .
\label{eq:app-interchiral-order}
\end{equation}
Grouping the fields by chirality requires
\begin{equation}
(4-k)(2+2)+2\cdot2
=
4(5-k)
\label{eq:app-interchiral-sign-count}
\end{equation}
interchiral transpositions, which is always even. Hence the block
factorization contributes no further sign.

\subsection{Canonical and redundant sector sums}

The sector sums entering the beta functions are canonical sums over
independent antisymmetric sector components,
\begin{equation}
\Sigma_{\mathcal S}^{(r)}
\equiv
\sum_{A\in\mathcal S}^{\rm can}
\sum_{B\in\{2L2R\}}^{\rm can}
\mathcal T_{ABB}^{(\alpha_r)\,A}.
\label{eq:app-canonical-Sigma-def}
\end{equation}
To extract the leading large-\(N\) coefficient, it is convenient to
evaluate a redundant unordered sum
\(\widetilde\Sigma_{\mathcal S}^{(r)}\), in which all flavor labels are
temporarily summed independently. On configurations with distinct labels
inside each antisymmetric block, each canonical component is counted
\[
\mathfrak m_{\mathcal S}\,
\mathfrak m_{\{2L2R\}}
\]
times. The relevant ordering multiplicities are
\begin{equation}
\mathfrak m_{\{4L\}}=\mathfrak m_{\{4R\}}=4!,
\qquad
\mathfrak m_{\{3L1R\}}=\mathfrak m_{\{1L3R\}}=3!,
\qquad
\mathfrak m_{\{2L2R\}}=2!\,2!.
\label{eq:app-ordering-multiplicities}
\end{equation}
When the external sector is itself \(\{2L2R\}\), the denominator is
\(\mathfrak m_{\{2L2R\}}^2\). Contributions supported on flavor-space
diagonals contain fewer independent flavor sums and are subleading.
Thus, for the leading \(N^7\) coefficient,
\begin{equation}
\Sigma_{\mathcal S}^{(r)}
=
\frac{
\widetilde\Sigma_{\mathcal S}^{(r)}
}{
\mathfrak m_{\mathcal S}\,
\mathfrak m_{\{2L2R\}}
}
+O(N^6),
\label{eq:app-canonical-from-unordered}
\end{equation}
with the obvious replacement by
\(\mathfrak m_{\{2L2R\}}^2\) when
\(\mathcal S=\{2L2R\}\).

\subsection{Universal chiral blocks}

Only two elementary chiral blocks are needed. The first is the
\((0,2,0)\) block. If the two positions at the \(Y\)- and \(Z\)-vertices
are denoted by
\[
Y:\ (u_1,u_2),
\qquad
Z:\ (v_1,v_2),
\]
the antisymmetric contraction is
\begin{equation}
\Omega_{YZ}
=
\delta^{u_1v_2}\delta^{u_2v_1}
-
\delta^{u_1v_1}\delta^{u_2v_2}.
\label{eq:app-Omega-YZ}
\end{equation}
After the covariance identifies \(v_1=u_1\) and \(v_2=u_2\), the
redundant unordered sum gives
\begin{equation}
\Omega
=
\sum_{u_1,u_2}
\left(
\delta^{u_1u_2}-1
\right)
=
N-N^2
=
-N(N-1).
\label{eq:app-Omega-result}
\end{equation}

The second block is the \((0,1,1)\) block. Suppose it acts on \(k\)
external positions
\[
X:\ (x_1,\ldots,x_k),
\]
and let
\[
Y:\ (a_1,a_2),
\qquad
Z:\ (b_1,b_2)
\]
be the two positions at the balanced vertices in the same chirality. If
\(x_m\) is the external position contracted with \(Z\), the four
elementary contractions are
\begin{align}
\Gamma_m^{\,a_\star}
&=
\delta^{x_m b_1}\delta^{a_1 b_2}\delta^{a_2 a_\star}
-\delta^{x_m b_2}\delta^{a_1 b_1}\delta^{a_2 a_\star}
\nonumber\\
&\quad
-\delta^{x_m b_1}\delta^{a_2 b_2}\delta^{a_1 a_\star}
+\delta^{x_m b_2}\delta^{a_2 b_1}\delta^{a_1 a_\star}.
\label{eq:app-Gamma-block}
\end{align}
The surviving \(Y\)-position \(a_\star\) must be restored to the
canonical external order. With the conventions used in the main text,
this projection gives
\begin{equation}
\sigma_m^{(k)}
=
(-1)^{k+3-m}.
\label{eq:app-sigma-km}
\end{equation}
After the covariance identification \(b_1=a_1\), \(b_2=a_2\), and after
imposing the projection constraints
\[
x_m=x_{m+1}=\cdots=x_k=a_\star,
\]
the four terms in Eq.~\eqref{eq:app-Gamma-block} contribute
\[
+N^m,\qquad -N^{m+1},\qquad -N^{m+1},\qquad +N^m .
\]
Thus the selected position contributes
\begin{equation}
\Xi_{k,m}
=
-2\,\sigma_m^{(k)}\,N^m(N-1).
\label{eq:app-Xi-km}
\end{equation}
Summing over \(m=1,\ldots,k\), one obtains the universal result
\begin{equation}
\Xi_k
=
2N(N-1)\frac{N^k-(-1)^k}{N+1}.
\label{eq:app-Xi-k}
\end{equation}
The values needed below are
\begin{align}
\Xi_4
&=
2N(N-1)^2(N^2+1),
&
\Xi_3
&=
2N(N-1)(N^2-N+1),
\nonumber\\
\Xi_2
&=
2N(N-1)^2,
&
\Xi_1
&=
2N(N-1).
\label{eq:app-Xi-values}
\end{align}

\subsection{Sector sums}

The leading tensor sums now follow by combining the two universal
blocks. Spectator external fields contribute one free flavor sum per
spectator. The results are collected in
Table~\ref{tab:app-leading-sector-sums}. The table displays the
redundant unordered tensor structure and the corresponding canonical
large-\(N\) sum.

\begin{table}[tbp]
\centering
\small
\renewcommand{\arraystretch}{1.35}
\begin{tabularx}{\textwidth}{
>{\raggedright\arraybackslash}p{2.5cm}
>{\centering\arraybackslash}p{1.7cm}
>{\centering\arraybackslash}p{3.4cm}
>{\centering\arraybackslash}X}
\toprule
\textbf{Final sector} &
\textbf{Topology} &
\(\boldsymbol{\widetilde\Sigma_{\mathcal S}^{(r)}}\) &
\(\boldsymbol{\Sigma_{\mathcal S}^{(r)}}\) \\
\midrule
\(\{4L\}\)
&
\(\alpha_1\)
&
\(\Omega\,\Xi_4\)
&
\(-\dfrac{1}{48}N^7+O(N^6)\)
\\[4pt]

\(\{4R\}\)
&
\(\alpha_2\)
&
\(\Omega\,\Xi_4\)
&
\(-\dfrac{1}{48}N^7+O(N^6)\)
\\[4pt]

\(\{3L1R\}\)
&
\(\alpha_1\)
&
\(\Omega\,N\,\Xi_3\)
&
\(-\dfrac{1}{12}N^7+O(N^6)\)
\\[4pt]

\(\{3L1R\}\)
&
\(\alpha_2\)
&
\(\Omega\,N^3\,\Xi_1\)
&
\(-\dfrac{1}{12}N^7+O(N^6)\)
\\[4pt]

\(\{1L3R\}\)
&
\(\alpha_1\)
&
\(\Omega\,N^3\,\Xi_1\)
&
\(-\dfrac{1}{12}N^7+O(N^6)\)
\\[4pt]

\(\{1L3R\}\)
&
\(\alpha_2\)
&
\(\Omega\,N\,\Xi_3\)
&
\(-\dfrac{1}{12}N^7+O(N^6)\)
\\[4pt]

\(\{2L2R\}\)
&
\(\alpha_1\)
&
\(\Omega\,N^2\,\Xi_2\)
&
\(-\dfrac{1}{8}N^7+O(N^6)\)
\\[4pt]

\(\{2L2R\}\)
&
\(\alpha_2\)
&
\(\Omega\,N^2\,\Xi_2\)
&
\(-\dfrac{1}{8}N^7+O(N^6)\)
\\
\bottomrule
\end{tabularx}
\caption{
Leading tensor sums for the two primitive topologies
\(\alpha_1=(0,1,1;\,0,2,0)\) and
\(\alpha_2=(0,2,0;\,0,1,1)\). The canonical sums are obtained from the
redundant unordered structures by dividing by the ordering
multiplicities in Eq.~\eqref{eq:app-ordering-multiplicities}. Only the
leading \(N^7\) coefficients are retained.
}
\label{tab:app-leading-sector-sums}
\end{table}

For completeness, let us spell out one representative entry. In the
\(\{4L\}\) sector, only \(\alpha_1\) contributes at leading order. The
left-moving block is \(\Xi_4\), while the right-moving block is
\(\Omega\), so
\begin{align}
\widetilde\Sigma_{\{4L\}}^{(1)}
&=
\Omega\,\Xi_4
\nonumber\\
&=
\left[-N(N-1)\right]
\left[2N(N-1)^2(N^2+1)\right]
\nonumber\\
&=
-2N^2(N-1)^3(N^2+1).
\label{eq:app-4L-representative}
\end{align}
Dividing by
\[
\mathfrak m_{\{4L\}}\mathfrak m_{\{2L2R\}}
=
4!(2!\,2!)
=
96
\]
gives
\[
\Sigma_{\{4L\}}^{(1)}
=
-\frac{1}{48}N^7+O(N^6).
\]
All other entries in Table~\ref{tab:app-leading-sector-sums} follow
from the same two blocks, with the indicated spectator factors.

\subsection{Normalization entering the sector beta functions}

The canonical sums do not enter the sector beta functions alone. They
are multiplied by the sector combinatorial factors
\begin{equation}
\kappa_{\{4L\}}=\kappa_{\{4R\}}=4!,
\qquad
\kappa_{\{3L1R\}}=\kappa_{\{1L3R\}}=3!,
\qquad
\kappa_{\{2L2R\}}=4.
\label{eq:app-kappa-values}
\end{equation}
Combining Eq.~\eqref{eq:app-kappa-values} with
Table~\ref{tab:app-leading-sector-sums}, one finds the common
large-\(N\) normalization
\begin{equation}
\kappa_{\mathcal S}
\Sigma_{\mathcal S}^{(r)}
=
-\frac12 N^7+O(N^6)
\label{eq:app-universal-kappa-Sigma}
\end{equation}
for every sector-compatible leading topology. This identity is the
tensor-sum origin of the uniform structure of the leading third-order
sector beta functions.

Collecting the canonical sums explicitly,
\begin{align}
\Sigma_{\{4L\}}^{(1)}
=
\Sigma_{\{4R\}}^{(2)}
&=
-\frac{1}{48}N^7+O(N^6),
\label{eq:app-summary-pure}
\\[4pt]
\Sigma_{\{3L1R\}}^{(1)}
=
\Sigma_{\{3L1R\}}^{(2)}
=
\Sigma_{\{1L3R\}}^{(1)}
=
\Sigma_{\{1L3R\}}^{(2)}
&=
-\frac{1}{12}N^7+O(N^6),
\label{eq:app-summary-imbalanced}
\\[4pt]
\Sigma_{\{2L2R\}}^{(1)}
=
\Sigma_{\{2L2R\}}^{(2)}
&=
-\frac{1}{8}N^7+O(N^6).
\label{eq:app-summary-balanced}
\end{align}
Only these \(N^7\) coefficients survive the overall \(N^{-7}\)
normalization in the strict large-\(N\) sector beta functions.
\section{Anomalous dimensions and disorder-strength beta functions}
\label{appendix:Anomalous-dimensions-and-disorder-strength-beta-functions}

For the third-order flow in section \ref{sec:RG-flow} we see the simple feature that for $J_{2L2R}\neq0$, every
sector beta function is proportional to \(J_{2L2R}^{2}\). This may tempt the reader to conclude that since every marginal deformation in the
quartic helical interaction space is built from four Majorana fields, one might interpret the effect of \(J_{2L2R}\) as a universal
external-leg dressing of the quartic monomials. If a monomial in sector
\(\mathcal S\) contains \(n_L\) left-moving and \(n_R\) right-moving
fermions, this reasoning would assign it the anomalous dimension
\begin{equation}
\gamma_{\mathcal O_{\mathcal S}}^{\rm ext}
=
n_L\gamma_\psi^L+n_R\gamma_\psi^R .
\label{eq:external-leg-gamma-naive}
\end{equation}
For an ensemble invariant under \(L\leftrightarrow R\), one has
\(\gamma_\psi^L=\gamma_\psi^R\equiv\gamma_\psi\), and since
\(n_L+n_R=4\) for every quartic sector, Eq.~\eqref{eq:external-leg-gamma-naive}
would give
\begin{equation}
\gamma_{\mathcal O_{\mathcal S}}^{\rm ext}
=
4\gamma_\psi ,
\label{eq:external-leg-gamma-four}
\end{equation}
independently of the chirality content. If this were the full
renormalization of the quartic deformation, one would then expect a universal
sector-independent coefficient $C$ for the beta-functions given by
\begin{equation}
\beta_{\mathcal S}^{(3)}
\equiv (\Delta_{\mathcal O}-2)J=
C\,J_{\mathcal S}J_{2L2R}^{2},
\label{eq:naive-sector-universal-beta}
\end{equation}
where $\Delta_\mathcal{O}$ is the scaling dimension of the operator $\mathcal{O}\sim\psi\psi\psi\psi$.
We want to clarify why this inference is not
valid for the sector beta functions computed here, and to identify
precisely what is the problem.

The form
\begin{equation}
\beta_J=(\Delta_{\mathcal O}-2)J
\label{eq:single-source-linear-beta}
\end{equation}
has a well-defined meaning in a different problem. It applies to a
two-dimensional CFT deformed by a source \(J\) for a \textbf{single} scaling
operator,
\begin{equation}
S
=
S_{\rm CFT}
+
J\int d^2x\,\mathcal O(x),
\label{eq:single-source-deformation}
\end{equation}
provided \(\mathcal O\) is an eigenoperator of the anomalous-dimension
matrix. In that setting, the linearized beta function near the fixed
point is controlled by the scaling dimension of the operator. More
generally, if one perturbs a theory by a small source \(J\) in the
presence of another coupling \(g\), then the linearized flow of \(J\)
can be written in the form
\begin{equation}
\beta_J
=
\left[\Delta_{\mathcal O}(g)-2\right]J
+
O(J^2),
\label{eq:linearized-source-beta-background}
\end{equation}
again only after \(\mathcal O\) has been chosen as an eigenoperator in
the background specified by \(g\). Equation~\eqref{eq:linearized-source-beta-background}
is a statement about the source of an operator insertion. It is not, by
itself, a statement about the width of a Gaussian disorder ensemble.

Such a distinction is essential in our work. The variable
\(J_{\mathcal S}\) is not introduced as the source of a single normalized
operator \(\mathcal O_{\mathcal S}\). It is the positive disorder
strength associated with many marginal couplings. By
definition,
\begin{equation}
\sum_{A\in\mathcal S}
\overline{J_A(\mu)^2}
=
d_{\mathcal S}C_{\mathcal S}
J_{\mathcal S}^2(\mu).
\label{eq:sector-strength-repeat}
\end{equation}
Thus \(J_{\mathcal S}\) is a radial statistical coordinate in the space
of random couplings \(J_A\), not a linear source multiplying one
operator. Its beta function is obtained by projecting the microscopic
running onto the second moment,
\begin{equation}
\beta_{\mathcal S}
=
\frac{1}{d_{\mathcal S}C_{\mathcal S}J_{\mathcal S}}
\sum_{A\in\mathcal S}
\overline{J_A\hat\beta_A}.
\label{eq:sector-projection-repeat}
\end{equation}
This projection is the point at which the source-renormalization
argument and the disorder-strength problem differ.

To see this difference without using any sector-specific result, suppose
that the microscopic third-order running generated by the
chirality-balanced disorder strength is linear in the final-sector
couplings. Then, within a fixed sector \(\mathcal S\), it has the
general form
\begin{equation}
\hat\beta_A^{(3)}
=
J_{2L2R}^{2}
\sum_{B\in\mathcal S}
M_{AB}^{(\mathcal S)}\,J_B,
\qquad
A\in\mathcal S .
\label{eq:general-linearized-micro-flow}
\end{equation}
The matrix \(M_{AB}^{(\mathcal S)}\) is the linearized RG kernel in the
sector \(\mathcal S\). It contains all local counterterm contributions
that map the sector back to itself in the background of the
chirality-balanced disorder. Substituting
Eq.~\eqref{eq:general-linearized-micro-flow} into
Eq.~\eqref{eq:sector-projection-repeat} gives
\begin{equation}
\beta_{\mathcal S}^{(3)}
=
\frac{J_{2L2R}^{2}}{d_{\mathcal S}C_{\mathcal S}J_{\mathcal S}}
\sum_{A,B\in\mathcal S}
\overline{J_AJ_B}\,
M_{AB}^{(\mathcal S)}.
\label{eq:sector-beta-general-M}
\end{equation}
For a sector-diagonal Gaussian ensemble, this is the covariance-weighted
trace of the linearized microscopic RG kernel. In particular, if the
covariance in the canonical independent coupling basis is written as
\begin{equation}
\overline{J_AJ_B}
=
J_{\mathcal S}^{2}\,
\mathcal C_{AB}^{(\mathcal S)},
\qquad
\sum_{A\in\mathcal S}
\mathcal C_{AA}^{(\mathcal S)}
=
d_{\mathcal S}C_{\mathcal S},
\label{eq:sector-covariance-metric}
\end{equation}
then
\begin{equation}
\beta_{\mathcal S}^{(3)}
=
J_{\mathcal S}J_{2L2R}^{2}\,
\frac{1}{d_{\mathcal S}C_{\mathcal S}}
\sum_{A,B\in\mathcal S}
\mathcal C_{AB}^{(\mathcal S)}
M_{AB}^{(\mathcal S)}.
\label{eq:sector-beta-trace-M}
\end{equation}
Therefore a universal coefficient would require
\begin{equation}
\frac{1}{d_{\mathcal S}C_{\mathcal S}}
\sum_{A,B\in\mathcal S}
\mathcal C_{AB}^{(\mathcal S)}
M_{AB}^{(\mathcal S)}
=
C
\label{eq:condition-sector-universal-M}
\end{equation}
with the same constant \(C\) for all five chirality sectors. This is the
actual condition behind Eq.~\eqref{eq:naive-sector-universal-beta}. It
is much stronger than the statement \(n_L+n_R=4\).

The external-leg argument corresponds to a very special case of
Eq.~\eqref{eq:general-linearized-micro-flow}. It assumes that the
linearized RG kernel is purely radial in each sector,
\begin{equation}
M_{AB}^{(\mathcal S)}
=
C_{\rm ext}\,\delta_{AB},
\label{eq:pure-radial-kernel}
\end{equation}
with the same \(C_{\rm ext}\) for all sectors. In that case,
Eq.~\eqref{eq:sector-beta-trace-M} indeed gives
\begin{equation}
\beta_{\mathcal S}^{(3)}
=
C_{\rm ext}\,
J_{\mathcal S}J_{2L2R}^{2}.
\label{eq:radial-flow-sector-beta}
\end{equation}
Thus there is a precise limit in which the disorder strength \(J_{\mathcal S}\) may be identified with
the radial amplitude of the microscopic coupling vector, and the
microscopic RG flow must be radial with a sector-independent eigenvalue.
In operator language, this is equivalent to assuming that the quartic
operator renormalizes only through the renormalization of
its four external fermion legs,
\begin{equation}
Z_{\mathcal O_{\mathcal S}}
=
Z_\psi^{\,4},
\label{eq:Z-O-equals-Zpsi4}
\end{equation}
or, equivalently,
\begin{equation}
\gamma_{\mathcal O_{\mathcal S}}
=
n_L\gamma_\psi^L+n_R\gamma_\psi^R .
\label{eq:gamma-O-only-external}
\end{equation}

However, Eq.~\eqref{eq:gamma-O-only-external} is not a general identity
for composite operators. A quartic Majorana monomial is a local
composite operator, and local composite operators can acquire anomalous
dimensions from contractions internal to the composite insertion and
from mixing with other operators of the same classical dimension. In a
basis of quartic monomials, the general structure is matrix-valued:
\begin{equation}
\gamma_{A}^{\ B}
=
\gamma_{A}^{\ B}\big|_{\rm ext}
+
\gamma_{A}^{\ B}\big|_{\rm conn/mix}.
\label{eq:gamma-matrix-split}
\end{equation}
The external-leg part is diagonal and depends only on the number of left
and right fields,
\begin{equation}
\gamma_{A}^{\ B}\big|_{\rm ext}
=
\left(
n_L(A)\gamma_\psi^L+n_R(A)\gamma_\psi^R
\right)\delta_A^{\ B}.
\label{eq:external-gamma-matrix}
\end{equation}
The connected and mixing part,
\(\gamma_{A}^{\ B}|_{\rm conn/mix}\), is not fixed by
\(n_L+n_R\). It depends on the local short-distance OPE of the
interaction insertions, on fermionic signs, on the flavor contractions,
and on the projection back to the quartic basis. Therefore the equality
\(\gamma_{\mathcal O_{\mathcal S}}=4\gamma_\psi\) is not a consequence
of the fact that the monomial has four fermions. It is an additional
assumption that the connected and mixing part vanishes, or at least has
a sector-independent covariance-weighted trace. No symmetry of the
helical quartic interaction space enforces this condition across all
five sectors.

The statistical \(L\leftrightarrow R\) symmetry is also not strong
enough to imply sector universality. It relates the purely chiral pair
\begin{equation}
\{4L\}\longleftrightarrow\{4R\},
\label{eq:LR-pure-relation}
\end{equation}
and the imbalanced pair
\begin{equation}
\{3L1R\}\longleftrightarrow\{1L3R\}.
\label{eq:LR-imbalanced-relation}
\end{equation}
It does not relate \(\{4L\}\) to \(\{3L1R\}\), nor either of these to
\(\{2L2R\}\). Hence \(L\leftrightarrow R\) symmetry might enforce
\begin{equation}
C_{4L}=C_{4R},
\qquad
C_{3L1R}=C_{1L3R},
\label{eq:LR-allowed-equalities}
\end{equation}
but it cannot enforce a single coefficient common to all sectors.

The CPT computation performed in section \ref{sec:RG-flow} evaluates
precisely the part discarded by the assumptions described in the previous paragraphs. At third
order, the microscopic beta function is generated by the local
logarithmic projection of three quartic insertions,
\begin{equation}
\left[
\frac{1}{3!}S_{\rm int}^{3}
\right]_{\rm local,log}.
\label{eq:third-order-local-log-repeat}
\end{equation}
This projection contains coordinate kernels, primitive shape-space
coefficients, and algebraic contraction tensors. After the disorder
average, the sector flow is not determined by a two-point wavefunction
renormalization factor, but by the fourth moment
\begin{equation}
\overline{J_AJ_BJ_CJ_D},
\label{eq:fourth-moment-repeat}
\end{equation}
together with the tensor structure that maps the ordered triple of
insertions back to the final monomial. Equivalently, the sector beta
function takes the form
\begin{equation}
\beta_{\mathcal S}^{(3)}
=
-\frac{1}{3!\,d_{\mathcal S}C_{\mathcal S}J_{\mathcal S}}
\sum_{A\in\mathcal S}
\sum_{B,C,D}
\sum_{\alpha\in\mathcal P_{BCD}^{A}}
\zeta_\alpha\,
\mathcal T_{BCD}^{(\alpha)\,A}
\,
\overline{J_AJ_BJ_CJ_D}.
\label{eq:sector-beta-fourpoint-repeat}
\end{equation}
This expression is the precise replacement for the single-source
formula \(\beta_J=(\Delta_{\mathcal O}-2)J\) in the present
disorder-averaged problem. The coefficient of
\(J_{\mathcal S}J_{2L2R}^{2}\) is obtained only after the fourth moment
has been factorized, the sector-compatible primitive kernels have been
selected, and the large-\(N\) flavor sums have been evaluated. The leading primitive coordinate integral has a universal
geometric weight, but the complete sector beta-function coefficient is
not universal. The full coefficient is a covariance-weighted,
sector-projected trace of the microscopic RG kernel, and that trace
depends on the chirality sector simply combinatorially through contributing topologies.

This also clarifies the relation to the anomalous dimension of the
fermion. The fermion anomalous dimensions
\(\gamma_\psi^L\) and \(\gamma_\psi^R\) are genuine RG data, but they
govern a different object. They control correlation functions with
external fermionic insertions. For example, the disorder-averaged
fermion two-point function satisfies the Callan--Symanzik equation of the
form
\begin{equation}
\left[
\mu\frac{\partial}{\partial\mu}
+
\sum_{\mathcal S}
\beta_{\mathcal S}
\frac{\partial}{\partial J_{\mathcal S}}
+
2\gamma_\nu
\right]
G_\nu(x;\{J_{\mathcal S}\},\mu)
=
0,
\qquad
\nu=L,R .
\label{eq:CS-fermion-general}
\end{equation}
The beta functions and the fermion anomalous dimensions therefore enter
the same RG equation, but they are independent RG functions. The former
describe the scale dependence of the disorder strengths that define the
ensemble of quartic couplings. The latter describe the scaling of
correlators with external fermionic fields.

In a one-parameter statistically symmetric random model, these two
pieces of RG data can sometimes be extracted from the same two-point
function analysis, because there is only one disorder strength and one
possible beta function. That special situation should not be confused
with the present multi-sector problem. Here the disorder ensemble has
several independent sector strengths, $J_{4L}, J_{3L1R}, J_{2L2R}, J_{1L3R}, J_{4R}$.
The single anomalous dimension of an external fermion cannot determine
the independent covariance-weighted traces of the microscopic RG kernel
in all these sectors. At most, external-leg dressing supplies a
universal diagonal contribution to the composite-operator
renormalization. It does not determine the connected local logarithms,
operator mixing, Gaussian pairings, or large-\(N\) flavor multiplicities
that enter Eq.~\eqref{eq:sector-beta-fourpoint-repeat}.

Therefore one may identify the
symbol \(J\) in a single-source anomalous-dimension argument with a
sector disorder strength \(J_{\mathcal S}\) only if the microscopic flow
inside the sector is radial and has a sector-independent eigenvalue.
This is equivalent to assuming that the quartic monomials renormalize
only by external-leg dressings. That assumption is not a
consequence of \(n_L+n_R=4\), nor of \(L\leftrightarrow R\) symmetry.
The CPT calculation shows that the leading
third-order sector flow is controlled by primitive local logarithms and
their sector-projected tensor sums. Thus, the fermion anomalous dimension remains an important part of the RG data, but it does not fix
the sector beta-function coefficients.

\bibliographystyle{JHEP}
\bibliography{biblio}

\end{document}